\documentclass{lmcs}
\pdfoutput=1
\usepackage[utf8]{inputenc}

\usepackage{lastpage}
\lmcsdoi{18}{1}{10}
\lmcsheading{}{\pageref{LastPage}}{}{}%
{Feb.~03,~2020}{Jan.~14,~2022}{}

\usepackage{amsmath}
\usepackage{cases}
\usepackage{amsthm}
\usepackage{amssymb}
\usepackage{amsfonts}
\usepackage{mathtools}
\usepackage{stmaryrd}
\usepackage{tikz}
\usepackage{xcolor}
\usepackage{soul}
\usepackage{graphicx}
\usepackage[shortlabels]{enumitem}
\usepackage{caption}
\usepackage{subcaption}
\usepackage{cmll}
\usepackage{enumerate}
\usepackage{bussproofs}
\usepackage{tabulary}
\usepackage{algorithm}
\usepackage{algorithmicx}
\usepackage{algpseudocode}
\usepackage{mdframed}
\usepackage{float}
\usepackage[hyperfootnotes=false]{hyperref}
\setlist[itemize]{leftmargin=13pt}

\newcommand{\N}[0]{\mathbb{N}}
\newcommand{\cC}[0]{\mathcal{C}}

\DeclareMathOperator{\Ob}{Ob}

\usetikzlibrary{calc,patterns,snakes,arrows,decorations.markings,shapes.misc,arrows.meta}

\def\R{\ensuremath{\mathrm{R}}}
\def\L{\ensuremath{\mathrm{L}}}
\def\Rarrow{{\, \rightarrow_\R \,}}
\def\Larrow{{\, \rightarrow_\L \,}}
\def\Rarrowstar{\, \rightarrow_\R^* \,}

\tikzstyle{morph}=[rectangle,draw]
\tikzstyle{wmorph}=[morph,minimum width=.7cm]
\newcommand\wirebundle[2]{
\draw ($(#1)+(-.3,0)$) -- ($(#2)+(-.3,0)$);
\draw ($(#1)+(.3,0)$) -- ($(#2)+(.3,0)$);
\node at ($(#1)!0.5!(#2)$) {$\dots$};}

\usepackage{ifthen}

\newcommand\storelabel[2]{\expandafter\xdef\csname label#1\endcsname{#2}}
\newcommand\getlabel[1]{\csname label#1\endcsname}

\newcommand\storecolor[2]{\expandafter\xdef\csname strandcolor#1\endcsname{#2}}
\newcommand\getcolor[1]{\csname strandcolor#1\endcsname}
\newcommand{\maybestorecolor}[2]{%
  \ifcsname strandcolor#1\endcsname%
  \else%
        \storecolor{#1}{#2}
  \fi%
}

\tikzset{
  zbraid/.style = {
    line width = .6,
    preaction={ shorten <= 0.5cm, shorten >= 0.5cm, draw, white, line width=8, shorten <= 0, shorten >= 0}
  }
}

\newcommand\abraid[4]{
    \begin{scope}[shift={(#2,#1)}]
      \draw[line width=.6,#3] (0,0) -- (.3,.3);
      \draw[line width=.6,#3] (.7,.7) -- (1,1);
      \draw[line width=.6,#4] (0,1) -- (1,0);
    \end{scope}
}

\newcommand\ainit[1]{
  \foreach \y in {0,...,#1}{
    \storelabel{\y}{0}
    \maybestorecolor{\y}{black}
    \draw[zbraid,\getcolor{\y}] (-.5,\y) -- (0,\y);
  }
  \xdef\nbbread{#1}
  \xdef\dmax{0}
}

\newcommand\aend{
  \foreach \y in {0,...,\nbbread}{
    \draw[zbraid,\getcolor{\y}] (\getlabel{\y},\y) -- (\dmax+.5,\y);
    \storecolor{\y}{black}
  }
}

\newcommand\aswap[1]{
  \pgfmathtruncatemacro\next{#1+1}
  \pgfmathsetmacro\dum{\dmax}
  \draw[zbraid,\getcolor{#1}] (\getlabel{#1},#1) -- (\dum,#1);
  \draw[zbraid,\getcolor{\next}] (\getlabel{\next},\next) -- (\dum,\next);

  \abraid{#1}{\dum}{\getcolor{#1}}{\getcolor{\next}}
  \pgfmathsetmacro\dum{\dum+1}
  \storelabel{#1}{\dum}
  \storelabel{\next}{\dum}
  \pgfmathsetmacro\dum{max(\dmax,\dum)}
  \xdef\dmax{\dum}

  \xdef\tmpcolor{\getcolor{#1}}
  \storecolor{#1}{\getcolor{\next}}
  \storecolor{\next}{\tmpcolor}

}

\def\swap#1#2;{
  \aswap{#1}\def\dum{#2}
  \ifx\dum\empty\else\swap#2;\fi
}

\usepackage{tikz}
\usepackage{ifthen}
\usetikzlibrary{calc,patterns,snakes,arrows,decorations.markings,shapes.misc}

\def\defaultfacecolor{green}

\tikzset{cross/.style={cross out, draw, 
         minimum size=2*(#1-\pgflinewidth), 
         inner sep=0pt, outer sep=0pt}}
\tikzstyle{bnode}=[draw,black,circle,fill=black,inner sep=1pt]
\tikzstyle{bplace}=[draw,blue,thick,cross,inner sep=2pt]
\tikzstyle{yspot}=[draw,\defaultfacecolor,circle,fill=\defaultfacecolor,inner sep=1pt]
\tikzstyle{eface}=[draw,\defaultfacecolor]
\tikzstyle{rnode}=[draw,red,circle,fill=red,inner sep=1pt]
\tikzstyle{dedge}=[postaction={nomorepostaction,decorate,
                    decoration={markings,mark=at position 0.5 with {\arrow{>}}}
                   }]
\makeatletter
\tikzset{nomorepostaction/.code={\let\tikz@postactions\pgfutil@empty}}
\makeatother

\newcommand\storeface[2]{\expandafter\xdef\csname faceid#1\endcsname{#2}}
\newcommand\getface[1]{\csname faceid#1\endcsname}
\newcounter{nextfaceid}
\newcounter{maxfaceseen}

\newcommand\storefinalface[2]{\expandafter\xdef\csname finalfaceid#1\endcsname{#2}}
\newcommand\getfinalface[1]{\csname finalfaceid#1\endcsname}

\newcommand\storefacecolor[2]{\expandafter\xdef\csname facecolor#1\endcsname{#2}}
\newcommand\getfacecolor[1]{\csname facecolor#1\endcsname}

\newcommand\storeminslice[2]{\expandafter\xdef\csname minfaceslice#1\endcsname{#2}}
\newcommand\getminslice[1]{\csname minfaceslice#1\endcsname}

\newcommand\storeedgeid[2]{\expandafter\xdef\csname edgeid#1\endcsname{#2}}
\newcommand\getedgeid[1]{\csname edgeid#1\endcsname}
\newcounter{nextedgeid}

\newcommand\storeedgepath[2]{\expandafter\xdef\csname edgepath#1\endcsname{#2}}
\newcommand\getedgepath[1]{\csname edgepath#1\endcsname}
\newcommand\expandedgepath[2]{\storeedgepath{\getedgeid{#1}}{\getedgepath{\getedgeid{#1}} #2}}

\newcommand\storeedgecolor[2]{\expandafter\xdef\csname edgecolor#1\endcsname{#2}}
\newcommand\getedgecolor[1]{\csname edgecolor#1\endcsname}

\newcommand\mergefaces[2]{
   \pgfmathtruncatemacro\valfa{\getfinalface{#1}}
   \pgfmathtruncatemacro\valfb{\getfinalface{#2}}
   \pgfmathtruncatemacro\newfaceid{min(\valfa, \valfb)}
   \pgfmathtruncatemacro\latestfaceid{\themaxfaceseen - 1}
   \foreach \x in {0,...,\latestfaceid}{
       \ifthenelse{\getfinalface{\x}=\valfa \OR \getfinalface{\x}=\valfb}{
          \storefinalface{\x}{\newfaceid}
       }{}
   }     
}

\newcommand\shiftfaces[3]{

}

\def\drawfaces{0}
\newcommand\enablefaces[0]{
   \def\drawfaces{1}
   \placesandspots
}

\newcommand\disablefaces[0]{
   \def\drawfaces{0}
}

\def\writefaceids{0}
\newcommand\enablefaceids[0]{
    \def\writefaceids{1}
}

\def\drawplaces{0}
\newcommand\enableplaces[0]{
   \def\drawplaces{1}
}

\newcommand\startdiagram[1]{
   \pgfmathtruncatemacro\maxstrandidx{#1 - 1}
   \pgfmathsetmacro\centeringoffset{-0.5 * #1}
   \pgfmathtruncatemacro\diagramlevel{0}

   \foreach \x in {0,...,\maxstrandidx}{
      \storeface{\x}{\x}
      \storefinalface{\x}{\x}
      \storefacecolor{\x}{\defaultfacecolor}
      \storeminslice{\x}{0}
   }
   \pgfmathtruncatemacro\tmpnextfaceid{\maxstrandidx + 1}
   \setcounter{nextfaceid}{\tmpnextfaceid}
   \setcounter{maxfaceseen}{\tmpnextfaceid}

   \setcounter{nextedgeid}{0}
}

\newcommand\drawinitialstrands[1]{
   \pgfmathtruncatemacro\maxstrandidx{#1 - 1}
   \pgfmathsetmacro\centeringoffset{-0.5 * #1}
   \pgfmathtruncatemacro\diagramlevel{0}

   \foreach \x in {0,...,\maxstrandidx}{
      \storeface{\x}{\x}
   }

  \begin{scope}[xshift=\centeringoffset cm]
   \ifthenelse{\maxstrandidx = 0}{}{
   \foreach \x in {1,...,\maxstrandidx}{

     \storeedgeid{\x}{\thenextedgeid}
     \storeedgepath{\thenextedgeid}{($(\x + \centeringoffset,.5)$) -- ($(\x+\centeringoffset,0)$)};
     \storeedgecolor{\thenextedgeid}{black}
     \stepcounter{nextedgeid}
   }
   }
   \end{scope}

    \foreach \x in {0,...,\maxstrandidx} {
        \node at ($(\x+\centeringoffset + .5, 0-\diagramlevel)$) (spot_\diagramlevel_\x) {};
    }
    \foreach \x in {1,...,\maxstrandidx} {
      \node at ($(\x+\centeringoffset, 0-\diagramlevel)$) (place_\diagramlevel_\x) {};
    }

   \pgfmathtruncatemacro\tmpnextfaceid{\maxstrandidx + 1}
   \setcounter{nextfaceid}{\tmpnextfaceid}

   \pgfmathtruncatemacro\latestfaceid{\themaxfaceseen - 1}

}

\newcommand\finishdiagram[0]{
   \ifthenelse{\maxstrandidx = 0}{}{
     \foreach \x in {1,...,\maxstrandidx}{
       \expandedgepath{\x}{ -- +(0,-.5)};
     
   }
   }

   \pgfmathtruncatemacro\maxedgeid{\thenextedgeid - 1}
   \ifthenelse{\thenextedgeid = 0}{}{
   \foreach \eid in {0,...,\maxedgeid}{
     \draw[rounded corners,\getedgecolor{\eid}] \getedgepath{\eid} ;
   }
   }
}

\newcommand\scanslice[3]{
    \def\wb{#1}
    \def\inputs{#2}
    \def\outputs{#3}
    \pgfmathtruncatemacro\nextdiaglevel{\diagramlevel+1}
    \pgfmathtruncatemacro\horizoffset{\outputs - \inputs}
    \pgfmathtruncatemacro\bottomrightcorner{\wb + \outputs + 1}
    \pgfmathtruncatemacro\innertoprightcorner{\wb + \inputs}

    \ifthenelse{\outputs=0 \AND \inputs>0}{
      \mergefaces{\getface{\wb}}{\getface{\innertoprightcorner}}
    }{}

   \ifthenelse{\horizoffset > 0}{
       \foreach \x in {\maxstrandidx,...,\innertoprightcorner}{
          \pgfmathtruncatemacro\offsetidx{\x + \horizoffset}
          \storeface{\offsetidx}{\getface{\x}}
       }
   }{
       \foreach \x in {\innertoprightcorner,...,\maxstrandidx}{
          \pgfmathtruncatemacro\offsetidx{\x + \horizoffset}
          \storeface{\offsetidx}{\getface{\x}}
       }
   }

   \pgfmathtruncatemacro\firstnewface{\wb + 1}
   \pgfmathtruncatemacro\finalnewface{\wb + \outputs - 1}
   \ifthenelse{\outputs > 1}{
     \foreach \x in {\firstnewface,...,\finalnewface}{
       \storeface{\x}{\thenextfaceid}

       \storeminslice{\thenextfaceid}{\nextdiaglevel}

       \ifthenelse{\themaxfaceseen > \thenextfaceid}{}{
         \storefinalface{\thenextfaceid}{\thenextfaceid}
         \storefacecolor{\thenextfaceid}{eface}
         \stepcounter{maxfaceseen}
       }
       \stepcounter{nextfaceid}

     }
   }{}

   \pgfmathtruncatemacro\latestfaceid{\themaxfaceseen - 1}

    \pgfmathtruncatemacro\maxstrandidx{\maxstrandidx + \horizoffset}
    \pgfmathtruncatemacro\diagramlevel{\diagramlevel + 1}

}

\newcommand\diagslicenovertex[3]{
    \def\wb{#1}
    \def\inputs{#2}
    \def\outputs{#3}
    \pgfmathtruncatemacro\nextdiaglevel{\diagramlevel+1}
    \pgfmathtruncatemacro\horizoffset{\outputs - \inputs}
    \pgfmathsetmacro\nextoffset{\centeringoffset - 0.5*\horizoffset}
    \pgfmathtruncatemacro\toprightcorner{\wb + \inputs + 1}
    \pgfmathtruncatemacro\bottomrightcorner{\wb + \outputs + 1}
    \pgfmathsetmacro\vertexpos{0.5* (\centeringoffset + \wb + 0.5 + 0.5*\inputs) + 0.5*(\nextoffset + \wb + 0.5 + 0.5*\outputs)}

    \node (curvertex) at ($(\vertexpos, -0.5*\diagramlevel - 0.5*\nextdiaglevel)$) {};
    \node at (curvertex) (v\diagramlevel) {};

    \ifthenelse{\wb=0}{}{
        \foreach \x in {1,...,\wb} {

          \expandedgepath{\x}{-- ($(\x + \nextoffset,0- \nextdiaglevel)$)};
        }
    }
    
    \ifthenelse{\inputs=0}{}{
      \foreach \x in {1,...,\inputs} {
         \pgfmathtruncatemacro\edgepos{\wb+\x}

         \expandedgepath{\edgepos}{-- ($(\vertexpos, -0.5*\diagramlevel - 0.5*\nextdiaglevel)$)};
       }
    }

    \ifthenelse{\toprightcorner > \maxstrandidx}{}{
       \foreach \x in {\toprightcorner,...,\maxstrandidx} {
         \expandedgepath{\x}{-- ($(\x + \horizoffset + \nextoffset, 0 - \nextdiaglevel)$)};
         }

         \ifthenelse{\inputs > \outputs}{
           \foreach \x in {\toprightcorner,...,\maxstrandidx}{
              \pgfmathtruncatemacro\edgepos{\x + \horizoffset}
              \storeedgeid{\edgepos}{\getedgeid{\x}}
           }
         }{}
         \ifthenelse{\outputs > \inputs}{
           \foreach \x in {\maxstrandidx,...,\toprightcorner}{
              \pgfmathtruncatemacro\edgepos{\x + \horizoffset}
              \storeedgeid{\edgepos}{\getedgeid{\x}}
           }           
         }{}
    }

    \ifthenelse{\outputs=0}{}{
      \foreach \x in {1,...,\outputs} {
        \pgfmathtruncatemacro\edgepos{\wb + \x}
        \storeedgeid{\edgepos}{\thenextedgeid}
        \storeedgepath{\thenextedgeid}{($(\vertexpos, -0.5*\diagramlevel - 0.5*\nextdiaglevel)$) -- ($(\edgepos + \nextoffset,0 - \nextdiaglevel)$)};
        \storeedgecolor{\thenextedgeid}{black}
        \stepcounter{nextedgeid}
       }
    }

    \ifthenelse{\drawfaces=0}{}{
         \ifthenelse{\inputs=0 \AND \outputs=0}{
            \pgfmathtruncatemacro\innertopleftcorner{\wb - 1}
            \pgfmathtruncatemacro\innertoprightcorner{\toprightcorner}

            \draw[eface] ($(\wb + \centeringoffset + .5, 0 - \diagramlevel)$) edge[bend left=40] ($(\wb + \nextoffset + .5, 0 - \nextdiaglevel)$);
            \draw[eface] ($(\wb + \centeringoffset + .5, 0 - \diagramlevel)$) edge[bend right=40] ($(\wb + \nextoffset + .5, 0 - \nextdiaglevel)$);
        }{
            \pgfmathtruncatemacro\innertopleftcorner{\wb}
            \pgfmathtruncatemacro\innertoprightcorner{\toprightcorner - 1}
        }

        \foreach \x in {0,...,\innertopleftcorner} {
                 \draw[eface,\getfacecolor{\getfinalface{\getface{\x}}}] ($(\x + \centeringoffset + .5, 0 - \diagramlevel)$) -- ($(\x + \nextoffset + .5, 0 - \nextdiaglevel)$);
        }
         
         \foreach \x in {\innertoprightcorner,...,\maxstrandidx} {
            \draw[eface,\getfacecolor{\getfinalface{\getface{\x}}}] ($(\x + \centeringoffset + .5, 0 - \diagramlevel)$) -- ($(\x + \horizoffset + \nextoffset + .5, 0 - \nextdiaglevel)$);
         }

    }

    \scanslice{#1}{#2}{#3}
    \pgfmathsetmacro\centeringoffset{\nextoffset}

    \foreach \x in {0,...,\maxstrandidx} {
        \node at ($(\x+\centeringoffset + .5, 0-\diagramlevel)$) (spot_\diagramlevel_\x) {};
    }
    \foreach \x in {1,...,\maxstrandidx} {
      \node at ($(\x+\centeringoffset, 0-\diagramlevel)$) (place_\diagramlevel_\x) {};
    }
    \ifthenelse{\drawfaces=0}{}{
       \placesandspots
    }

}

\newcommand\diagslice[3]{
  \diagslicenovertex{#1}{#2}{#3}
  \node[bnode] at (curvertex) {};
}

\newcommand\reddiagslice[3]{
  \diagslicenovertex{#1}{#2}{#3};
  \node[rnode] at (curvertex) {};
}

\newcommand\placesandspots[0]{
    \foreach \x in {0,...,\maxstrandidx} {
        \node[yspot,\getfacecolor{\getfinalface{\getface{\x}}}] at (spot_\diagramlevel_\x) {};
        
       \ifthenelse{\writefaceids=1 \AND \diagramlevel=\getminslice{\getface{\x}} \AND \getfinalface{\getface{\x}}=\getface{\x}}{
            \node[node distance=.25cm,above of=spot_\diagramlevel_\x] {\small \getface{\x}};
        }{}
    }

    \ifthenelse{\maxstrandidx=0 \OR \drawplaces=0}{}{
       \foreach \x in {1,...,\maxstrandidx} {
          \node[bplace] at (place_\diagramlevel_\x) {};
       }
    }
}

\def\drawswaps{1}
      \def\drawdartnumbers{1}

      \tikzstyle{dart}=[arrows = {Stealth[harpoon,line width=1pt,scale=0.5]-Stealth[harpoon,line width=1pt,scale=0.5]}, line width=4pt]
          \tikzstyle{halfdart}=[arrows = {-Stealth[harpoon,line width=1pt,scale=1]}, line width=1pt]
          \tikzstyle{ylink}=[thick,blue,->]
          \tikzstyle{helper}=[inner sep=1.5pt,line width=0pt,circle]

    \newcounter{dartcounter}
    \newcommand\dart[2]{
      \stepcounter{dartcounter}
      \edef\firstdart{\arabic{dartcounter}}
      \stepcounter{dartcounter}
      \edef\seconddart{\arabic{dartcounter}}
      \draw[line width=4pt,shorten <= 7pt,shorten >= 7pt] (#1) -- (#2);
      \draw[dart] (#1) -- (#2);
      \draw[white,line width=2pt] (#1) -- (#2)
      node[midway,anchor=mid,auto=left,black] (dart\firstdart) {\ifthenelse{\drawdartnumbers = 1}{\firstdart}{}}
      node[pos=0.5,anchor=south,auto=right,node distance=10pt,black] (dart\seconddart) {\ifthenelse{\drawdartnumbers = 1}{\seconddart}{}};
      \ifthenelse{\drawswaps = 1}{
        \draw[thick,red,<->,shorten <=-4pt,shorten >=-4pt] (dart\firstdart) edge[bend right=90] (dart\seconddart);
      }{}
    }

    \newcommand\bdart[2]{
      \stepcounter{dartcounter}
      \edef\firstdart{\arabic{dartcounter}}
      \stepcounter{dartcounter}
      \edef\seconddart{\arabic{dartcounter}}
      \path (#1) -- (#2)
      node[pos=0.2,auto=left,inner sep=15pt] (helper1) {}
      node[pos=0.8,auto=left,inner sep=15pt] (helper2) {};
      
      \draw[green] (#1) .. controls (helper1) and (helper2) .. (#2)
       node[midvay,auto=left,blue,draw] {};
      
      node[midway,anchor=mid,auto=left,black] (dart\firstdart) {\ifthenelse{\drawdartnumbers = 1}{\firstdart}{}}
      node[pos=0.5,anchor=south,auto=right,node distance=10pt,black] (dart\seconddart) {\ifthenelse{\drawdartnumbers = 1}{\seconddart}{}};
      \ifthenelse{\drawswaps = 1}{
        \draw[thick,red,<->,shorten <=-4pt,shorten >=-4pt] (dart\firstdart) edge[bend right=90] (dart\seconddart);
      }{}
    }

   \newcommand\ddart[2]{
      \stepcounter{dartcounter}
      \edef\firstdart{\arabic{dartcounter}}
      \stepcounter{dartcounter}
      \edef\seconddart{\arabic{dartcounter}}
      \path[line width=2pt] (#1) -- (#2)
      node[midway,anchor=mid,auto=left,black] (dart\firstdart) {\ifthenelse{\drawdartnumbers = 1}{\firstdart}{}}
      node[pos=0.5,anchor=south,auto=right,node distance=10pt,black] (dart\seconddart) {\ifthenelse{\drawdartnumbers = 1}{\seconddart}{}}
      node[pos=0,auto=right,helper] (p0) {}
      node [pos=1,auto=right,helper] (p1) {}
      node[pos=0,auto=left,helper] (p2) {}
      node[pos=1,auto=left,helper] (p3) {};
      \draw[halfdart,dashed] (p1) -- (p0);
      \draw[halfdart] (p2) -- (p3);
      \ifthenelse{\drawswaps = 1}{
        \draw[thick,red,<->,shorten <=-4pt,shorten >=-4pt] (dart\firstdart) edge[bend right=90] (dart\seconddart);
      }{}
    }

\usepackage{braids}

\begin{document}

\title[Normalization for planar string diagrams]{Normalization for planar string diagrams and a quadratic equivalence algorithm}

\author[A.~Delpeuch]{Antonin Delpeuch\rsuper{a}}

\author[J.~Vicary]{Jamie Vicary\rsuper{b}}

\address{Department of Computer Science, University of Oxford, Wolfson Building, Parks Road, Oxford, OX1 3QD, UK}
\email{antonin.delpeuch@cs.ox.ac.uk}
\thanks{This work was supported by an EPSRC scholarship.}

\address{Computer Laboratory, University of Cambridge, J. J. Thomson Avenue, Cambridge, CB3 0FD, UK}
\email{jamie.vicary@cl.cam.ac.uk}

\begin{abstract}
  In the graphical calculus of planar string diagrams, equality is generated by exchange moves, which swap the heights of adjacent vertices. We show that  left- and right-handed exchanges each give strongly normalizing rewrite strategies for connected string diagrams. We use this result to give a linear-time solution to the equivalence problem in the connected case, and a quadratic solution in the general case. We also give a stronger proof of the Joyal-Street coherence theorem, settling Selinger's conjecture on recumbent isotopy.
\end{abstract}

\maketitle

\section*{Introduction}
\label{sec:introduction}

\subsection{Motivation and summary}

\begin{figure}[b]
  \centering
  \begin{subfigure}{0.2\textwidth}
      \centering
    \begin{tikzpicture}[scale=0.4]
      \startdiagram{2}
      \drawinitialstrands{2}
      \diagslice{0}{0}{2}
      \diagslice{2}{1}{0}
      \diagslice{1}{0}{1}
      \diagslice{1}{1}{0}
      \diagslice{0}{2}{0}
      \finishdiagram
    \end{tikzpicture}
    \caption{Disconnected}
  \end{subfigure}
  \qquad
  \begin{subfigure}{0.2\textwidth}
    \centering
        \vspace{.2cm}
    \begin{tikzpicture}[scale=0.4]
      \startdiagram{1}
      \diagslice{0}{0}{3}
      \diagslice{0}{1}{2}
      \diagslice{2}{1}{2}
      \diagslice{3}{2}{0}
      \diagslice{0}{1}{0}
      \finishdiagram
    \end{tikzpicture}
    \caption{Connected}
  \end{subfigure}
  \qquad
  \begin{subfigure}{0.25\textwidth}
    \centering
    \vspace{.7cm}
    \begin{tikzpicture}[scale=0.4]
      \startdiagram{4}
      \drawinitialstrands{4}
      \diagslice{0}{0}{1}
      \diagslice{1}{3}{0}
      \diagslice{1}{0}{2}
      \finishdiagram
    \end{tikzpicture}
    \caption{Boundary-connected}
  \end{subfigure}
  \caption{Connectedness for string diagrams}
  \label{fig:boundary-connectedness}
\end{figure}
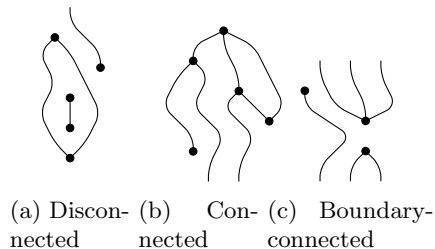

String diagrams are a geometrical notation for the mathematical theory of monoidal categories, a logical toolkit for describing the algebra of compositional systems. Examples are given in Figure~\ref{fig:boundary-connectedness}; a standard interpretation of such a diagram is that wires represent systems storing computational data, and vertices represent processes taking place over time (read from top to bottom), with each process having input and output data represented by the wires attached above and below the vertex respectively. Over the last 10 years, string diagrams have found increasingly broad application across theoretical computer science, in areas including quantum computation~\cite{backens2014zxcalculus,duncan2014verifying,dixon2010open}, natural language processing~\cite{clark2008compositional}, interacting agents~\cite{mellies2012game,ghani2016compositional},  circuit design~\cite{ghica2017diagrammatic}, and rewriting~\cite{mimram20143dimensional}; a key survey paper by Selinger~\cite{selinger2011survey} has received over 400 citations in 10 years, with two-thirds of those in the last 5 years.

Despite this significant activity, there are no general results\footnote{There are a variety of interesting results about the equivalence problem over some specific signatures, which we survey in Section~\ref{sec:relatedwork}.} about the complexity of deciding \textit{equivalence} of string diagrams, an important question if the theory is to become a mainstream logical technique that can form part of real-world systems. Equivalence of string diagrams is a geometrical notion,  with two string diagrams being equivalent (that is, representing equal morphisms of the corresponding monoidal category) just when their string diagram representations are related by a \textit{recumbent isotopy}~\cite[Theorem~3, Caveat~9]{selinger2011survey}. Here, \emph{isotopy} means that one diagram can be deformed into the other without breaking,  intersecting or reordering input or output wires, and \emph{recumbent} means that inputs cannot exchange with outputs, and wires and vertices remain essentially `upright' throughout the isotopy~\cite{joyal1991geometry}. This equality relation is then sometimes collapsed further by adding additional axioms, in a way that suits the application.

In this paper we take a first look at the complexity of the general equivalence problem for {planar} string diagrams\footnote{By the work of Joyal and Street~\cite{joyal1991geometry}, this corresponds to the word problem for monoidal categories which are free on a given generating set of objects and morphisms. Furthermore, our results extend immediately to bicategories which are free on a given set of generating 1- and 2-morphisms, but we prefer to keep the discussion at the level of monoidal categories.} (henceforth simply \emph{diagrams}), without additional axioms.  This does not include all the features used by some applications of string diagrams (for example, braided or symmetric monoidal structure), but it is already a nontrivial setting, and seems a suitable place to begin building the theory.

Our main results are as follows. We write $v$ for the number of vertices in a diagram, and $e$ for the number of edges; also, we say that a diagram is \textit{connected} when there is a path in the diagram between any two vertices, and \emph{boundary-connected} when it is either connected, or every vertex has a path in the diagram to a boundary. See Figure~\ref{fig:boundary-connectedness} for examples of these notions.

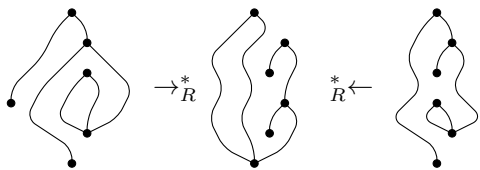
\begin{figure}[b]
  \centering
    \begin{tikzpicture}[scale=0.4]
      \startdiagram{1}
      \diagslice{0}{0}{2}
      \diagslice{1}{1}{2}
      \diagslice{2}{0}{2}
      \diagslice{0}{1}{0}
      \diagslice{1}{3}{0}
      \diagslice{0}{1}{0}
      \finishdiagram

      \node at (3.4,-3) {$\rightarrow_R^*$};

      \begin{scope}[xshift=6cm]
        \startdiagram{1}
        \diagslice{0}{0}{2}
        \diagslice{2}{0}{2}
        \diagslice{2}{1}{0}
        \diagslice{2}{1}{2}
        \diagslice{2}{1}{0}
        \diagslice{0}{3}{0}
        \finishdiagram
      \end{scope}

      \node at (9.2,-3) {${}_R^*$\hspace{0cm}$\leftarrow$};

      \begin{scope}[xshift=12cm]
        \startdiagram{1}
        \diagslice{0}{0}{2}
        \diagslice{1}{1}{2}
        \diagslice{1}{1}{0}
        \diagslice{1}{0}{2}
        \diagslice{1}{3}{0}
        \diagslice{0}{1}{0}
        \finishdiagram
      \end{scope}
    \end{tikzpicture}
    \caption{Two connected diagrams with the same right normal form}
    \label{fig:normalization-connected-diagrams}
\end{figure}
\begin{itemize}
\item For boundary-connected diagrams, we build a rewrite strategy that generates the equality relation, show it is strongly normalizing~(Theorem~\ref{thm:termination-connected}), show it terminates after $O(v^3)$ steps~(Theorem~\ref{thm:termination-connected-bound}), show the normal forms can be constructed in $O(ve)$ time (Theorem~\ref{thm:efficient-normalizing}), and show equality can be decided in $O(v+e)$ time (Corollary~\ref{coro:linear-time-sd-word-problem}).
\item For general diagrams, with no constraints on connectivity, we use the above results to derive a scheme that decides equality in $O(ve)$ time (Theorem~\ref{thm:quadratic-word-problem-disconnected}).
\item We show that the recumbency property listed above is unnecessary; that is, we show that two diagrams are recumbent isotopic, and hence equal, just when they are isotopic (Theorem~\ref{thm:selinger}).\footnote{This has a nice expression in categorical terms (Corollary~\ref{cor:pivotal}): it says that for a monoidal signature $\Sigma$, the embedding functor from the free monoidal category on $\Sigma$ to the free pivotal category on $\Sigma$ is faithful.} This proves a conjecture of Selinger~\cite{selinger2011survey}.
\end{itemize}
This final result is attractive, since in practice the recumbency property is highly constraining, forcing the entire diagram to remain essentially ``vertical'' throughout the isotopy.

\subsection{Related work}
\label{sec:relatedwork}

The use of rewriting techniques on diagrams is ubiquitous in the communities which use monoidal or higher categories. Diagrammatic rewriting has been studied in detail for particular signatures, such as those of boolean circuits~\cite{lafont2003algebraic, lafont2009diagram} or the ZX-calculus~\cite{duncan2014verifying, dixon2010open}. More generally, rewriting theory of 2-polygraphs was developed by Guiraud and Malbos~\cite{guiraud2018polygraphs}, extending classical results on monoids. In these approaches, the goal is to decide equality of diagrams up to the axioms in the signature, and structural equalities such as the exchange law or even symmetry are strict. Our results focus instead on the structural equalities, and do not allow equalities in the signature.

The word problem for the structural equalities has attracted attention in higher categories, but with no complexity result so far. The foundational work of Burroni~\cite{burroni1993higherdimensional} establishes the link between the word problem for an algebraic structure and the path problem in the next dimension. Later, Makkai~\cite{makkai2005word} showed decidability of the word problem for higher cells in strict $\omega$-categories. Our work refines this result at dimension 2 by giving a complexity bound at dimension two. We also relate these computational results to the well established theory of embedded graphs (also called \emph{maps}) \cite{hopcroft1974linear,deverdiere2013testing}, via reductions which bridge the differences in the notions of isotopy used.

The study of equivalence in category theory often takes the form of coherence results. These state that all morphisms between given source and targets and built from a particular signature are equal. These results often rely on rewriting techniques, the spirit of which was already present since Mac Lane's coherence theorem for monoidal categories~\cite{maclane1963natural}. More recently, Forest and Mimram~\cite{forest2018coherence} use rewriting to prove coherence for Gray monoids. They use similar techniques, with a focus on coherence of reductions rather than their length.

There has recently been activity in the development of computer proof assistants for string diagrams, including \emph{Quantomatic}~\cite{dixon2010open}, \emph{Globular}~\cite{globular,Bar2017} and its successor \emph{homotopy.io}~\cite{homotopyio-tool}. Our string diagram isotopy algorithm could yield a geometrical notion of ``tactic'' for such a proof assistant, automatically finding isotopies between diagrams, or rearranging diagrams to normal form.

\subsection{Outline}

\noindent
This paper has the following structure. We first introduce our formalism, defining diagrams and a rewriting relation on them. In Section~\ref{sec:termination}, we show that the rewriting relation terminates on connected diagrams and derive an asymptotic upper bound on reduction length. Section~\ref{sec:confluence} shows confluence of the rewriting relation, which gives a simple algorithm to normalize connected diagrams. Section~\ref{sec:algorithm} analyses the structure of right normal forms and shows how to compute them more efficiently. Section~\ref{sec:disconnected} extends these results to disconnected diagrams and Section~\ref{sec:linear-time} improves the complexity in the connected case by reducing the problem to the more widespread notion of planar map isotopy.

\section{Monoidal categories and string diagrams} \label{sec:formalism}

We first introduce monoidal categories and show why morphism expressions written as terms
are ill-suited to reason about equivalence. We then introduce string diagrams, which
offer a more intuitive graphical representation of these morphisms, at the expense of
requiring the manipulation of elaborate topological objects. Finally, we show
how string diagrams can be encoded combinatorially, providing an efficient
representation to detect equivalence.

\subsection{Monoidal categories}

\begin{defi}
  A \emph{monoidal category} $\cC$ is a category equiped with a bifunctor $\_ \otimes \_ : \cC \times \cC \rightarrow \cC$ such that $\otimes$ is strictly associative and has a unit $I \in \cC$.
\end{defi}
In this work, the monoidal categories considered are strict, meaning
that the associativity and unitality of the monoidal product are
equalities rather than isomorphisms. The latter case corresponds to
the notion of \emph{weak monoidal category}. Mac Lane's coherence
theorem states that any weak monoidal category is equivalent to a
strict monoidal category.

Morphisms in monoidal categories can be composed in two ways.
Given $f : A \rightarrow B$ and $g : B \rightarrow C$, we get $g \circ f : A \rightarrow C$
as in any category. This composition is associative and has units $1_A : A \to A$. This is intuitively a sequential composition, as $g$ is executed after $f$.

Given $u : A \rightarrow B$ and $v : C \rightarrow D$, we can use the
bifunctor $\otimes$ to form $u \otimes v : A \otimes C \to B \otimes
D$. By interpreting the monoidal product $\otimes$ as a pairing
operation, this is intuitively a parallel composition, as $u$ and $v$
are executed independly of each other, simultaneously.

The functoriality of $\otimes$ requires a compatibility between the two composition operations, called the \emph{exchange law} or \emph{bifunctoriality equation}:
\[ (g \otimes k) \circ (f \otimes j) = (g \circ f) \otimes (k \circ j) \]
when both sides of the equation are defined.

The bifunctoriality equation above, combined with
the associativity and unitality of $\otimes$ and $\circ$, gives rise to a rich equational theory.
For instance, given morphisms $a, b : I \rightarrow I$, we have the following derivation, known as the Eckmann-Hilton argument:

\begin{align}
  a \circ b &= (a \otimes 1_I) \circ b &\text{(unitality of $\otimes$)} \\
  &= (a \otimes 1_I) \circ (1_I \otimes b) &\text{(unitality of $\otimes$)} \\
  &= (a \circ 1_I) \otimes (1_I \circ b) &\text{(bifunctoriality)} \\
  &= a \otimes (1_I \circ b) &\text{(unitality of $\circ$)} \\
  & = a \otimes b &\text{(unitality of $\circ$)} \\
 &= (1_I \circ a) \otimes b &\text{(unitality of $\circ$)} \\
 &= (1_I \circ a) \otimes (b \circ 1_I) &\text{(unitality of $\circ$)} \\
 &= (1_I \otimes b) \circ (a \otimes 1_I) &\text{(bifunctoriality)} \\
 &= b \circ (a \otimes 1_I) &\text{(unitality of $\otimes$)} \\
 &= b \circ a &\text{(unitality of $\otimes$)}
\end{align}

The derivation above shows multiple issues with the representation of
morphisms as terms to reason about equivalence. First, the rewriting
strategy used to derive the equality is not obvious: one needs to
introduce identities by unitality in creative ways in steps 1, 2, 6
and 7. Therefore it seems difficult to obtain a terminating and
confluent rewriting system in this presentation.  Second, the
bifunctoriality equation only holds when the domains and codomains of
the morphisms involved are compatible: one cannot, in general, replace
any expression $(g \otimes k) \circ (f \otimes j)$ by $(g \circ f)
\otimes (k \circ j)$. Thus we are required to keep track of the domains
and codomains of all sub-expressions involved to understand which axiom
can be applied. In the example above all domains and codomains are $I$ so the bifunctoriality equation could always be applied, but this is not true in general.

\subsection{String diagrams} \label{sec:string-diagrams}

String diagrams are graphical representations of morphisms in a monoidal
category. They were proposed independently by~\cite{hotz1965algebraisierung} and~\cite{joyal1988planar,joyal1991geometry}.

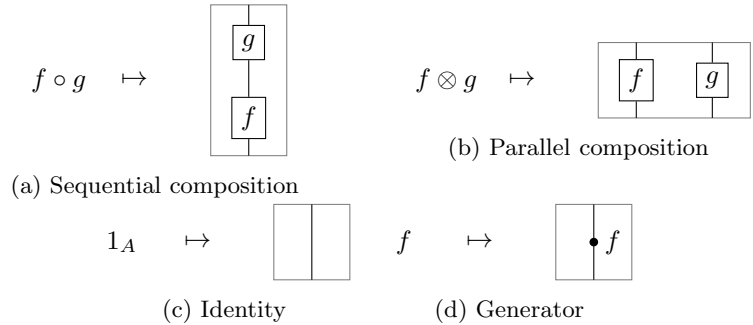
\begin{figure}[b]
  \centering
  \begin{subfigure}{0.45\textwidth}
    \centering
    \begin{tikzpicture}
      \node at (0,0) {$f \circ g$};
      \node at (1,0) {$\mapsto$};
      \draw[gray] (2,-1) rectangle (3,1);
      \node[rectangle,draw] at (2.5,-.5) (f) {$f$};
      \node[rectangle,draw] at (2.5,.5) (g) {$g$};
      \draw (2.5,1) -- (g) -- (f) -- (2.5,-1);
    \end{tikzpicture}
    \caption{Sequential composition}
  \end{subfigure}
  \begin{subfigure}{0.45\textwidth}
    \centering
    \begin{tikzpicture}
      \node at (0,0) {$f \otimes g$};
      \node at (1,0) {$\mapsto$};
      \draw[gray] (2,-.5) rectangle (4,.5);
      \node[rectangle,draw] at (2.5,0) (f) {$f$};
      \node[rectangle,draw] at (3.5,0) (g) {$g$};
      \draw (2.5,.5) -- (f) -- (2.5,-.5);
      \draw (3.5,.5) -- (g) -- (3.5,-.5);
    \end{tikzpicture}
    \caption{Parallel composition}
  \end{subfigure}
  \begin{subfigure}{0.3\textwidth}
    \centering
    \begin{tikzpicture}
      \node at (0,0) {$1_A$};
      \node at (1,0) {$\mapsto$};
      \draw[gray] (2,-.5) rectangle (3,.5);
      \draw (2.5,.5) -- (2.5,-.5);
    \end{tikzpicture}
    \caption{Identity}
  \end{subfigure}
  \begin{subfigure}{0.3\textwidth}
    \centering
    \begin{tikzpicture}
      \node at (0,0) {$f$};
      \node at (1,0) {$\mapsto$};
      \draw[gray] (2,-.5) rectangle (3,.5);
      \node[bnode] at (2.5,0) (f) {};
      \draw (2.5,.5) -- (2.5,-.5);
      \node[node distance=.25cm,right of=f] {$f$};
    \end{tikzpicture}
    \caption{Generator}
  \end{subfigure}
  \caption{Recursive translation of a morphism expression to a string diagram}
  \label{fig:def-sd}
\end{figure}
Any term representing a morphism in a monoidal
category can be inductively translated to a string diagram as shown in Figure~\ref{fig:def-sd}.
The reverse translation can be defined too. This first
requires defining a class of well-behaved diagrams to ease the
analysis: for instance, \cite{joyal1991geometry} require diagrams to be ``recumbent'':

\begin{defiC}[\cite{joyal1991geometry}]\label{defi:recumbent-plane-diagram}
  A \emph{recumbent} (or \emph{progressive}) \emph{plane diagram} is an embedded graph $\Gamma \hookrightarrow [a,b] \times \mathbb{R}$ (see \cite{joyal1991geometry}) such that the projection of any edge on the vertical axis is injective.
\end{defiC}

In the definition above, $\Gamma$ is a topological graph and $[a,b]
\times \mathbb{R}$ is a section of the plane it gets embedded into.
We can then associate to any recumbent string diagram a corresponding
morphism in the free monoidal category on the generators involved.
This is done by breaking down a diagram into basic blocks of either
generators or identities and composing them sequentially or in
parallel according to how these basic blocks are laid out. The
requirement for edges to be injective along the vertical axis ensures
that there are no ``cups'' or ``caps'' which would not be
interpretable without adjoints.  The bifunctoriality equation and the
associativity and unitality of compositions ensure that the resulting
morphism does not depend on the order in which the blocks are
composed, as shown in Figure~\ref{fig:bifunctoriality}.  The details
of this construction can be found in
\cite{joyal1988planar,joyal1991geometry}.
\begin{figure}[b]

    \centering
\begin{tikzpicture}
\node[draw,rectangle] (a1) {$f_2$};
\node[draw,rectangle,right of=a1] (b1) {$g_2$};
\node[draw,rectangle,below of=a1] (c1) {$f_1$};
\node[draw,rectangle,below of=b1] (d1) {$g_1$};
\node[above of=a1,node distance=0.6cm] (p1) {};
\node[above of=b1,node distance=0.6cm] (p2) {};
\node[below of=c1,node distance=0.6cm] (p3) {};
\node[below of=d1,node distance=0.6cm] (p4) {};
\draw (p1) -- (a1) -- (c1) -- (p3);
\draw (p2) -- (b1) -- (d1) -- (p4);
\node[left of=a1,node distance=0.5cm] {$\Big($};
\node[left of=c1,node distance=0.5cm] {$\Big($};
\node[right of=b1,node distance=0.5cm] {$\Big)$};
\node[right of=d1,node distance=0.5cm] {$\Big)$};

\node[right of=b1] (blank) {};
\node[below of=blank,node distance=0.5cm] (egal) {$=$};

\node[draw,rectangle,right of=blank] (a1) {$f_2$};
\node[draw,rectangle,right of=a1,node distance=1.62cm] (b1) {$g_2$};
\node[draw,rectangle,below of=a1] (c1) {$f_1$};
\node[draw,rectangle,below of=b1] (d1) {$g_1$};
\node[above of=a1,node distance=0.6cm] (p1) {};
\node[above of=b1,node distance=0.6cm] (p2) {};
\node[below of=c1,node distance=0.6cm] (p3) {};
\node[below of=d1,node distance=0.6cm] (p4) {};
\draw (p1) -- (a1) -- (c1) -- (p3);
\draw (p2) -- (b1) -- (d1) -- (p4);
\node[right of=egal,node distance=0.45cm] (p1) {\Large $\Bigg($};
\node[right of=p1,node distance=1.10cm] (p2) {\Large $\Bigg)$};
\node[right of=p1,node distance=1.60cm] (p3) {\Large $\Bigg($};
\node[right of=p3,node distance=1.15cm] (p4) {\Large $\Bigg)$};

\end{tikzpicture}
\caption{The bifunctoriality equation in string diagrams}
\label{fig:bifunctoriality}
\end{figure}

The main result establishing the usefulness
of string diagrams is the invariance of the
interpretation as a morphism up to topological
deformations.

\begin{defiC}[{\cite[Definition~5]{joyal1988planar}}]
  A \emph{deformation of recumbent graphs} is a deformation $h : \Gamma \times [0,1] \rightarrow [a,b] \times \mathbb{R}$ of planar graphs (see Definition~\ref{defi:recumbent-plane-diagram}) such that the image $\Gamma(t)$ of $h(-,t)$ is recumbent for all $t \in [0,1]$.
\end{defiC}

\begin{thmC}[{\cite[Theorem~3]{joyal1988planar}}] \label{thm:joyal-street}
  If $h: \Gamma \times [0,1] \rightarrow [a,b] \times \mathbb{R}$ is a deformation of recumbent diagrams then the value $v(\Gamma(t))$ is independent of $t \in [0,1]$.
\end{thmC}

This means that instead of manipulating
terms to represent morphisms, we can simply
rely on string diagrams to derive equalities
in monoidal categories. For instance, our example
derivation of $a \circ b = b \circ a$ for scalars
is represented in string diagrams by Figure~\ref{fig:eckmann-hilton}.
The argument can simply be understood as the rotation of two vertices
around each other in the plane.

\begin{figure}
  \centering
  \begin{tikzpicture}[every node/.style={node distance=.25cm}]
    \node[bnode] at (0,-.25) (a) {};
    \node[bnode] at (0,.25) (b) {};
    \node[left of=a] {$a$};
    \node[left of=b] {$b$};

    \node at (1,0) {$=$};

    \begin{scope}[xshift=2cm]
      \node[bnode] at (-.25,-.25) (a) {};
      \node[bnode] at (.25,.25) (b) {};
      \node[left of=a] {$a$};
      \node[left of=b] {$b$};
    \end{scope}

    \node at (3,0) {$=$};
    \begin{scope}[xshift=4cm]
      \node[bnode] at (-.3,0) (a) {};
      \node[bnode] at (.3,0) (b) {};
      \node[left of=a] {$a$};
      \node[left of=b] {$b$};
    \end{scope}

    \node at (5,0) {$=$};
    \begin{scope}[xshift=6cm]
      \node[bnode] at (-.25,.25) (a) {};
      \node[bnode] at (.25,-.25) (b) {};
      \node[left of=a] {$a$};
      \node[left of=b] {$b$};
    \end{scope}

    \node at (7,0) {$=$};
    \begin{scope}[xshift=8cm]
      \node[bnode] at (0,.25) (a) {};
      \node[bnode] at (0,-.25) (b) {};
      \node[left of=a] {$a$};
      \node[left of=b] {$b$};
    \end{scope}

  \end{tikzpicture}
  \caption{The Eckmann-Hilton argument in string diagrams}
  \label{fig:eckmann-hilton}
\end{figure}

However, while this representation is easy to manipulate at an intuitive level, one can argue that
the topological notions are relatively involved and describing precisely the objects involved is tedious. These graphical objects also seem harder to encode in a computer, making them of little use to solve the word problem.

In fact, it is possible to encode a string diagram more efficiently than by listing the explicit positions of its vertices and the trajectories of its edges. This relies on the notion of general position:

\begin{defi}
  A string diagram is in \emph{general position} when none of its vertices share the same height.
\end{defi}

Any string diagram can be deformed slightly to be in a general position.

\begin{lemC}[\cite{joyal1988planar}]
  Given a string diagram $\Gamma$, there exists a diagram $\Gamma'$ in general position and
  a deformation of recumbent diagrams between $\Gamma$ and $\Gamma'$.
\end{lemC}

\begin{proof}
  If two vertices are at the same height, then they can be slightly
  perturbated such that one is above the other. This can be done for
  each pair of vertices with recumbent transformations.
\end{proof}

Any string diagram in general position can be cut up in slices, each of which contains exactly one generator.

\begin{lem} \label{lemma:slice-decomposition}
  Given a string diagram in general position $\Gamma$ containing $n$
  vertices, there exists heights $h_1, \dots, h_{n+1}$ such that there
  is exactly one vertex between $h_i$ and $h_{i+1}$.
\end{lem}

\begin{proof}
  Let us call $y_1 < \dots < y_n$ the heights of the $n$ vertices in $\Gamma$. Any choice of $h_i$ such that $h_1 < y_1 < h_2 < \dots < h_n < y_n < h_{n+1}$ satisfies the property.
\end{proof}

\subsection{Combinatorial encoding of string diagrams}
\label{sec:combinatorialfoundation}

Our aim is to give complexity results about problems involving string
diagrams. We must therefore propose an efficient encoding for them and
define the computational model that we use to measure time complexity
of our algorithms. We will use the RAM model, used for instance
by~\cite{hopcroft1974linear}, which assumes constant time random
access to the working memory and constant time arithmetic operations
on integers. This is a widespread model, which closely matches the
architecture of today's computers, despite the fact that arithmetic
operations on unbounded integers would not be implementable in
constant time. We discuss the implications of this feature
in Sections~\ref{sec:disconnected} and~\ref{sec:linear-time}.

Lemma~\ref{lemma:slice-decomposition} provides the basis for a compact
encoding of string diagrams. The main idea is that a diagram is cut
into slices given by the lemma, and each slice can be described by the
following data:
\begin{itemize}
\item the number of wires at the top and bottom of the slice (which we will not directly encode as they are redundant with the neighbouring slices);
\item the number of input and output wires for the generator in the slice;
\item the horizontal position of the generator, described for instance by the number of wires passing
  to the left of the generator;
\item the generator morphism itself, denoted by an identifier taken from the signature. For our purposes this will be omitted to simplify the presentation, as our results are applicable to any signature.
\end{itemize}

This encoding scheme is essentially identical to that used by the proof assistant \textit{Globular}~\cite{bar2016data}, although the result in this section is new, and is not implied by the existing literature. This encoding scheme serves as a formal combinatorial foundation for our results, although we will build most of our arguments at a more intuitive level with the corresponding graphical diagrams.

\begin{figure}[b]
\tikzset{box/.style={draw, circle, minimum width=2pt, inner sep=2pt, fill=black}}
  \centering
  \begin{tikzpicture}[scale=.5]
    \startdiagram{3}
    \drawinitialstrands{3}
    \diagslice{1}{1}{2}
    \diagslice{2}{0}{1}
    \diagslice{0}{1}{0}
    \finishdiagram
    \node [anchor=west] at (2.8,-2) {$D = (2,3,[1,2,0],[1,0,1],[2,1,0])$};
  \end{tikzpicture}
  \caption{\em Example of a diagram $D$ together with its encoding.}
  \label{fig:examplediagram}
\end{figure}
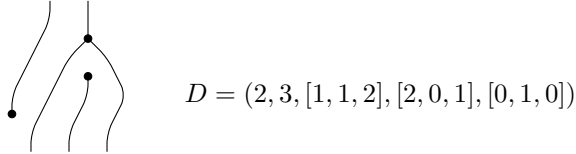

We give an example of a diagram, together with its encoding, in Figure~\ref{fig:examplediagram}. Note that in this example diagram, and in the other diagrams in the paper, we use small circles for the vertices, rather than boxes which are sometimes seen.

\begin{defi}
For a natural number $n \in \N$, we define the total order $[n] = \{0, \ldots, n-1\}$.
\end{defi}

\begin{defi}
A \textit{diagram} $D = (D.S, D.N, D.H, D.I, D.O)$ comprises $D.S \in \N$, the number of \textit{source edges}; $D.N \in \N$, the diagram \textit{height}; and functions \mbox{$D.H, D.I, D.O : [D.N] \to \N$} of \textit{vertex horizontal positions}, \textit{vertex source size} and \textit{vertex target size} respectively.
\end{defi}

Given a diagram, we can compute the number of edges that exist at level just above each vertex, by starting with the number of source edges $D.S$, and then supposing that each vertex $n \in [D.N]$ removes $D.I(n)$ wires and adds $D.O(n)$ wires. We develop that formally as follows.

\begin{defi}
\label{def:delta}
For a diagram $D$, we define $D.\Delta:[D.N] \to \N$ as $D.\Delta(n) = D.O(n) - D.I(n)$.
\end{defi}

\begin{defi}[Wires at each level]
\label{def:wiresatlevel}
For a diagram $D$, we define $D.W:[D.N + 1] \to \N$ as $D.W(0) = D.S$, and for $n \in [D.N]$ as $D.W(n+1) = D.W(n) + D.\Delta(n)$.
\end{defi}

Not all diagrams will be geometrically meaningful, and we give validity conditions which check that there are ``enough edges'' above each vertex to serve as its source edges.

\begin{defi}
A diagram $D$ is \emph{valid} when for all $n \in [D.N]$, we have \\ \mbox{$D.W(n) \geq D.H(n) + D.I(n)$}.
\end{defi}

\begin{figure}[b]
\[
\begin{tikzpicture}[scale=.4]

  \startdiagram{11}
  \drawinitialstrands{11}
  \diagslice{6}{2}{2}
  \diagslice{2}{2}{2}
  \finishdiagram
  \begin{scope}[every node/.style={scale=0.6}]
    \node at (0,-1) {$\dots$};
    \node at (-4,-1) {$\dots$};
    \node at (4,-1) {$\dots$};
    \node at (-2,0) {$\dots$};
    \node at (-2,-2.2) {$\dots$};
    \node at (2,0.2) {$\dots$};
    \node at (2,-2) {$\dots$};
  \end{scope}

  \node at (6,-.5) {$\Rarrow$};
  \node at (6,-1.5) {${} _\L {}\hspace{-3pt}\leftarrow$};

  \begin{scope}[xshift=12cm]
  \startdiagram{11}
  \drawinitialstrands{11}
  \diagslice{2}{2}{2}
  \diagslice{6}{2}{2}
    \begin{scope}[every node/.style={scale=0.6},xscale=-1]
    \node at (0,-1) {$\dots$};
    \node at (-4,-1) {$\dots$};
    \node at (4,-1) {$\dots$};
    \node at (-2,0) {$\dots$};
    \node at (-2,-2.2) {$\dots$};
    \node at (2,0.2) {$\dots$};
    \node at (2,-2) {$\dots$};
  \end{scope}

  \finishdiagram
  \end{scope}

    \end{tikzpicture}
\]
\caption{Right and left exchanges as rewrites on diagrams.}
\label{fig:rightleftexchange}
\end{figure}
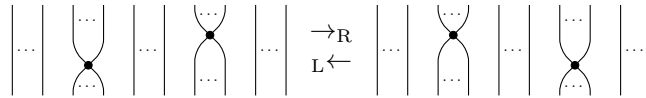

We now formalize the right and left exchange moves illustrated in Figure~\ref{fig:rightleftexchange}. All that needs to be checked is that the two vertices with adjacent heights share no common edges.

\begin{defi}
For $n \in [D.N - 1]$, a diagram $D$ \emph{admits a left exchange move} at height $n$ when $D.H(n+1) \geq D.H(n) + D.O(n)$, and \textit{admits a right exchange move} at height $n$ when $D.H(n) \geq D.H(n+1) + D.I(n+1)$. A \emph{right reduction} is a series of right exchange moves.
\end{defi}

\begin{defi}
\label{def:rightleftexchange}
For a diagram $D$ which admits a left or right exchange move at height $n \in [D.N-1]$, its \textit{left exchange} $D_{\L,n}$ or \textit{right exchange} $D_{\R,n}$, respectively, is defined to be identical to $D$, except at heights $n$, $n+1$ as follows:
{\small 
\begin{align*}
D_{\L,n}.H(n) &= D.H(n+1) - D.\Delta(n)
\\ D_{\L,n}.I(n) &= D.I(n+1)
\\ D_{\L,n}.O(n) &= D.O(n+1)
\\ D_{\L,n}.H(n+1) &= D.H(n)
\\ D_{\L,n}.I(n+1) &= D.I(n)
\\ D_{\L,n}.O(n+1) &= D.O(n)
\\ D_{\R,n}.H(n)&= D.H(n+1)
\\ D_{\R,n}.I(n) &= D.I(n+1)
\\ D_{\R,n}.O(n) &= D.O(n+1)
\\ D_{\R,n}.H(n+1) &= D.H(n) + D.\Delta(n+1)
\\ D_{\R,n}.I(n+1) &= D.I(n)
\\ D_{\R,n}.O(n+1) &= D.O(n)
\end{align*}
}
\end{defi}

\begin{lem}
\label{lem:exchangevalid}
For a valid diagram $D$ which admits a right (or left) exchange move at height~$n$, its right exchange $D_{\R,n}$ (or left exchange $D _{\L,n}$) is a valid diagram.
\end{lem}

\begin{proof}
  Let $D$ be a valid diagram which admits a right exchange at height $n$. We prove that $D_{R,n}$ is valid again. The case of left exchanges is symmetrical.

  We need to check that for each height $k \in [D_{R,n}.N]$, we have $D_{R,n}.H(k) + D.I_{R,n}(k) \leq D_{R,n}.W(k)$.

  For $k < n$ or $k > n+1$, $D_{R,n}.H(k) = D.H(k)$, $D_{R,n}.I(k) = D.I(k)$ and $D_{R,n}.W(k) = D.W(k)$ so by validity of $D$ the inequality holds.

  For $k = n$, by definition of $D_{R,n}$ we have $D_{R,n}.H(n) + D_{R,n}.I(n) = D.H(n+1) + D.I(n+1)$. As $D$ admits a right exchange at height $n$, this is bounded by $D.H(n)$, so a fortiori $D.W(n)$.

  For $k = n+1$, $D_{R,n}.H(n+1) + D_{R,n}.I(n+1) = D.H(n) + D.O(n+1) - D.I(n+1) + D.I(n)$. By validity of $D$, this is bounded by $D.W(n) + D.O(n+1) - D.I(n+1) = D_{R,n}.W(n+1)$.
\end{proof}

It is a consequence of Theorem~\ref{thm:joyal-street} that right and left exchanges preserve the meaning of the diagram.
With respect to our data structure described here, it is clear that the following operations can be performed in constant time, since they involve computing fixed formulae over the natural numbers, and testing a fixed number of inequalities:
\begin{itemize}
\item checking whether a left or right exchange is admissible at a given height;
\item given an admissible left or right exchange, computing the rewritten diagram in place.
\end{itemize}
Furthermore, the memory space needed to represent a diagram is linear
in the number of vertices. We will use these
observations as building blocks for our complexity arguments in the
main part of the paper.

\begin{figure}[b]
  \centering
  \begin{subfigure}{0.45\textwidth}
    \centering
    \begin{tikzpicture}[scale=0.5]
      \storecolor{3}{red}
      \storecolor{4}{red}
      \ainit{6}
      \swap 2 4 1 3 5 4 2;
      \aend
    \end{tikzpicture}
    \caption{A funnel with collapsible source}
  \end{subfigure}
    \begin{subfigure}{0.45\textwidth}
    \centering
    \begin{tikzpicture}[scale=0.5]
      \storecolor{0}{red}
      \storecolor{5}{red}
      \ainit{6}
      \swap 0 2 4 1 3;
      \aend
    \end{tikzpicture}
    \caption{A funnel with collapsible target}
  \end{subfigure}
  \caption{Example of funnels} \label{fig:example-funnels}
\end{figure}
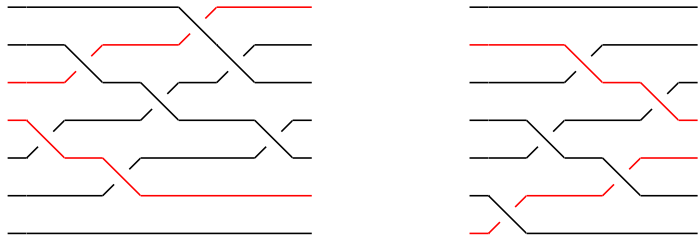

As indicated in Figure~\ref{fig:rightleftexchange}, we write $\Rarrow$ and $\Larrow$ for the relations on diagrams given by a single right exchange and left exchange respectively.
We illustrate some interesting cases of these exchange moves. In degenerate cases where $u$ and $v$ have no inputs or outputs, it can be possible to apply two right exchanges in sequence to the same pair of vertices:
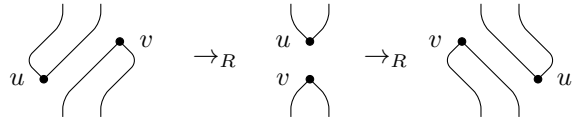
\begin{figure}[H]
  \centering
  \begin{tikzpicture}[scale=0.5, every node/.style={node distance=0.35cm}]
    \startdiagram{3}
    \drawinitialstrands{3}
    \diagslice{2}{0}{2}
    \diagslice{0}{2}{0}
    \finishdiagram
    \node[left of=v1] {$u$};
    \node[right of=v0] {$v$};

    \node at (3.5,-1) {$\rightarrow_R$};

    \begin{scope}[xshift=6cm]
      \startdiagram{3}
      \drawinitialstrands{3}
      \diagslice{0}{2}{0}
      \diagslice{0}{0}{2}
      \finishdiagram
      \node[left of=v1] {$v$};
      \node[left of=v0] {$u$};

    \end{scope}

    \node at (8,-1) {$\rightarrow_R$};

    \begin{scope}[xshift=11cm]
      \startdiagram{3}
      \drawinitialstrands{3}
      \diagslice{0}{0}{2}
      \diagslice{2}{2}{0}
      \finishdiagram
      \node[right of=v1] {$u$};
      \node[left of=v0] {$v$};

    \end{scope}
    \end{tikzpicture}
\end{figure}
\noindent Furthermore, if there are no edges at all, then right exchanges can be applied indefinitely,
which corresponds to the Eckmann-Hilton argument shown in Section~\ref{sec:string-diagrams}.
\begin{figure}[H]
  \centering

  \begin{tikzpicture}[scale=0.5,every node/.style={node distance=0.3cm}]

    \startdiagram{1}
    \drawinitialstrands{1}
    \diagslice{0}{0}{0}
    \diagslice{0}{0}{0}
    \finishdiagram
    \node[left of=v0] {$v$};
    \node[left of=v1] {$u$};

    \node at (1.8,-1) {$\rightarrow_R$};

    \begin{scope}[xshift=4cm]
    \startdiagram{1}
    \diagslice{0}{0}{0}
    \diagslice{0}{0}{0}
    \finishdiagram
    \node[left of=v0] {$u$};
    \node[left of=v1] {$v$};

    \end{scope}

    \node at (5.8,-1) {$\rightarrow_R$};
    \node at (10,-1) {$\rightarrow_R \quad \dots$};

    \begin{scope}[xshift=8cm]
    \startdiagram{1}
    \diagslice{0}{0}{0}
    \diagslice{0}{0}{0}
    \finishdiagram
    \node[left of=v0] {$v$};
    \node[left of=v1] {$u$};

    \end{scope}
    \end{tikzpicture}
\end{figure}
Throughout this article we will use a braid notation to represent right reductions (series of right exchanges), such as in Figure~\ref{fig:example-collapsible-reduction} or Figure~\ref{fig:example-funnels}. These braidings represent the trajectory of the vertices as the reduction progresses, seen from the right-hand side of the diagram. Each crossing in the braid diagram corresponds to an exchange of two nodes in the string diagram.

\subsection{Converting between representations of morphisms} \label{sec:representation-conversion}

This combinatorial encoding given in this section can be rendered into an actual diagram. Generating such a representation from our encoding involves computing suitable planar layouts for the vertices and edges respecting all the properties of this class of topological graphs. Various algorithms can be used to this end. In this work we use a simple layout strategy that enforces a constant vertical spacing between diagram levels and a constant horizontal spacing between the wires at each level.\footnote{It is simple enough to be programmed in \LaTeX, so that our string diagrams are generated with this rendering process, directly from their combinatorial encodings.} Each level is horizontally centered based on the number of wires that cross it. Vertices are horizontally centered between their input and output ports. An example of such a rendered diagram can be found in Figure~\ref{fig:examplediagram}.

It is also possible to convert a morphism expression into a combinatorial encoding of its string diagram in linear time.

\begin{thm}
  Two morphisms expressions denote the same morphism if and only if the
  corresponding diagrams are related by a series of exchanges.
\end{thm}

\begin{proof}
  Theorem~\ref{thm:joyal-street} shows that two morphism expressions
  denote the same morphism if and only if their string diagrams are
  related by a deformation of recumbent graphs. Therefore we only need to show
  that string diagrams are related by a deformation if and only
  if their combinatorial encodings are related by a series of exchanges.

  If two diagrams are related by a series of exchanges, they represent
  the same morphism as exchanges preserve denotation.

  Conversely, let $h$ be a deformation between diagrams $\Gamma$ and $\Gamma'$ in
  generic position, with $h(0) = \Gamma$ and $h(1) = \Gamma'$.  We can
  assume that $h(t)$ is always in generic position except for a finite
  number of $t \in (0,1)$: if it is not the case, translate each
  vertex $v_i$ vertically by $\epsilon_i$, uniformly for all $t \in [0,1]$.
  The $\epsilon_i$ can be chosen to make sure no vertices remain at
  the same height for a non-trivial interval $t \in [u,v]$.

  Furthermore, we can make sure that when $h(t)$ is not in generic
  position, only two vertices in $h(t)$ are at the same height. If
  this is not the case, the deformation can be modified to satisfy
  this condition by picking delays $\eta_i$ for each vertex $v_i$, and
  delaying the movement of each vertex $v_i$ by $\eta_i$ over the
  course of the transformation. Again, the delays can be chosen
  collectively to ensure that at most two vertices occupy the same
  height at a given time.
  
  Let $t_1 < \dots < t_k$ be the instants at which $h(t)$ is not in
  generic position. For any other $t \in [0,1]$ the combinatorial
  encoding of $h(t)$ is defined. By connectedness, the combinatorial
  encoding of $h(t)$ is constant for $t \in (t_i, t_{i+1})$ so this defines a
  sequence of diagrams $D_0, \dots, D_k$.  Since at most one pair of
  vertices exchange heights around instant $t_i$, $D_i$ and $D_{i+1}$
  are related by a single exchange move or are equel. This gives the required
  sequence of exchanges between the source and target diagrams.
\end{proof}

\begin{figure}[b]
  \centering
  \begin{subfigure}{0.45\textwidth}
    \centering
    \begin{tikzpicture}[scale=0.3]
      \begin{scope}[yshift=6.5cm,xshift=-3cm]
        \startdiagram{1}

        \diagslice{0}{0}{2}
        \diagslice{1}{0}{2}
        \diagslice{2}{2}{0}
        \reddiagslice{2}{0}{1}
        \reddiagslice{2}{1}{1}
        \diagslice{1}{1}{0}
        \diagslice{0}{2}{0}
        \storeedgecolor{4}{red}
        \finishdiagram
        \node[rnode] at (v3) {};
        \node[rnode] at (v4) {};
      \end{scope}
      
      \storecolor{2}{red}
      \storecolor{3}{red}
      \ainit{6}
      \swap 1 2 3 2 1 3 2 4 3;
      \aend

      \begin{scope}[yshift=6.5cm,xshift=12cm]
        \startdiagram{1}

        \diagslice{0}{0}{2}
        \reddiagslice{1}{0}{1}
        \reddiagslice{1}{1}{1}
        \diagslice{2}{0}{2}
        \diagslice{2}{1}{0}
        \diagslice{2}{2}{0}
        \diagslice{0}{2}{0}
        \storeedgecolor{2}{red}
        \finishdiagram
        \node[rnode] at (v1) {};
        \node[rnode] at (v2) {};
      \end{scope}

    \end{tikzpicture}
  \end{subfigure}
  \begin{subfigure}{0.45\textwidth}
    \centering
    \begin{tikzpicture}[scale=0.3]
      \begin{scope}[yshift=5.5cm,xshift=-3cm]
        \startdiagram{1}
        \diagslice{0}{0}{2}
        \diagslice{1}{0}{2}
        \diagslice{2}{2}{0}
        \reddiagslice{2}{0}{1}
        \diagslice{1}{1}{0}
        \diagslice{0}{2}{0}
        \finishdiagram
        \node[rnode] at (v3) {};
      \end{scope}
      
      \storecolor{2}{red}
      \ainit{5}
      \swap 1 2 1 2 3;
      \aend

      \begin{scope}[yshift=5.5cm,xshift=8cm]
        \startdiagram{1}
        \diagslice{0}{0}{2}
        \reddiagslice{1}{0}{1}
        \diagslice{2}{0}{2}
        \diagslice{2}{1}{0}
        \diagslice{2}{2}{0}
        \diagslice{0}{2}{0}
        \finishdiagram
        \node[rnode] at (v1) {};
      \end{scope}

    \end{tikzpicture}
  \end{subfigure}
  \caption{A collapsible reduction and its collapsed counterpart}
  \label{fig:example-collapsible-reduction}
\end{figure}

Therefore, solving the word problem for free monoidal categories can
be done by providing an algorithm to determine if two diagrams can be
related by a series of exchanges.  We will first show that right
reductions form a terminating and confluent rewriting strategy on
connected diagrams. Termination will be shown in Section~\ref{sec:termination}
and confluence in Section~\ref{sec:confluence}.

\section{Termination} \label{sec:termination}

To prove termination of right reductions on connected diagrams, we
first introduce the class of linear diagrams, which we will study
before tackling the general case. We will see in
Lemma~\ref{lemma:spirals} that they exhibit the longest reductions.

\begin{defi}
  A diagram with $n$ vertices is \emph{linear} if it is connected, acyclic and has only two leaves (vertices connected to only one edge). We identify its vertices with the indices $1, \dots, n$ such that $1$ and $n$ are the leaves, and $k$ is connected to $k-1$ and $k+1$ for all $1 < k < n$. 
\end{defi}

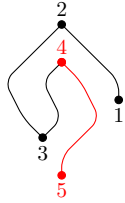
\begin{figure}
  \centering
  \begin{tikzpicture}[scale=0.5]
        \startdiagram{1}
        \diagslice{0}{0}{2}
        \reddiagslice{1}{0}{2}
        \diagslice{3}{1}{0}
        \diagslice{0}{2}{0}
        \reddiagslice{0}{1}{0}
        \storeedgecolor{3}{red}
        \finishdiagram
        \node[rnode] at (v1) {};

        \begin{scope}[every node/.style={node distance=.25cm,scale=0.8}]
          \node[above of=v0] {$2$};
          \node[above of=v1,red] {$4$};
          \node[below of=v2] {$1$};
          \node[below of=v3] {$3$};
          \node[below of=v4,red] {$5$};
        \end{scope}
  \end{tikzpicture}
  \caption{Example of a linear diagram, with final vertices in red}
  \label{fig:linear}
\end{figure}

As the name indicates, linear diagrams have a path-like shape (since
they are connected and cannot contain any
branching). Figure~\ref{fig:linear} gives an example of a linear
diagram.  The choice of the start and end of the indexing is
arbitrary, as it can be reversed. We therefore assume that linear
diagrams come with a chosen order.

\begin{defi}
  In a linear diagram with $n$ vertices, $n \geq 2$, the \emph{final} vertices are the vertices $n-1$ and $n$.
\end{defi}

\begin{defi}
  In a linear diagram, the \emph{final interval} is the set of vertices whose height is between the heights of the final vertices, including the final vertices themselves. If the final interval only consists of the final vertices, the diagram is \emph{collapsible}.
\end{defi}

In Figure~\ref{fig:linear}, vertices 1 and 3 are in the final interval, as well as the final vertices
themselves, vertices 4 and 5.

\begin{defi}
  A right reduction is \emph{collapsible} when its source and target are collapsible, and any exchange between a non-final vertex $v$ and a final vertex $f_1$ is immediately followed or preceded by an exchange between $v$ and the other final vertex $f_2$. In other words, all non-collapsible steps of the reduction are isolated.
\end{defi}

We call these reductions collapsible because as the final vertices move synchronously, they can be merged together: this defines a reduction on a shorter linear diagram. Figure~\ref{fig:example-collapsible-reduction} shows an example of a collapsible reduction with the final vertices in red, and the corresponding collapsed reduction on the shorter diagram.

\begin{defi}
  Given a collapsible reduction on a linear diagram $l$ of size $n$, the corresponding \emph{collapsed
    reduction} is obtained by erasing vertex number $n$ in $l$.
\end{defi}

\begin{defi}
  A right reduction of string diagrams $r : A \Rarrowstar B$ is called a \emph{funnel} when:
  \begin{itemize}
    \item each non-final vertex is exchanged at most once with a final vertex.
    \item if an exchange involves non-final vertices $u$ and $v$, then both $u$ and $v$ are exchanged with a final vertex in the course of the rewrite, and these two final vertices are different.
  \end{itemize}
\end{defi}

\noindent We are especially interested in the cases where the source or target of the funnel is collapsible, as in Figure~\ref{fig:example-funnels}. The name \emph{funnel} comes from the shape of these reductions when depicted as braids: these are reductions where the final vertices converge or diverge from each other.

The following lemmas will establish various properties of funnels that we will need for the decomposition of Lemma~\ref{lemma:funnel-collapsible-decomposition}.

\begin{lem} \label{lemma:simple-final-commutation}
  Let $r : A \Rarrowstar B$ be a funnel with $A$ collapsible and $e : B \Rarrow C$ be a right exchange of two non-final vertices $u$ and $v$ that are not touched by $r$. Then the reduction $r; e : A \Rarrowstar B \Rarrow C$ can be rearranged as $e' ; r' : A \Rarrow B' \Rarrowstar C$, where $e'$ exchanges $u$ and $v$ in $A$, and $r'$ is a funnel.
\end{lem}

\begin{proof}
  As $u$ and $v$ are not touched by $r$, the two reductions commute directly.
\end{proof}

\begin{lem} \label{lemma:final-monotone}
  Let $r : A \Rarrowstar B$ be a funnel reduction where $A$ or $B$ is collapsible. Then, the trajectory of all non-final vertices is monotone in $r$.
\end{lem}

\begin{proof}
  Let us assume by symmetry that the source $A$ of the reduction is collapsible. Consider an exchange of non-final vertices $u$ and $v$ in $r$. By definition, $u$ and $v$ are exchanged with two different final vertices over the course of $r$. Because $A$ is collapsible, this means that both $u$ and $v$ have entered the final interval earlier in the reduction, by being exchanged with the bottom and top final vertices (respectively). Figure~\ref{fig:horizontal-position-funnel} shows the general position of such an exchange.
  
  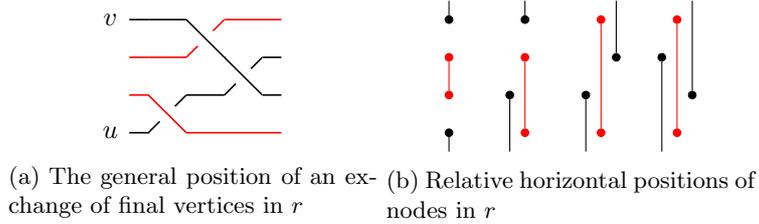
\begin{figure}[hb]
    \centering
    \begin{subfigure}{0.4\textwidth}
      \centering
    \begin{tikzpicture}[scale=0.5]
      \node at (-1,0) {$u$};
      \node at (-1,3) {$v$};
    \storecolor{1}{red}
    \storecolor{2}{red}
    \ainit{3}
    \swap 0 2 1;
    \aend
  \end{tikzpicture}
  \caption{The general position of an exchange of final vertices in $r$}
\end{subfigure}
\begin{subfigure}{0.4\textwidth}
  \centering
  \begin{tikzpicture}[yscale=0.5] 
  \node[bnode] at (0,0) (u) {};
  \node[rnode] at (0,1) (f) {};
  \node[rnode] at (0,2) (g) {};
  \node[bnode] at (0,3) (v) {};

  \draw (u) -- (0,-.5);
  \draw (v) -- (0,3.5);
  \draw[red] (f) -- (g);

  \node[bnode] at (0.8,1) (u) {};
  \node[rnode] at (1,0) (f) {};
  \node[rnode] at (1,2) (g) {};
  \node[bnode] at (1,3) (v) {};

  \draw (u) -- (.8,-.5);
  \draw (v) -- (1,3.5);
  \draw[red] (f) -- (g);

  \node[bnode] at (1.8,1) (u) {};
  \node[rnode] at (2,0) (f) {};
  \node[rnode] at (2,3) (g) {};
  \node[bnode] at (2.2,2) (v) {};

  \draw (u) -- (1.8,-.5);
  \draw (v) -- (2.2,3.5);
  \draw[red] (f) -- (g);

  \node[bnode] at (2.8,2) (u) {};
  \node[rnode] at (3,0) (f) {};
  \node[rnode] at (3,3) (g) {};
  \node[bnode] at (3.2,1) (v) {};

  \draw (u) -- (2.8,-.5);
  \draw (v) -- (3.2,3.5);
  \draw[red] (f) -- (g);
\end{tikzpicture}
\caption{Relative horizontal positions of nodes in $r$}
\end{subfigure}
\caption{Horizontal position of non-final nodes in a funnel} \label{fig:horizontal-position-funnel}
\end{figure}

  As all the exchanges involved are right exchanges, $u$ and $v$ are on different sides of the final edge when they are exchanged: $u$ is on the left and $v$ is on the right of the final edge. This means that $u$ necessarily goes up and $v$ goes down. As this applies to all exchanges of non-final vertices, this means that the trajectory of both vertices is monotone.  
\end{proof}

\begin{defi}
  An \emph{interval right exchange} $i : A \Rarrowstar B$ is a right reduction swapping a vertex $x$ and a set of consecutive vertices $v_1, \dots, v_k$ which is adjacent to $x$ in $A$ and $B$. The vertex $x$ is exchanged first with $v_1$, then $v_2$, up to~$v_k$.
\end{defi}

\noindent An interval right exchange looks like this:
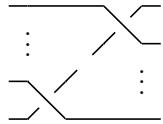
\begin{figure}[H]
  \centering
\begin{tikzpicture}[scale=0.5]
  \storecolor{2}{white}
  \ainit{3}

  \swap 0 1 2;
  \aend
  \node at (0,2.2) {$\vdots$};
  \node at (3,1.2) {$\vdots$};
\end{tikzpicture}
\end{figure}


\begin{lem} \label{lemma:final-interval-collect}
  Let $r: A \Rarrowstar B$ be a funnel reduction with $A$ collapsible and $e : B \Rarrow C$ be an exchange of a non-final vertex $v$ with a final vertex $f_2$, such that $v$ is exchanged with the other final vertex $f_1$ in $r$. This gives a reduction path $r ; e : A \Rarrowstar B \Rarrow C$. A reduction of the same length can be obtained: $i ; r' : A \Rarrowstar D \Rarrowstar C$ where $r'$ is a funnel reduction and $i$ exchanges $v$ with the final interval in $A$.
\end{lem}

\begin{proof}
  By symmetry let us assume that $f_1$ is the highest final vertex, and $f_2$ is the lowest. Somewhere in $r$, $v$ enters the final interval by being exchanged with $f_1$. By Lemma~\ref{lemma:final-monotone}, the trajectory of $v$ in $r$ is monotone. In fact, because $v$ ends up being adjacent to $f_2$ in $B$, $v$ is exchanged exactly once with each non-final vertex that is exchanged with $f_2$ over the course of $r$.

  Exchanges that do not involve $v$ can be divided in two blocks: the ones that are on the right of the trajectory of $v$, and the ones that are on the left. The block on the right commutes with $e$ because the vertices they exchange are disjoint, so we can permute the two.

  \begin{figure}[H]
  \centering

  \begin{tikzpicture}[scale=0.5]
    \storecolor{0}{red}
    \storecolor{1}{white}
    \storecolor{2}{white}
    \storecolor{3}{white}
    \storecolor{4}{red}
    \ainit{5}
    \swap 4 1 3 1 2 1 0;
    \aend
    \draw[zbraid] (-.5,5) -- (-3.5,5);
    \draw[zbraid,red] (-.5,0) -- (-3,3) -- (-3.5,3);
    \draw[zbraid,red] (-.5,4)  -- (-3.5,4);

    \draw[dashed,blue] (-.5,-.2) -- (6,-.2) -- (6,.8) -- (4,2.8) -- (3,2.8) -- (2,3.8) -- (-3,3.8) -- (-3,2.8) -- (-.5,-.2);
    \draw[dashed,blue] (6,1.2) -- (4,3.2) -- (3,3.2) -- (2,4.2) -- (1,4.2) -- (1,5.2) -- (6,5.2) -- (6,1.2);
    \node[blue] at (1,2) {\small left block};
    \node[blue] at (4.2,4) {\small right block};

    \node at (-4,5) {$v$};
    \node at (-4,4) {$f_1$};
    \node at (-4,3) {$f_2$};
    \node at (6.5,-.5) {$e$};
  \end{tikzpicture}
\end{figure}

  We now need to pull the block on the left through the exchanges
  involving $v$. Notice that $v$ is the first vertex to be exchanged
  with $f_1$ over the course of $r$. This is because all other such
  vertices cannot be exchanged with $v$ in $f$ and $v$ is adjacent to
  $f_2$ in $B$. Thus, the block on the left does not contain any
  exchange involving $f_1$: it only contains exchanges involving
  non-final vertices or $f_2$. By successive application of the
  Reidemeister type III move (which pulls one exchange through two
  other exchanges), we can therefore pull the left block through the
  trajectory of $v$.
\end{proof}

\begin{lem} \label{lemma:same-side-pull-through} Let $r : A \Rarrowstar B$ be a funnel reduction with $A$ collapsible, followed by an exchange $e : B \Rarrow B'$ of two non-final vertices $u$, $v$ such that both vertices are exchanged with the same final vertex $f$ in $r$.  Then, the sequence $r; e$ can be rewritten as $e' ; r' : A \Rarrow A' \Rarrowstar B'$ where $e'$ exchanges $u$ and $v$ in $A$, and $r'$ is a funnel.
\end{lem}

\begin{proof}
  We show that $e$ can be pulled through all exchanges involving $u$ or $v$ in $r$. By symmetry, we will assume that the final vertex $f$ exchanged with $u$ and $v$ is the lowest one, and that $u$ is the vertex below $v$ in $B$.

  By induction, consider the last exchange in $r$ that involves one of $u$ or $v$ and another vertex $x$. Because the trajectories of $u$ and $v$ always go up by Lemma~\ref{lemma:final-monotone}, the trajectory of $x$ goes down. As $u$ and $v$ are adjacent in $B$, this last exchange must be between $u$ and $x$, and $x$ must have been exchanged previously with $v$. Moreover, this previous exchange is necessarily the last one involving $v$ (otherwise any later exchange with $y$ would require a later exchange between $y$ and $u$). Therefore, $e$ can be pulled through the last exchanges involving $u$ and $v$.

    \begin{figure}[H]
    \centering
  \begin{tikzpicture}[scale=0.5]
    \storecolor{2}{red}
    \storecolor{3}{red}
    \ainit{5}
    \swap 1 0 3 4 2 1 3 2 3;
    \aend
    \node at (9.8,2) {$x$};
    \node at (9.8,3) {$v$};
    \node at (9.8,4) {$u$};
    \draw[dashed,blue,opacity=0.5] (0,5) -- (0,-1);
    \draw[dashed,blue,opacity=0.5] (8,5) -- (8,-1);
    \draw[dashed,blue,opacity=0.5] (9,5) -- (9,-1);
    \draw[blue,decoration={brace},decorate] (8,-1) -- node[below] {$r$} (0,-1);
    \draw[blue,decoration={brace},decorate] (9,-1) -- node[below] {$e$} (8,-1);

    \node at (10.5,2.5) {$\rightarrow$};

    \begin{scope}[xshift=12cm]
    \storecolor{2}{red}
    \storecolor{3}{red}
    \ainit{5}
    \swap 0 1 0 3 4 2 1 3 2;
    \aend
    \node at (9.8,2) {$x$};
    \node at (9.8,3) {$v$};
    \node at (9.8,4) {$u$};
    \draw[dashed,blue,opacity=0.5] (0,5) -- (0,-1);
    \draw[dashed,blue,opacity=0.5] (1,5) -- (1,-1);
    \draw[dashed,blue,opacity=0.5] (9,5) -- (9,-1);
    \draw[blue,decoration={brace},decorate] (1,-1) -- node[below] {$e'$} (0,-1);
        \draw[blue,decoration={brace},decorate] (9,-1) -- node[below] {$r'$} (1,-1);
        
    \end{scope}
  \end{tikzpicture}
\end{figure}
  \noindent We perform these pull-through moves inductively, which eventually moves $e'$ at the beginning of the reduction. The subsequent exchange the same nodes as $r$ in the same order, so they form a funnel.
\end{proof}

Finally, the following lemma decomposes reductions on linear diagrams into two parts: a collapsible part and a funnel part. This decomposition is illustrated by Figure~\ref{figure:collapsible-funnel-decomposition}. As a collapsible reduction can be seen as a reduction on a shorter linear diagram, this will let us work inductively on the size of the linear string diagram.

\begin{lem} \label{lemma:funnel-collapsible-decomposition}
  Let $r : A \Rarrowstar B$ be a reduction with $A$ collapsible. Then $r$ can be rearranged and decomposed as \[ c ; f : A \Rarrowstar X \Rarrowstar B \] with $c$ collapsible and $f$ a funnel.
\end{lem}

\begin{proof}
  We construct the decomposition into collapsible and funnel parts by induction on the length of the rewrite $r$. For length $0$, the result is clear. For length $1$, there are two cases: if the exchange touches a final vertex, then it goes in the funnel part of the decomposition, otherwise it forms the collapsible part.

  Assume we have a rewrite of length $k + 1$. Use the induction hypothesis to decompose the first $k$ exchanges: \[ c ; f ; z : A \Rarrowstar X \Rarrowstar B' \Rarrow B \] with $c$ collapsible and $f$ a funnel.

  If $f ; z$ is also a funnel, then this gives us the required decomposition. Otherwise, this funnelity can fail for multiple reasons.

  First, it can be that $z$ exchanges a final vertex $v$ with a non-final vertex $w$ that is already exchanged with a final vertex in $f$.  In this case, by Lemma~\ref{lemma:final-interval-collect}, we can rearrange $f; z$ into $i; f'$ where $f'$ is a funnel and $i$ exchanges $v$ with the final interval. As the domain of $i$ is collapsible, $i$ is collapsible itself so we have the required decomposition.

  Second, it can be that $z$ exchanges two non-final vertices that are not exchanged with any final vertex in $f$. In this case, by Lemma~\ref{lemma:simple-final-commutation}, $z$ commutes with $f$: we obtain $c ; z ; f : A \Rarrowstar X \Rarrow X' \Rarrowstar B$, and $c ; z$ is collapsible so we have the required decomposition.

  It cannot be the case that only one of the two non-final vertices $z$ exchanges has been previously exchanged with a final vertex in $f$. This is because the heights of all vertices which have been exchanged with a final vertex lie in the final interval, and all other non-final vertices are outside the final interval.

  Third, it can be that $z$ exchanges two non-final vertices that are both exchanged in $f$ with a final vertex. In this case, as we have assumed that $f ; z$ is not final, it must be the vertices were exchanged with the same final vertex. We can therefore apply Lemma~\ref{lemma:same-side-pull-through} and rearrange the rewrite into $e' ; f'$ with $e'$ exchanging the same non-final vertices as $z$ and $f'$ funnel. As $e'$ is collapsible, this gives the required decomposition.

  Finally, it cannot be the case that $z$ exchanges the two final vertices, as final vertices can never be exchanged together since they are connected by an edge.
\end{proof}

\begin{lem} \label{lemma:extend-to-collapsible}
  Let $r : A \Rarrowstar B$ be a right reduction on a linear diagram. Then $r$ can be extended on some side such that its domain or codomain is collapsible.
\end{lem}

  \begin{proof}
  Our strategy to extend $r$ depends on the topology of the final vertices. We know that vertex $n$ is connected solely to $n-1$ and that $n-1$ is connected to both $n-2$ and $n$. Here are the possible ways these connections can happen:

  
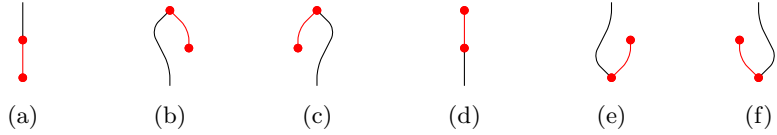
\begin{figure}[H]
  \centering
  \begin{subfigure}{0.15\textwidth}
    \centering
    \begin{tikzpicture}[scale=0.5,yscale=-1]
      \startdiagram{1}
      \reddiagslice{0}{0}{1}
      \reddiagslice{0}{1}{1}
      \storeedgecolor{0}{red}
      \finishdiagram
      \node[rnode] (a) at (v0) {};
      \node[rnode] (b) at (v1) {};
    \end{tikzpicture}
    \caption{}
  \end{subfigure}
  \begin{subfigure}{0.15\textwidth}
    \centering
    \begin{tikzpicture}[scale=0.5]
      \startdiagram{1}
      \reddiagslice{0}{0}{2}
      \reddiagslice{1}{1}{0}
      \storeedgecolor{1}{red}
      \finishdiagram
      \node[rnode] (a) at (v0) {};
      \node[rnode] (b) at (v1) {};
    \end{tikzpicture}
    \caption{}
  \end{subfigure}
    \begin{subfigure}{0.15\textwidth}
    \centering
    \begin{tikzpicture}[scale=0.5]
      \startdiagram{1}
      \reddiagslice{0}{0}{2}
      \reddiagslice{0}{1}{0}
      \storeedgecolor{0}{red}
      \finishdiagram
      \node[rnode] (a) at (v0) {};
      \node[rnode] (b) at (v1) {};
    \end{tikzpicture}
    \caption{}
  \end{subfigure}
  \begin{subfigure}{0.15\textwidth}
    \centering
    \begin{tikzpicture}[scale=0.5]
      \startdiagram{1}
      \reddiagslice{0}{0}{1}
      \reddiagslice{0}{1}{1}
      \storeedgecolor{0}{red}
      \finishdiagram
      \node[rnode] (a) at (v0) {};
      \node[rnode] (b) at (v1) {};
    \end{tikzpicture}
    \caption{}
  \end{subfigure}
  \begin{subfigure}{0.15\textwidth}
    \centering
    \begin{tikzpicture}[scale=0.5,yscale=-1]
      \startdiagram{1}
      \reddiagslice{0}{0}{2}
      \reddiagslice{1}{1}{0}
      \storeedgecolor{1}{red}
      \finishdiagram
      \node[rnode] (a) at (v0) {};
      \node[rnode] (b) at (v1) {};

    \end{tikzpicture}
    \caption{}
  \end{subfigure}
  \begin{subfigure}{0.15\textwidth}
    \centering
    \begin{tikzpicture}[scale=0.5,yscale=-1]
      \startdiagram{1}
      \reddiagslice{0}{0}{2}
      \reddiagslice{0}{1}{0}
      \storeedgecolor{0}{red}
      \finishdiagram
      \node[rnode] (a) at (v0) {};
      \node[rnode] (b) at (v1) {};
    \end{tikzpicture}
    \caption{}
  \end{subfigure}

\end{figure}


  \noindent The orientation of the edges involved is preserved by the reductions so the same situation is observed in both $A$ and $B$.

  Consider situation (a). If the terminal layout $B$ is not collapsible, non-final nodes are present between $n$ and $n-1$. Some of them are on the left side of the edge connecting the final vertices and the others are on the right-hand side. Any two such nodes which are not on the same side of the final edge can be exchanged, so by appending a right reduction to $r$ we can ensure that all the ones on the left are just below $n-1$, and all the ones on the right are just above $n$. Then, by adding further right exchanges, we can move these non-final nodes outside the final interval, leading to a collapsible configuration. This is illustrated in Figure~\ref{fig:straight_to_collapsible_first}. In the situation illustrated in Figure~\ref{fig:straight_to_collapsible_second}, we choose instead to prepend right exchanges before $r$: this is necessary to expell vertices nested inside the cap outside the final interval. The other cases are similar: in each of them, we can either prepend or append right exchanges to obtain a collapsible configuration.
  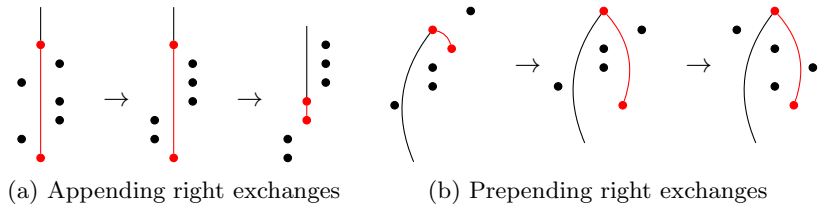
\begin{figure}[b]
  \centering
  \begin{subfigure}{0.45\textwidth}
    \centering
    \begin{tikzpicture}[scale=.5]
      \tikzstyle{bnode}=[draw,circle,fill=black,inner sep=1pt]
      \node[rnode] (a) at (0,0) {};
      \node[rnode] (b) at (0,3) {};
      \draw[red] (a) -- (b);
      \draw (b) -- +(0,1);
      \node[bnode] at (-.5,.5) {};
      \node[bnode] at (-.5,2) {};
      \node[bnode] at (.5,1) {};
      \node[bnode] at (.5,1.5) {};
      \node[bnode] at (.5,2.5) {};

      \node at (2,1.5) {$\rightarrow$};

      \begin{scope}[xshift=3.5cm]
        \node[rnode] (a) at (0,0) {};
        \node[rnode] (b) at (0,3) {};
        \draw[red] (a) -- (b);
        \draw (b) -- +(0,1);
        \node[bnode] at (-.5,.5) {};
        \node[bnode] at (-.5,1) {};
        \node[bnode] at (.5,1.5) {};
        \node[bnode] at (.5,2) {};
        \node[bnode] at (.5,2.5) {};

        \node at (2,1.5) {$\rightarrow$};

        \begin{scope}[xshift=3.5cm]
          \node[rnode] (a) at (0,1) {};
          \node[rnode] (b) at (0,1.5) {};
          \draw[red] (a) -- (b);
          \draw (b) -- +(0,2);
          \node[bnode] at (-.5,0) {};
          \node[bnode] at (-.5,.5) {};
          \node[bnode] at (.5,3) {};
          \node[bnode] at (.5,2) {};
          \node[bnode] at (.5,2.5) {};
        \end{scope}
      \end{scope}
    \end{tikzpicture}
    \caption{Appending right exchanges}
    \label{fig:straight_to_collapsible_first}
  \end{subfigure}
  \begin{subfigure}{0.45\textwidth}
    \centering
    \begin{tikzpicture}[scale=0.5]
      \tikzstyle{bnode}=[draw,circle,fill=black,inner sep=1pt]
      \node[rnode] (a) at (0,0) {};
      \node[rnode] (b) at (-.5,2.5) {};
      \draw[red] (a) edge[bend right] (b);
      \draw (b) edge[bend right] +(-.5,-3.5);
      \node[bnode] at (-.5,.5) {};
      \node[bnode] at (.5,1) {};
      \node[bnode] at (-.5,1.5) {};
      \node[bnode] at (-1.5,2) {};

      \node at (-2.5,1) {$\rightarrow$};

      \begin{scope}[xshift=-4.5cm]
        \node[rnode] (a) at (0,0) {};
        \node[rnode] (b) at (-.5,2.5) {};
        \draw[red] (a) edge[bend right] (b);
        \draw (b) edge[bend right] +(-.5,-3.5);
        \node[bnode] at (-.5,1.5) {};
        \node[bnode] at (-.5,1) {};
        \node[bnode] at (.5,2) {};
        \node[bnode] at (-1.7,.5) {};

        \node at (-2.5,1) {$\rightarrow$};

        \begin{scope}[xshift=-4.5cm]
          \node[bnode] at (-1.5,0) {};
          \node[bnode] at (-.5,0.5) {};
          \node[bnode] at (-.5,1) {};
          \node[rnode] (a) at (0,1.5) {};
          \node[rnode] (b) at (-.5,2) {};
          \draw[red] (a) edge[bend right] (b);
          \draw (b) edge[bend right] +(-.5,-3.5);
          \node[bnode] at (.5,2.5) {};

        \end{scope}
      \end{scope}
    \end{tikzpicture}
    \caption{Prepending right exchanges} \label{fig:straight_to_collapsible_second}
  \end{subfigure}
  \caption{Extending a reduction so that one end is collapsed}
\end{figure}
\end{proof}

We can now show termination of right reductions. A finer analysis of the bound obtained on the length of reductions is presented in Section~\ref{sec:complexity}.

\begin{thm} \label{thm:termination-connected}
  Right reductions are terminating on connected diagrams.
\end{thm}

\begin{proof}
  We first show termination for linear diagrams. Notice that the length of a funnel reduction on a linear diagram of size $n$ is bounded by $F(n) = O(n^2)$. This is because exchanges involving final vertices happen at most $O(n)$ times and exchanges involving only non-final vertices happen at most once per pair of non-final vertices by Lemma~\ref{lemma:final-monotone}.

  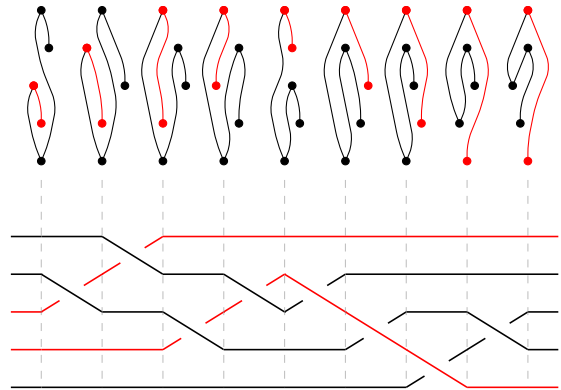
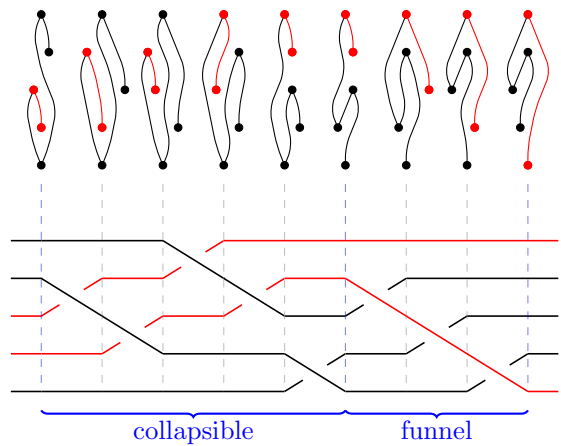
\begin{figure}[H]
  \vspace{-.3cm}
  \centering
  \begin{subfigure}{0.45\textwidth}
    \centering
    \begin{tikzpicture}[yscale=0.5,xscale=0.8]
        \node[draw,circle,white] at (0,-7) {};

        \begin{scope}[xscale=0.25,yshift=4.5cm]
        \startdiagram{1}
        \diagslice{0}{0}{2}
        \diagslice{1}{1}{0}
        \reddiagslice{0}{0}{2}
        \reddiagslice{1}{1}{0}
        \diagslice{0}{2}{0}
        \storeedgecolor{3}{red}
        \finishdiagram
        \node[rnode] at (v2) {};
        \end{scope}
        
        \begin{scope}[xshift=1cm,xscale=0.25,yshift=4.5cm]
        \startdiagram{1}
        \diagslice{0}{0}{2}
        \reddiagslice{0}{0}{2}
        \diagslice{3}{1}{0}
        \reddiagslice{1}{1}{0}
        \diagslice{0}{2}{0}
        \storeedgecolor{3}{red}
        \finishdiagram
        \node[rnode] at (v1) {};
        \end{scope}

        \begin{scope}[xshift=2cm,xscale=0.25,yshift=4.5cm]
        \startdiagram{1}
        \reddiagslice{0}{0}{2}
        \diagslice{2}{0}{2}
        \diagslice{3}{1}{0}
        \reddiagslice{1}{1}{0}
        \diagslice{0}{2}{0}
        \storeedgecolor{1}{red}
        \finishdiagram
        \node[rnode] at (v0) {};
        \end{scope}

        \begin{scope}[xshift=3cm,xscale=0.25,yshift=4.5cm]
        \startdiagram{1}
        \reddiagslice{0}{0}{2}
        \diagslice{2}{0}{2}
        \reddiagslice{1}{1}{0}
        \diagslice{2}{1}{0}
        \diagslice{0}{2}{0}
        \storeedgecolor{1}{red}
        \finishdiagram
                \node[rnode] at (v0) {};
        \end{scope}

        \begin{scope}[xshift=4cm,xscale=0.25,yshift=4.5cm]
        \startdiagram{1}
        \reddiagslice{0}{0}{2}
        \reddiagslice{1}{1}{0}
        \diagslice{1}{0}{2}
        \diagslice{2}{1}{0}
        \diagslice{0}{2}{0}
        \storeedgecolor{1}{red}
        \finishdiagram
                \node[rnode] at (v0) {};
        \end{scope}

        \begin{scope}[xshift=5cm,xscale=0.25,yshift=4.5cm]
        \startdiagram{1}
        \reddiagslice{0}{0}{2}
        \diagslice{1}{0}{2}
        \reddiagslice{3}{1}{0}
        \diagslice{2}{1}{0}
        \diagslice{0}{2}{0}
        \storeedgecolor{1}{red}
        \finishdiagram
                \node[rnode] at (v0) {};
        \end{scope}

        \begin{scope}[xshift=6cm,xscale=0.25,yshift=4.5cm]
        \startdiagram{1}
        \reddiagslice{0}{0}{2}
        \diagslice{1}{0}{2}
        \diagslice{2}{1}{0}
        \reddiagslice{2}{1}{0}
        \diagslice{0}{2}{0}
        \storeedgecolor{1}{red}
        \finishdiagram
                \node[rnode] at (v0) {};
        \end{scope}
                
        \begin{scope}[xshift=7cm,xscale=0.25,yshift=4.5cm]
        \startdiagram{1}
        \reddiagslice{0}{0}{2}
        \diagslice{1}{0}{2}
        \diagslice{2}{1}{0}
        \diagslice{0}{2}{0}
        \reddiagslice{0}{1}{0}
        \storeedgecolor{1}{red}
        \finishdiagram
                \node[rnode] at (v0) {};
        \end{scope}

        \begin{scope}[xshift=8cm,xscale=0.25,yshift=4.5cm]
        \startdiagram{1}
        \reddiagslice{0}{0}{2}
        \diagslice{1}{0}{2}
        \diagslice{0}{2}{0}
        \diagslice{0}{1}{0}
        \reddiagslice{0}{1}{0}
        \storeedgecolor{1}{red}
        \finishdiagram
        \node[rnode] at (v0) {};
        \end{scope}

  \begin{scope}[yshift=-6cm]
  \storecolor{1}{red}
  \storecolor{2}{red}
  \ainit{4}
  \swap 2 3 1 2 2 1 0 1;
  \aend
  \end{scope}
  \foreach \x in {0,...,8} {
    \draw[gray,dashed,opacity=.5] (\x,-.5) -- (\x,-6);
    }
    \end{tikzpicture}
    \vspace{-.5cm}
  \caption{Reducing a diagram to its normal form}
  \end{subfigure}

  \begin{subfigure}{0.45\textwidth}
    \centering
    \hspace{-.5cm}
    \begin{tikzpicture}[yscale=0.5,xscale=0.8]
  \tikzstyle{bnode}=[draw,black,circle,fill=black,inner sep=1pt]
  \tikzstyle{rnode}=[draw,red,circle,fill=red,inner sep=1pt]
    \node[draw,circle,white] at (0,-7) {};

    \begin{scope}[xscale=0.25,yshift=4.5cm]
      \startdiagram{1}
      \diagslice{0}{0}{2}
      \diagslice{1}{1}{0}
      \reddiagslice{0}{0}{2}
      \reddiagslice{1}{1}{0}
      \diagslice{0}{2}{0}
      \storeedgecolor{3}{red}
      \finishdiagram
      \node[rnode] at (v2) {};
      \node[rnode] at (v3) {};
    \end{scope}

    \begin{scope}[xshift=1cm,xscale=0.25,yshift=4.5cm]
      \startdiagram{1}
      \diagslice{0}{0}{2}
      \reddiagslice{0}{0}{2}
      \diagslice{3}{1}{0}
      \reddiagslice{1}{1}{0}
      \diagslice{0}{2}{0}
      \storeedgecolor{3}{red}
      \finishdiagram
      \node[rnode] at (v1) {};
      \node[rnode] at (v3) {};
    \end{scope}

    \begin{scope}[xshift=2cm,xscale=0.25,yshift=4.5cm]
      \startdiagram{1}
      \diagslice{0}{0}{2}
      \reddiagslice{0}{0}{2}
      \reddiagslice{1}{1}{0}
      \diagslice{2}{1}{0}
      \diagslice{0}{2}{0}
      \storeedgecolor{3}{red}
      \finishdiagram
      \node[rnode] at (v1) {};
      \node[rnode] at (v2) {};
    \end{scope}

    \begin{scope}[xshift=3cm,xscale=0.25,yshift=4.5cm]
      \startdiagram{1}
      \reddiagslice{0}{0}{2}
      \diagslice{2}{0}{2}
      \reddiagslice{1}{1}{0}
      \diagslice{2}{1}{0}
      \diagslice{0}{2}{0}
      \storeedgecolor{1}{red}
      \finishdiagram
      \node[rnode] at (v0) {};
      \node[rnode] at (v2) {};
    \end{scope}

    \begin{scope}[xshift=4cm,xscale=0.25,yshift=4.5cm]
      \startdiagram{1}
      \reddiagslice{0}{0}{2}
      \reddiagslice{1}{1}{0}
      \diagslice{1}{0}{2}
      \diagslice{2}{1}{0}
      \diagslice{0}{2}{0}
      \storeedgecolor{1}{red}
      \finishdiagram
      \node[rnode] at (v0) {};
      \node[rnode] at (v1) {};
    \end{scope}

    \begin{scope}[xshift=5cm,xscale=0.25,yshift=4.5cm]
      \startdiagram{1}
      \reddiagslice{0}{0}{2}
      \reddiagslice{1}{1}{0}
      \diagslice{1}{0}{2}
      \diagslice{0}{2}{0}
      \diagslice{0}{1}{0}
      \storeedgecolor{1}{red}
      \finishdiagram
      \node[rnode] at (v0) {};
      \node[rnode] at (v1) {};
    \end{scope}

   \begin{scope}[xshift=6cm,xscale=0.25,yshift=4.5cm]
      \startdiagram{1}
      \reddiagslice{0}{0}{2}
      \diagslice{1}{0}{2}
      \reddiagslice{3}{1}{0}
      \diagslice{0}{2}{0}
      \diagslice{0}{1}{0}
      \storeedgecolor{1}{red}
      \finishdiagram
      \node[rnode] at (v0) {};
      \node[rnode] at (v2) {};
    \end{scope}

   \begin{scope}[xshift=7cm,xscale=0.25,yshift=4.5cm]
     \startdiagram{1}
     \reddiagslice{0}{0}{2}
     \diagslice{1}{0}{2}
     \diagslice{0}{2}{0}
     \reddiagslice{1}{1}{0}
     \diagslice{0}{1}{0}
     \storeedgecolor{1}{red}
     \finishdiagram
     \node[rnode] at (v0) {};
     \node[rnode] at (v3) {};
   \end{scope}

     \begin{scope}[xshift=8cm,xscale=0.25,yshift=4.5cm]
     \startdiagram{1}
     \reddiagslice{0}{0}{2}
     \diagslice{1}{0}{2}
     \diagslice{0}{2}{0}
     \diagslice{0}{1}{0}
     \reddiagslice{0}{1}{0}
     \storeedgecolor{1}{red}
     \finishdiagram
     \node[rnode] at (v0) {};
     \node[rnode] at (v4) {};
   \end{scope}

  \begin{scope}[yshift=-6cm]
  \storecolor{1}{red}
  \storecolor{2}{red}
  \ainit{4}
  \swap 2 1 3 2 0 2 1 0;
  \aend
  \end{scope}
  \foreach \x/\c in {0/blue,1/gray,2/gray,3/gray,4/gray,5/blue,6/gray,7/gray,8/blue} {
    \draw[dashed,opacity=.5,\c] (\x,-.5) -- (\x,-6);
  }
  \draw[thick,blue,decoration={brace,mirror},decorate] (0,-6.5) -- node[below] {collapsible} (5,-6.5);
  \draw[thick,blue,decoration={brace,mirror},decorate] (5,-6.5) -- node[below] {funnel} (8,-6.5);

  \end{tikzpicture}
  \caption{Decomposition from Lemma~\ref{lemma:funnel-collapsible-decomposition}}
\end{subfigure}
\caption{Decomposition into collapsible and funnel reductions} \label{figure:collapsible-funnel-decomposition}
\end{figure}

  
  We can now define a bound $B(n)$ on the length of right reductions on linear diagrams of size $n$, by induction on $n$. Consider such a reduction $r$. By Lemma~\ref{lemma:extend-to-collapsible}, we can assume that one end of $r$ is collapsible (by making $r$ potentially longer). By Lemma~\ref{lemma:funnel-collapsible-decomposition},
we can decompose $r$ into a funnel part $f$ and a collapsible part $c$.  The collapsible part $c$ gives rise to a collapsed reduction $c'$, whose length is bounded by $B(n-1)$ by induction. Because an exchange involving the last vertex in the shorter diagram
corresponds to two exchanges in the longer diagram, we obtain $|c| \leq 2 B(n-1)$. By the observation above, $|f| \leq F(n)$. Hence, $|r| \leq 2 B(n-1) + F(n) \eqqcolon B(n)$.
This shows termination of right reductions on linear diagrams.

  We now move to the general case of connected diagrams. Assume by contradiction that there is an infinite reduction on a connected diagram. By the pigeonhole principle, there is a pair of vertices that are exchanged infinitely often. Consider a simple path between these two vertices and erase all vertices not visited by this path. The infinite reduction on the connected diagram induces an infinite reduction on the linear diagram, which contradicts termination on linear diagrams.
\end{proof}

\noindent Some diagrams are not connected as graphs but all their vertices are connected to a boundary. Theorem~\ref{thm:termination-connected} can be extended to these cases.

\begin{defi}
  A diagram $D$ is \emph{boundary-connected} if it is connected or all vertices in $D$
  are connected to one of the two boundaries of the diagram.
\end{defi}

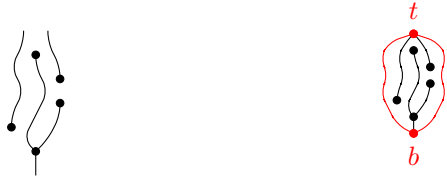
\begin{figure}[b]
  \centering
    \begin{subfigure}{0.4\textwidth}
    \centering
    \begin{tikzpicture}[scale=0.32]
      \startdiagram{3}
      \drawinitialstrands{3}
      \diagslice{1}{0}{1}
      \diagslice{2}{1}{0}
      \diagslice{2}{0}{1}
      \diagslice{0}{1}{0}
      \diagslice{0}{2}{1}
      \finishdiagram
    \end{tikzpicture}
    \caption{The original diagram $D$}
    \label{fig:connected-via-boundary-a}
  \end{subfigure}
    \begin{subfigure}{0.4\textwidth}
      \centering
      \vspace{-.5cm}
    \begin{tikzpicture}[scale=0.22]
      \startdiagram{1}
      \reddiagslice{0}{0}{4}
      \diagslice{2}{0}{1}
      \diagslice{3}{1}{0}
      \diagslice{3}{0}{1}
      \diagslice{1}{1}{0}
      \diagslice{1}{2}{1}
      \reddiagslice{0}{3}{0}
      \storeedgecolor{0}{red}
      \storeedgecolor{3}{red}
      \finishdiagram
      \node[rnode] at (v0) {};
      \node[rnode] at (v6) {};
      \node[node distance=.3cm,above of=v0,red] {$t$};
      \node[node distance=.3cm,below of=v6,red] {$b$};
    \end{tikzpicture}
    \caption{The transformed diagram $D'$}
  \end{subfigure}
  \caption{Adding nodes on the boundaries to make a diagram connected} \label{fig:connected-via-boundary}
\end{figure}

Figure~\ref{fig:connected-via-boundary-a} shows a diagram that is not connected
(it has three connected components) but which is boundary-connected, since each
component contains an open wire. Each vertex is therefore connnected to either the
top or bottom boundary of the diagram via these open wires.

\begin{cor} \label{corollary:boundary}
  Right reductions on boundary-connected diagrams are terminating.
\end{cor}

\begin{proof}
  Let $D$ be boundary-connected. Consider the diagram $D'$ obtained from $D$ by adding two vertices $b, t$ at the bottom and top boundaries, and adding two edges from $b$ to $t$ on each side of the diagram, as in Figure~\ref{fig:connected-via-boundary}. Every edge connected to the boundary in $D$ is connected to one of $b, t$ in $D'$, so $D'$ is connected. Any right reduction on $D$ induces a reduction of the same length on $D'$, therefore right reductions on $D$ terminate.
\end{proof}

\section{Upper bound on reduction length} \label{sec:complexity}

\noindent Beyond termination, we can use the same proof techniques
to derive an asymptotic bound on reduction length. We first introduce a parametric cost
on exchanges of linear diagrams:

\begin{defi}
  Given a reduction $r$ on a linear diagram of size $n$ and an integer $w$, the \emph{cost} of $r$ at weight $w$ is $X + w Y$, where $X$ is the number of exchanges not involving vertex number $n$ in $r$ and $Y$ is the number of exchanges involving vertex $n$ in~$r$.
\end{defi}

\begin{lem} \label{lemma:funnel-cost}
  The maximum cost at weight $w$ of a funnel with a collapsible end is $f(n,w) = O(n^2 + w n)$, where $n$ is the length of the linear diagram.
\end{lem}

\begin{proof}
  A funnel contains two types of exchanges. Those with final vertices account for at most $n - 2$ exchanges, because there is at most one for each non-final vertex. The ones with only non-final vertices are bounded by $O(n^2)$ as any pair of non-final vertices is exchanged at most once by Lemma~\ref{lemma:final-monotone}. The bound follows from the definition of the cost.
\end{proof}

\begin{thm} \label{thm:max-linear-cost}
  The maximum cost of a right reduction on a linear diagram is $O(n^3 + w \cdot n^2)$, where $n$ is the size of the diagram.
\end{thm}

\begin{proof}
  Let $g(n,w) = \sum_{k=1}^{n} f(k,w+n-k)$. We show that $g(n,w)$ bounds the cost of any right reduction on a linear diagram of size $n$. By Lemma~\ref{lemma:funnel-cost}, the desired bound will follow.
   We work by induction on $n$. For $n \leq 1$, no right exchanges can be performed, so the bound holds.  Consider a reduction $r : A \Rarrowstar B$ on a linear diagram of size $n$. By Lemma~\ref{lemma:extend-to-collapsible}, we can assume that $A$ or $B$ is collapsible (up to an extension which increases the cost of $r$). By Lemma~\ref{lemma:funnel-collapsible-decomposition}, we can rearrange the exchanges in $r$ to obtain a funnel and a collapsible reduction. By definition, the cost of the funnel part is bounded by $f(n,w)$. For the collapsible part, consider the reduction induced by merging the final vertices together: this gives a reduction on a diagram of size $n-1$. Each exchange involving the last vertex in this induced reduction corresponds to an exchange of both final vertices in the original reduction, which has cost $w+1$. Therefore, by induction, the cost of the collapsible part is bounded by $g(n-1,w+1)$. We therefore obtain the bound $g(n-1,w+1) + f(n,w) = g(n,w)$ on the cost of $r$ at weight $w$.
\end{proof}

\begin{thm} \label{thm:termination-connected-bound}
  The maximum length of a reduction on a diagram of size $n$ vertices is $O(n^3)$.
\end{thm}

\begin{proof}
  By the same argument as Corollary~\ref{corollary:boundary} we can assume that the diagram is connected.
  Consider a connected string diagram $D$ with $v$ vertices. Pick a spanning tree on $D$ and let $D'$ be the string diagram obtained from $D$ by removing all edges which are not in the spanning tree. Any reduction on $D$ induces a reduction of the same length on $D'$, so it is enough to bound the length of reductions on $D'$.

  Pick an arbitrary vertex of $D'$ as root for the tree and consider a depth-first search of $D'$ from that root.  This defines an envelope on the tree, which can be seen as a linear diagram $L$ if we duplicate the nodes every time they are visited (see Figure~\ref{fig:unfold-spanning-tree}). The length of this diagram is linear in the number of edges in $D'$, which is linear in the number of vertices in~$D$ (since $D'$ is a tree, it has one less edge than vertices, and its vertices are those of $D$).

  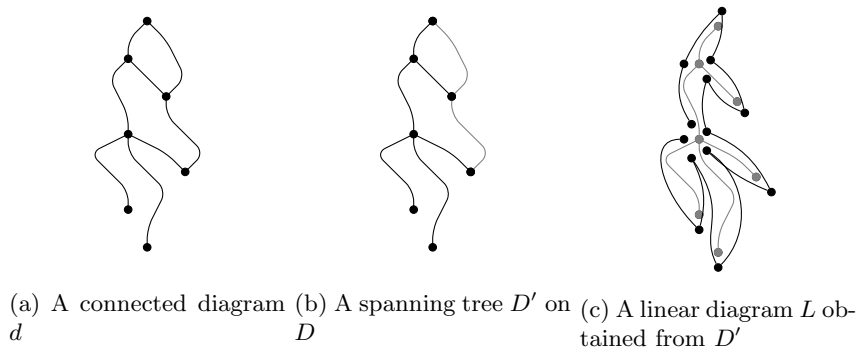
\begin{figure}[b]
  \centering
  \begin{subfigure}{0.3\textwidth}
    \centering
    \begin{tikzpicture}[scale=0.5]
      \startdiagram{1}
      \diagslice{0}{0}{2}
      \diagslice{0}{1}{2}
      \diagslice{1}{2}{1}
      \diagslice{0}{1}{3}
      \diagslice{2}{2}{0}
      \diagslice{0}{1}{0}
      \diagslice{0}{1}{0}
      \finishdiagram
  \end{tikzpicture}
  \caption{A connected diagram $D$}
\end{subfigure}
\begin{subfigure}{0.3\textwidth}
  \centering
  \begin{tikzpicture}[scale=0.5]
      \startdiagram{1}
      \diagslice{0}{0}{2}
      \diagslice{0}{1}{2}
      \diagslice{1}{2}{1}
      \diagslice{0}{1}{3}
      \diagslice{2}{2}{0}
      \diagslice{0}{1}{0}
      \diagslice{0}{1}{0}
      \storeedgecolor{0}{red}
      \storeedgecolor{2}{red}
      \storeedgecolor{3}{red}
      \storeedgecolor{5}{red}
      \storeedgecolor{6}{red}
      \storeedgecolor{7}{red}
      \finishdiagram
      \node[bnode,red] at (v0) {};
      \node[bnode,red] at (v1) {};
      \node[bnode,red] at (v2) {};
      \node[bnode,red] at (v3) {};
      \node[bnode,red] at (v4) {};
      \node[bnode,red] at (v5) {};
      \node[bnode,red] at (v6) {};

  \end{tikzpicture}
  \caption{A spanning tree $D'$ on $D$}
\end{subfigure}
\begin{subfigure}{0.3\textwidth}
  \centering
  \begin{tikzpicture}[scale=0.5]
      \startdiagram{1}
      \diagslice{0}{0}{2}
      \diagslice{0}{1}{2}
      \diagslice{1}{2}{1}
      \diagslice{0}{1}{3}
      \diagslice{2}{2}{0}
      \diagslice{0}{1}{0}
      \diagslice{0}{1}{0}
      \storeedgecolor{1}{white}
      \storeedgecolor{4}{white}
      \finishdiagram

    \node[bnode] at (v0) {};
    \node[bnode] at (v1) {};
    \node[bnode] at (v2) {};
    \node[bnode] at (v3) {};
    \node[bnode] at (v4) {};
    \node[bnode] at (v5) {};
    \node[bnode] at (v6) {};

    \node[bnode,red] at ($(v6)+(0,-.4)$) (0v0) {};
    \node[bnode,red] at ($(v5)+(0,-.4)$) (1v0) {};
    \node[bnode,red] at ($(v4)+(0.4,-.4)$) (2v0) {};
    \node[bnode,red] at ($(v3)+(.2,-.3)$) (3v0) {};
    \node[bnode,red] at ($(v3)+(-.2,-.5)$) (3v1) {};
    \node[bnode,red] at ($(v3)+(.2,.2)$) (3v2) {};
    \node[bnode,red] at ($(v3)+(-.2,.4)$) (3v3) {};
    \node[bnode,red] at ($(v3)+(-.4,0)$) (3v4) {};
    \node[bnode,red] at ($(v2)+(.2,-.3)$) (4v0) {};
    \node[bnode,red] at ($(v1)+(.2,-.4)$) (5v0) {};
    \node[bnode,red] at ($(v1)+(.3,.1)$) (5v1) {};
    \node[bnode,red] at ($(v1)+(-.4,0)$) (5v2) {};
    \node[bnode,red] at ($(v0)+(0.1,.4)$) (6v0) {};

    \draw[red] (3v4) .. controls ($(3v4)+(-1,0)$) and ($(1v0)+(-1,1)$) .. (1v0);
    \draw[red] (1v0) edge[bend right=20] (3v1);
    \draw[red] (3v1) .. controls ($(3v1)+(1,-1.5)$) and ($(0v0)+(-.5,1)$) .. (0v0);
    \draw[red] (0v0) edge[bend right=50] (3v0);
    \draw[red] (3v0) edge[bend right=10] (2v0);
    \draw[red] (2v0) edge[bend right=20] (3v2);
    \draw[red] (3v2) edge[bend left=15] (5v0);
    \draw[red] (5v0) edge[bend right=20] (4v0);
    \draw[red] (4v0) edge[bend right=20] (5v1);
    \draw[red] (5v1) edge[bend right=15] (6v0);
    \draw[red] (6v0) edge[bend right=10] (5v2);
    \draw[red] (5v2) edge[bend right=25] (3v3);

  \end{tikzpicture}
  \caption{A linear diagram $L$ obtained from $D'$}
  
\end{subfigure}
\caption{Transforming a connected diagram to a linear diagram \label{fig:unfold-spanning-tree}}
\end{figure}

  Any right reduction on $D'$ translates to a right reduction on $L$, where exchanging vertices $x$ and $y$ corresponds to exchanging all the copies of $x$ and $y$ in the same way.  The reduction on $L$ is therefore at least as long as the reduction on $D'$. By Theorem~\ref{thm:max-linear-cost} and because the number of vertices in $L$ is linear in $v$, the reduction on $L$ has length $O(v^3)$. This bound also applies the original reduction on $D'$ and hence on $D$.
\end{proof}

\noindent This asymptotic bound on reduction length is attained by a class of spiral-shaped diagrams:

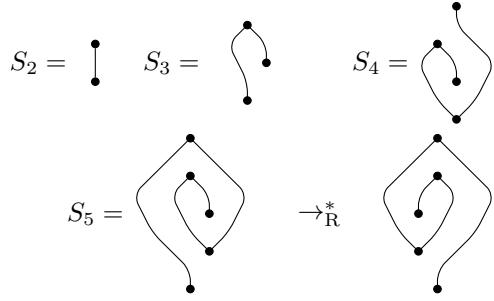
\begin{figure}[H]
  \centering
  \begin{tikzpicture}[scale=0.5]
    \begin{scope}[xshift=1cm]
    \node[bnode] at (0,-1) (a) {};
    \node[bnode] at (0,0) (b) {};
    \draw (a) -- (b);
    \node at (-1.5,-.5) {$S_2 = $};
    \end{scope}

    \begin{scope}[xshift=5cm,yshift=1cm]
        \startdiagram{1}
        \diagslice{0}{0}{2}
        \diagslice{1}{1}{0}
        \diagslice{0}{1}{0}
        \finishdiagram
    \end{scope}

    \begin{scope}[xshift=10.5cm,yshift=1.5cm]
      \startdiagram{1}
      \diagslice{0}{0}{1}
      \diagslice{0}{0}{2}
      \diagslice{1}{1}{0}
      \diagslice{0}{2}{0}
      \finishdiagram
    \end{scope}
    
    \foreach \n in {3,4} {%
      \pgfmathsetmacro\sft{-11+5.5*\n}
      \begin{scope}[xshift=\sft cm]
        \node at (-2.5,-.5) {$S_\n = $};
    \end{scope}}

    \begin{scope}[yshift=-2cm,xshift=-13cm]
      \begin{scope}[xshift=16.5cm]
        \node at (-2.5,-2.5) {$S_5 = $};
        \startdiagram{1}
        \diagslice{0}{0}{2}
        \diagslice{1}{0}{2}
        \diagslice{2}{1}{0}
        \diagslice{1}{2}{0}
        \diagslice{0}{1}{0}
        \finishdiagram
    \end{scope}
    \begin{scope}[xshift=23cm,xscale=-1]
        \startdiagram{1}
        \diagslice{0}{0}{2}
        \diagslice{1}{0}{2}
        \diagslice{2}{1}{0}
        \diagslice{1}{2}{0}
        \diagslice{0}{1}{0}
        \finishdiagram
      \node at (3.1,-2.5) {$\Rarrowstar$};
    \end{scope}
    \end{scope}
  \end{tikzpicture}
\end{figure}
  

\begin{lem} \label{lemma:spirals}
  For all $n$, the diagram $S_n$ right reduces to its normal form in $\binom{n}{3}$ steps.
\end{lem}

\begin{proof}
  A reduction of $S_n$ to its normal form starts with $n-2$ exchanges of one end with the rest, followed by the reduction for $S_{n-1}$ where the end weighs one more vertex. Therefore, the cost of a right reduction of $S_n$ to its normal form is $s(n,w) = w(n-2) + s(n-1,w+1)$. We also have $s(2,w) = 0$ for all $w$. From this we obtain \[ s(n,w) = \frac{(n-1)(n-2)(n-3+3w)}{6} \] which gives $\binom{n}{3}$ for $w=1$.
\end{proof}

\section{Confluence} \label{sec:confluence}

\begin{lem} \label{lemma:locally-confluent}
  The right reduction relation is locally confluent.
\end{lem}

\begin{proof}
  Let $F, G, H$ be diagrams with $G {}_R\leftarrow F \rightarrow_R  H$.
  If the two pairs of nodes exchanged in the two branches are disjoint, then the exchanges commute and we can close the diagram in one step: we have $H \Rarrow K$ and $G \Rarrow K$. Otherwise, the rewriting patterns overlap. There are nodes $u$, $v$ and $w$ in $F$, such that $u$ and $v$ are adjacent and are exchanged to obtain $G$, and $v$ and $w$ are adjacent and are exchanged to obtain $H$. The situation looks like this:

  
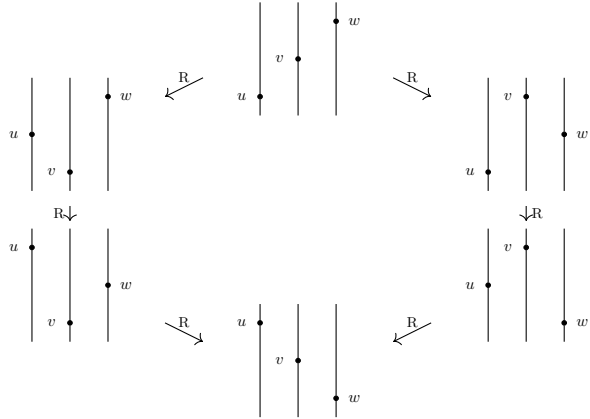
\begin{figure}[H]
  \centering
\begin{tikzpicture}[scale=0.5,every node/.style={scale=0.6}]
  \foreach \yu/\yv/\yw/\xs/\ys in {
    0/1/2/0/0,
    1/0/2/-6cm/-2cm,
    0/2/1/6cm/-2cm,
    2/0/1/-6cm/-6cm,
    1/2/0/6cm/-6cm,
    2/1/0/0cm/-8cm
  } {
    \begin{scope}[xshift=\xs,yshift=\ys,node distance=0.4cm]
      \node[bnode] at (0,\yu) (u) {};
      \node[bnode] at (1,\yv) (v) {};
      \node[bnode] at (2,\yw) (w) {};
      \node[left of=u] {$u$};
      \node[left of=v] {$v$};
      \node[right of=w] {$w$};
      \draw (0,-.5) -- (u) -- (0,2.5);
      \draw (1,-.5) -- (v) -- (1,2.5);
      \draw (2,-.5) -- (w) -- (2,2.5);
    \end{scope}
  }
  \begin{scope}[yshift=-1cm]
  \draw[->] (-1.5,1.5) -- node[above] {\small R} (-2.5,1); 
  \draw[->] (3.5,1.5) -- node[above] {\small R} (4.5,1); 
  \end{scope}

\begin{scope}[rotate=180,xshift=-2cm,yshift=5cm]
  \draw[<-] (-1.5,1.5) -- node[above] {\small R} (-2.5,1); 
  \draw[<-] (3.5,1.5) -- node[above] {\small R} (4.5,1); 
\end{scope}
\draw[<-] (-5,-3.3) -- node[left] {\small R} (-5,-2.9);
\draw[<-] (7,-3.3) -- node[right] {\small R} (7,-2.9);

\end{tikzpicture}
\end{figure}


  As $u$ and $v$ can be exchanged in $F$, there is no edge from the output of $v$ to the input of $u$, and any edge going from the output of $w$ to the input of $u$ has to pass to the left of $v$. As $v$ and $w$ can be exchanged in $F$, there is no edge from the output of $w$ to the input of $v$, and any edge going from the output of $w$ to the input of $u$ has to pass to the right of $v$, which is impossible by the previous observation, so there is no edge from $w$ to $u$. Therefore, $w$ and $u$ can be exchanged both in $G$ and $H$. In the resulting diagrams, we can then exchange $(v,w)$ and $(u,v)$ respectively, which closes the diagram. Note that the braids representation of both sides of the diagram correspond to the Reidemeister type 3 move.
\end{proof}

\begin{thm}
\label{thm:stronglynormalizing}
  Right reductions are confluent and therefore define normal forms for diagrams under the equivalence relation induced by exchanges.
\end{thm}

\begin{proof}
  By Theorem~\ref{thm:termination-connected-bound} the reduction is terminating and by Lemma~\ref{lemma:locally-confluent} it is locally confluent, so by Newman's lemma right reductions are confluent. Therefore, the right normal form for a given diagram can be obtained by applying any legal right exchanges until a normal form is reached.
\end{proof}

\section{Computing normal forms} \label{sec:algorithm}

It follows from Theorem~\ref{thm:stronglynormalizing} that applying the right reduction rewrite strategy allows us to find normal forms in $O(v^4)$ time, where $v$ is the number of vertices: we perform $O(v^3)$ exchanges, each of which can be found and performed in $O(v)$ time. In this section we show that this complexity can be improved, giving a procedure which constructs the normal form directly in $O(ve)$ time, where $e$ is the number of edges.

Let $D$ be a connected diagram in right normal form and $v \in D$ be a vertex.
We analyze how a new vertex $l$ can be added to $D$ by connecting it to $v$ only, such that $l$ becomes a leaf in the new diagram.
First, we need to choose whether to connect $l$ to the domain or codomain of $v$.
Assume for instance that we connect it to the domain of $v$. If $v$ has $k$ edges in its
domain before the addition, there are $k+1$ possible positions for the new edge between $l$ and $v$.
Assume that such a position is chosen. The height of the vertex $l$ in the new diagram must also be chosen, as shown in Figure~\ref{fig:growing-leaf}. The following lemma shows that there is only one such choice such that the new diagram is in right normal form.

\begin{figure}[b]
  \centering
  \begin{tikzpicture}[scale=0.6,every node/.style={node distance=.25cm}]

    \definecolor{dgreen}{RGB}{0,180,0}
    \def\defaultfacecolor{dgreen}

    \begin{scope}[xshift=0cm]
      \startdiagram{1}
      \diagslice{0}{0}{1}
      \diagslice{1}{0}{1}
      \diagslice{1}{0}{1}
      \diagslice{0}{3}{1}
      \diagslice{0}{1}{0}
      \storeedgecolor{2}{dgreen}
      \finishdiagram
      \node[bnode,fill=dgreen,draw=dgreen] at (v2) {};
      \node[right of=v2,dgreen] {$l$};
      \node[bnode] at (v3) {};
      \node[below right of=v3] {$v$};
    \end{scope}
    \begin{scope}[xshift=4cm]
      \startdiagram{1}
      \diagslice{0}{0}{1}
      \diagslice{1}{0}{1}
      \diagslice{2}{0}{1}
      \diagslice{0}{3}{1}
      \diagslice{0}{1}{0}
      \storeedgecolor{1}{dgreen}
      \finishdiagram
      \node[bnode,fill=dgreen,draw=dgreen] at (v1) {};
      \node[bnode] at (v3) {};
      \node[right of=v1,dgreen] {$l$};
      \node[below right of=v3] {$v$};
    \end{scope}
    \begin{scope}[xshift=8cm]
      \startdiagram{1}
      \diagslice{0}{0}{1}
      \diagslice{0}{0}{1}
      \diagslice{2}{0}{1}
      \diagslice{0}{3}{1}
      \diagslice{0}{1}{0}
      \storeedgecolor{0}{dgreen}
      \finishdiagram
      \node[bnode,fill=dgreen,draw=dgreen] at (v0) {};
      \node[bnode] at (v3) {};
      \node[right of=v0,dgreen] {$l$};
      \node[below right of=v3] {$v$};
    \end{scope}
  \end{tikzpicture}
  \caption{Possible vertical positions to grow a leaf $l$ on $v$. Only the central diagram is in right normal form.}
  \label{fig:growing-leaf}
\end{figure}


\begin{lem} \label{lemma:growing-leaf}
  Let $D'$ be a diagram obtained from $D$ by adding a leaf $l$ connected to a vertex $v \in D$, at
  a determined side (domain or codomain) and position between existing edges on that side.
  There is a unique vertical position of $l$ such that $D'$ is in right normal form. Furthermore the horizontal position of $l$ at this height is determined by the position between the existing edges of $v$ that we grow it from.
\end{lem}

\begin{proof}
  Let us first show that there is a vertical position for $l$ such that $D'$ is in right normal form. First, pick an initial vertical position for $l$, such as the position immediately above or below $v$ (depending on the orientation of the connection between $v$ and $l$). Then, normalize by applying right exchanges. All the right exchanges involve $l$: otherwise, by contradiction, consider the first exchange not involving $l$. Removing $l$ from its domain gives us $D$ again (because the relative positions of vertices in $D$ has not changed), and the exchange still applies to this diagram, which contradicts normality of $D$. This shows the existence of the vertical position and uniqueness follows from confluence.
\end{proof}

This observation already gives us a way to construct the right normal form of any acyclic connected diagram. For any tree, we can remove one leaf, compute the right normal form of the remaining tree recursively, and add the leaf at the height given by the lemma. However, this does not let us normalize cycles yet.

\begin{defi}
  A \emph{simple face} in a string diagram is a simple edge loop whose inner region does not contain any other vertex or edge. (An edge loop is simple when no edge appears twice in the loop, and each vertex is visited at most once in the loop.)
\end{defi}

\begin{defi}
  Let $p$ be an oriented path in a diagram. For each vertex $v$ visited by $p$, we define the winding number of $v$ as follows:
  $$
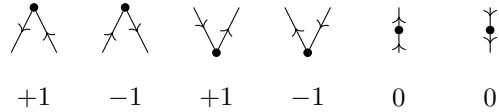
\begin{tikzpicture}[scale=0.6]

    \node[bnode] at (0,2) (n) {};
    \draw[dedge] (.5,1) --  (n);
    \draw[dedge] (n) -- (-.5,1);
    \node at (0,0) {$+1$};

    \node[bnode] at (2,2) (n) {};
    \draw[dedge] (1.5,1) --  (n);
    \draw[dedge] (n) -- (2.5,1);
    \node at (2,0) {$-1$};

    \node[bnode] at (4,1) (n) {};
    \draw[dedge] (3.5,2) --  (n);
    \draw[dedge] (n) -- (4.5,2);
    \node at (4,0) {$+1$};

    \node[bnode] at (6,1) (n) {};
    \draw[dedge] (6.5,2) --  (n);
    \draw[dedge] (n) -- (5.5,2);
    \node at (6,0) {$-1$};

    \node[bnode] at (8,1.5) (n) {};
    \draw[dedge] (8,1) -- (n);
    \draw[dedge] (n) -- (8,2);
    \node at (8,0) {$0$};

    \node[bnode] at (10,1.5) (n) {};
    \draw[dedge] (10,2) -- (n);
    \draw[dedge] (n) -- (10,1);
    \node at (10,0) {$0$};
  \end{tikzpicture}
$$
\end{defi}

\begin{defi}
  Given a simple face in a diagram $D$ and an edge $e$ in the face, the \emph{mountain range} starting on $e$ is the sequence of partial sums of winding numbers when visiting the face in direct rotation, starting from $e$.
\end{defi}

\begin{figure}[b]
  \centering
  \begin{subfigure}{0.45\textwidth}
    \centering
    \begin{tikzpicture}[scale=0.4]
      \startdiagram{1}
      \diagslice{0}{0}{2}
      \diagslice{1}{0}{2}
      \diagslice{0}{2}{0}
      \diagslice{2}{0}{2}
      \diagslice{1}{2}{0}
      \diagslice{1}{1}{1}
      \diagslice{0}{2}{0}
      \tikzstyle{redarrow}=[red,latex-]
      \storeedgecolor{1}{redarrow}
      \finishdiagram
    \end{tikzpicture}
    \caption{An edge in a face}
  \end{subfigure}
  \begin{subfigure}{0.45\textwidth}
    \centering
    \begin{tikzpicture}[scale=0.5]
      \node (p) at (0,0) {};
      \foreach \y in {0,...,3} {
        \draw[dashed,blue,opacity=0.5] (0,\y) -- (7,\y);
      }
      \foreach \x/\y in {0/0,1/1,2/2,3/1,4/2,5/2,6/3,7/2} {
        \draw (p.center) -- (\x,\y);
        \node (p) at (\x,\y) {};
      }
    \end{tikzpicture}
    \caption{The montain range for this edge}
  \end{subfigure}
  \caption{Example of a chosen edge in a face and its mountain range} \label{fig:example-mountain-range}
\end{figure}

Figure~\ref{fig:example-mountain-range} gives an example of a mountain range for an edge in a simple face. Because a cycle forms a closed loop in the plane, the winding numbers of its vertices sums up to two when visited in direct rotation. This means that a mountain range always stops two levels higher than it started.

\begin{defi}
  An edge in a simple face is \emph{eliminable} if the mountain range starting from it never reaches $0$ after the first step.
\end{defi}

For instance, the edge above is eliminable, but its predecessor is not because the montain range starts with a valley that goes at level $-1$ and then $0$.

\begin{lem} \label{lemma:two-eliminable}
  In any simple face there are exactly two eliminable edges.
\end{lem}

\begin{proof}
  Pick an edge in the face and draw the mountain range for it.  Let $m$ be the minimum level it reaches after the first step. As the mountain range starts at $0$ and ends at $2$, $m \leq 1$. Consider the last edge to reach $m$, we will denote it by $e_1$. The mountain range on the right of $e_1$ never goes below $m+1$ by definition. When drawing the mountain range for $e_1$, the left part of the range is shifted upwards by $2$, so this part never goes below $2-m \geq 1$ when drawn as part of the mountain range for $e_1$. So $e_1$ is eliminable. Similarly, consider the last edge $e_2$ to reach $1$ in the mountain range starting from $e_1$: it is also eliminable for the same reason. These are the only two edges which satisfy the criterion.
\end{proof}

\begin{lem} \label{lemma:eliminable-normalized}
  Let $D$ be a connected diagram in right normal form and $e$ be an eliminable edge in a simple face of $D$. Then the diagram $D'$ obtained from $D$ by removing $e$ is in right normal form.
\end{lem}

\begin{proof}
  Consider such an edge. We first analyze what it means to be eliminable in geometrical terms. Let us call $u$ the starting point of $e$ and $v$ its end point. We know that $e$ is immediately followed by a left turn (winding number $+1$) at $v$. The next vertex where a rotation happens $w$ also has winding number $+1$ (otherwise the number of rotations from $e$ to the edge after $w$ would be null). By symmetry let us assume that $e$ points upwards when travelling in the direct orientation on the face.

  There are three sorts of right exchanges that could potentially be enabled by removing~$e$.

  \paragraph{Exchanging $u$ and $v$}
  The first one would be exchanging the endpoints of $e$ together, but this is impossible because of the left turn on $v$ which imposes a horizontal ordering: no such right exchange can be made.

  \paragraph{Exchanging $u$ or $v$ with another vertex $x$}
  The second one would be exchanging one of the endpoints of $e$ with another vertex. This other vertex must be in the interval between the endpoints (otherwise the exchange was already possible before). That is not possible for $v$ because of the left turn on this vertex. For $u$, this would require having another vertex $x$ immediately to the left of $e$ with no edge linked from below. We will see in a later paragraph that this is not possible.

\paragraph{Exchanging two vertices $x$, $y$ distinct from $u$ and $v$}
Finally, the third case consists in exchanging two nodes $x$ and $y$ between $u$ and $v$, $x$ immediately to the left of $e$ with no edge linked from below, and $y$ immediately to the right of $e$ with no edge from above. We will show that no such $x$ exists.

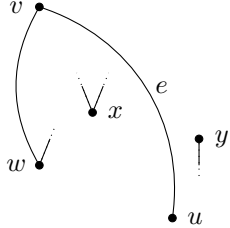
\begin{figure}[H]
  \centering
  \begin{tikzpicture}[scale=0.7,every node/.style={node distance=.3cm}]
    \node[bnode] at (1.5,0) (u) {};
    \node[right of=u] {$u$};
    \node[bnode] at (-1,1) (w) {};
    \node[left of=w] {$w$};
    \node[bnode] at (-1,4) (v) {};
    \node[left of=v] {$v$};

    \draw (u.center) edge[bend right=40] node [right] {$e$} (v.center);
    \draw (v.center) edge[bend right=30] (w.center);
    \draw (w.center) to +(.2,.5);
    \draw [dotted] (w.center) to +(.3,.75);

    \node[bnode] at (-.0,2) (x) {};
    \node[right of=x] {$x$};
    \node[bnode] at (2,1.5) (y) {};
    \node[right of=y] {$y$};

    \draw (x) -- +(-.2,.5);
    \draw [dotted] (x) -- +(-.3,.75);
    \draw (x) -- +(.2,.5);
    \draw [dotted] (x) -- +(.3,.75);
    \draw (y) -- +(0,-.5);
    \draw [dotted] (y) -- +(0,-.75);
  \end{tikzpicture}
\end{figure}

\paragraph{Ruling out the existence of $x$}

Because $e$ is the right boundary of the face, such an $x$ must be a part of the boundary of the face. As part of this cycle, it has two edges coming from above. Browsing the cycle in the direct orientation can visit $x$ in two directions: from left to right or from right to left.

  If $x$ is visited from left to right, this contradicts the fact that $x$ is immediately to the left of $e$, because the interior of the face is contained between the two edges linked to $x$.

  If $x$ is visited from right to left, consider the path from $w$ to $x$.  It starts upwards and ends downwards, so it has odd winding number.  As $x$ itself is a right turn, this number cannot be negative: otherwise, travelling from $e$ to the edge following $x$ would have null or negative winding number, contradicting the assumption that $e$ is eliminable. So, the path from $w$ to $x$ has positive winding number, and therefore one edge in this path is located between $x$ and $e$, which contradicts the fact that $x$ is immediately to the left of $e$.
\end{proof}

\begin{thm} \label{thm:efficient-normalizing}
  The right normal form of a boundary-connected string diagram in free monoidal categories can be decided in time $O(v e)$ where $v$ is the number of vertices and $e$ is the number of edges.
\end{thm}

\begin{proof}
  Again we can restrict our attention to the case of connected diagrams thanks to the reduction of Figure~\ref{fig:connected-via-boundary}. We construct the right normal form of any connected string diagram by induction on the number of edges. The initial case (no edge) is clear.

  Given a diagram $D$, there are two cases. We can check in $O(v)$ if $D$ has a leaf, in which case we remove this leaf and obtain a diagram $D'$ with one less edge that we can inductively normalize. Then, by Lemma~\ref{lemma:growing-leaf}, we can deduce the right normal form for $D$, by inserting back the leaf at the unique spot which makes the diagram normalized. Such a spot can be found in $O(v)$ by applying right exchanges on the leaf as long as they are admissible. If $D$ does not have any leaf, then it has a simple face. In that case, by Lemma~\ref{lemma:two-eliminable}, there are two eliminable edges in this face. These can be identified in $O(v)$ thanks to the characterization via mountain ranges. We can remove one of them, obtaining diagram $D''$, and inductively normalize $D''$. By uniqueness of the normal form for $D''$ and by Lemma~\ref{lemma:eliminable-normalized}, the normal form for $D''$ can be obtained by normalizing $D$ and then removing the edge. So the normal form for $D$ can be reconstructed from the normal form for $D''$ by adding the edge back. This can also be computed in $O(v)$. We therefore obtain a normalizing algorithm with $e$ induction steps, each of which takes $O(v)$ time, so the overall complexity is $O(ve)$.
\end{proof}

\section{Extension to disconnected diagrams} \label{sec:disconnected}

The connectivity requirement is crucial to obtain termination of right reductions and therefore the right normal forms on which we relied on for our results. In this section, we extend our results to arbitrary diagrams. Our approach is to define a complete invariant for the exchange rule.

\begin{figure}[b]
\centering
\begin{subfigure}{0.23\textwidth}

      \definecolor{dgreen}{RGB}{0,180,0}
    \def\defaultfacecolor{dgreen}

    \centering
  \begin{tikzpicture}[scale=.4,every node/.style={scale=.4}]
  \startdiagram{1}

  \scanslice{0}{0}{2}
  \scanslice{0}{0}{2}
  \scanslice{1}{1}{2}

  \drawinitialstrands{1}
  \enablefaceids
  \enablefaces
  \diagslice{0}{0}{2}
  \diagslice{0}{0}{2}
  \diagslice{1}{1}{2}
  \diagslice{0}{2}{1}
  \diagslice{2}{2}{0}
  \diagslice{0}{2}{0}
  \finishdiagram

  \begin{scope}[every node/.style={node distance=.25cm,scale=.6}]
    \node[above of=v0] {\small A};
    \node[above of=v1] {\small B};
  \end{scope}

  \end{tikzpicture}
  \end{subfigure}
  \begin{subfigure}{0.23\textwidth}
    \begin{tikzpicture}[scale=0.6,every node/.style={scale=0.6,node distance=2cm}]
      \node[circle,draw] (0) {$0$};
      \node[rectangle,draw,below left of=0] (A) {$A$};
      \node[rectangle,draw,below right of=0] (B) {$B$};
      \node[circle,draw,node distance=1.414cm,below of=A] (1) {$1$};
      \node[circle,draw,below left of=B] (2) {$2$};
      \node[circle,draw,below right of=B] (3) {$3$};
      \draw (3) -- (B) -- (0) -- (A) -- (1);
      \draw (2) -- (B);
      
    \end{tikzpicture}
  \end{subfigure}
  \caption{A diagram and its structural tree}
  \label{fig:structural-tree}
\end{figure}
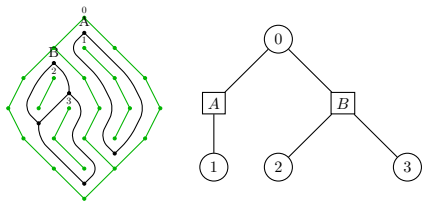

In general, a digram can contain multiple connected
components. Because we are dealing here with non-symmetric monoidal
categories, the way these components nest into each other's faces
matters as this tree structure is preserved by exchanges. In
Appendix~\ref{app:disconnected-extension}, we define notions of faces,
components and enclosure relations between them from the combinatorial
representation of diagrams. Each diagram is then represented by a structural tree
as in Figure~\ref{fig:structural-tree}, where face nodes have an unordered
set of component children and component nodes have an ordered list of face children.
We show in Appendix~\ref{app:completeness-structural-tree} that such a tree
is a complete invariant for exchanges.

\subsection{Word problem}

We show how to compute the structural tree of a diagram, and therefore solve the word
problem in the general case.
Algorithm~\ref{alg:structural-tree} scans the diagram in one pass and computes
simultaneously the components and faces of the diagram, as well as the
inclusion relation between them. Appendix~\ref{app:disconnected-extension} defines
components and faces as equivalence classes of \emph{places} and \emph{spots} under
an adjacency relation, so we use two union-find data structures to represent them.

\begin{algo}
  The following algorithm computes the faces, components, and relations between them.
\begin{algorithmic}
 \State initialize union-find data structures $F$ for faces and $C$ for components
 \State initialize parent pointer arrays $PF$ for faces and $PC$ for components
 \For{$h = 0$ to $D.N$}
   \For{$k = 0$ to $D.W(h)$}
   \If{$s_{h,k}$ adjacent to $s_{h-1,k}$ and to $s_{h-1,k-D.\Delta(h-1)}$}
   
         \State \Call{union}{$F(h-1,k),F(h-1,k-D.\Delta(h-1))$}
         \State $F(h,k) \gets F(h-1,k)$
      \ElsIf{$s_{h,k}$ adjacent to $s_{h-1,k}$ only}
          \State $F(h,k) \gets F(h-1,k)$
      \ElsIf{$s_{h,k}$ adjacent to $s_{h-1,k-D.\Delta(h-1)}$ only}
          \State $F(h,k) \gets F(h-1,k-D.\Delta(h-1))$
      \Else
          \State $F(h,k) \gets \text{a fresh face id}$
          \State $PF(h,k) \gets (h-1,D.H(h-1))$
      \EndIf
   \EndFor
   \For{$k = 0$ to $D.W(h)-1$}
       \State Update components similarly
   \EndFor
 \EndFor
\end{algorithmic}
\label{alg:structural-tree}
\end{algo}

Unions of faces are performed when scanning vertices with no output. Each of them costs $O(\log^* f)$, where $f$ is the number of faces of the diagram and $\log^*$ is the log-star function. So total cost of all unions of faces is $O(v \log^* f)$.
Scanning the diagram with the two loops takes $O(v e)$ operations, and checking if two spots or places are adjacent takes constant time. Therefore, the computation of the faces, components and their relations can be done in $O(v e)$.

Then, we apply the algorithm of
Theorem~\ref{thm:efficient-normalizing} to compute the right normal
form of each component, which can be done again in quadratic time.

Finally, the structural tree of the diagram is converted to an integer recursively in Algorithm~\ref{alg:coding}, where we assume a coding function $\chi$ injectively mapping any tuple of integers to an integer (using an appropriate encoding).
As the structural tree is a complete invariant for diagram equivalence, we obtain the following theorem.

A note of caution about this last step is that it does rely pretty
crucially on the encoding of large data structures into integers and
exploits the fact that the computational model lets us compare
unbounded integers for free in constant time, which could be
unrealistic for practical purposes.  For implementations, this can be
mitigated using hashing techniques which let compare large
datastructures quickly on average, rejecting different structures quickly
and only resorting to full comparison when the hashes are equal.

\begin{algo}
   The following algorithm recursively computes an integer representation for a structural tree.
  \begin{algorithmic}
    \If{$n$ is a face node}
    \State compute the integer representation of its children components recursively;
    \State sort the list of children components as $l$
    \Return $\chi(l)$
    \EndIf
    \If{$n$ is a component node with normalized root component $c$}
    \State sort the children faces by order of introduction in the normalized component $c$
    \State compute the integer representation of the children faces recursively as $l$, preserving the order;
    \Return $\chi(c,l)$
    \EndIf
  \end{algorithmic}
  \label{alg:coding}
\end{algo}

\begin{thm} \label{thm:quadratic-word-problem-disconnected}
  The word problem for string diagrams in a monoidal category can be solved in $O(v e)$, where
  $v$ is the number of vertices and $e$ is the number of edges. \qed
\end{thm}

\section{Linear-time solution to the word problem in the connected case} \label{sec:linear-time}

In this section we show how the word problem can be solved in linear
time for boundary-connected diagrams via a reduction to the problem of map isomorphism.
In the disconnected case, the components enclosed in a face can spin around each other, so comparing two
faces amounts to comparing their sets of components. Therefore, there is little hope to extend this result to the disconnected case.

We first recall some background notions of topological graph theory. We refer
the interested reader to \cite{mohar2001graphs} for a more in-depth treatment of
these notions.

\subsection{Background on planar maps}

A multigraph is a set of vertices $V$ and of edges $E$ where each edge
$e \in E$ is associated with a set of one or two vertices $V(e)$. In
other words it is an undirected graph where multiple edges can exist
between two vertices, and loops are allowed.

A planar map is a discrete representation of the embedding of a connected multigraph
(seen as a topological space) in a surface.

\begin{defi}
  A \emph{map} is a set $\Omega$ of \emph{darts} (or half-edges) and
  two permutations $x$ and $y$ of $\Omega$ such that $x^2 = 1$, $x$
  has no stationary point, and the permutation group $G$ generated by
  $x$ and $y$ is transitive (for any $a,b \in \Omega$ there is $g \in
  G$ such that $g(a) = b$).
\end{defi}

\begin{figure}[H]
  \hfill
    \begin{subfigure}[b]{0.35\textwidth}
    \[
    x = (\begin{smallmatrix}1 & 2\end{smallmatrix}) (\begin{smallmatrix}3 & 4\end{smallmatrix}) (\begin{smallmatrix}5 & 6\end{smallmatrix}) (\begin{smallmatrix}7 & 8\end{smallmatrix}) (\begin{smallmatrix}9 & 10\end{smallmatrix})
    \]
    \[
    y = (\begin{smallmatrix}2 & 10 & 7\end{smallmatrix}) (\begin{smallmatrix}4 & 8 & 6\end{smallmatrix}) (\begin{smallmatrix}1 & 3 & 5 & 9\end{smallmatrix})
    \]
    \caption{A planar map given by two permutations}
  \end{subfigure}\hfill
    \begin{subfigure}[b]{0.4\textwidth}
      \centering
  \begin{tikzpicture}[scale=0.7,every node/.style={scale=0.7}]

    \node at (-3,0) (a) {$a$};
    \node at (3,0) (c) {$c$};
    \node at (0,2) (b) {$b$};
    \node at (0,-2) (d) {$d$};

    \dart{a}{b};
    \dart{b}{c};
    \dart{c}{d};
    \dart{d}{b};
    \dart{d}{a};

    \foreach \x/\y in {10/7,7/2,2/10,6/4,4/8,8/6} {
      \draw[ylink] (dart\x) edge[bend right] (dart\y);
    }
    \draw[ylink] (dart1) edge[bend left=55] (dart3);
    \draw[ylink] (dart5) edge[bend left=55] (dart9);
    \draw[ylink] (dart3) .. controls (4,1) and (4,-1) .. (dart5);
        \draw[ylink] (dart9) .. controls (-4,-1) and (-4,1) .. (dart1);
 
  \end{tikzpicture}
  \caption{Graphical representation where darts are numbered half-edges}
  \end{subfigure} \hfill\null
\end{figure}


Two maps are isomorphic when there is a bijection between their sets
of darts respecting the permutations $x$ and $y$ of both maps.

In a map $m$, the cycles of $x$ are called edges of $m$. The cycles of
$y$ are called faces of $m$. The cycles of $xy$ are called vertices. The Euler characteristic
of $m$ is
\[ \chi(m) = v - e + f \] where $v$ is the number of vertices, $e$ of
edges and $f$ of faces. A map $m$ is \emph{planar} if $\chi(m) = 2$.

Any embedding of a multigraph in the plane gives rise to a planar map.
\begin{thmC}[\cite{jones1978theory}] \label{thm:jones}
Any two embeddings of a multigraph in the plane are isotopic if and only if the corresponding
planar maps are isomorphic.
\end{thmC}

\begin{thmC}[\cite{hopcroft1974linear}] \label{thm:linear-time-map-iso}
  Determining if two planar maps are isomorphic can be decided in linear time.
\end{thmC}

Again, the same note of caution about the comparison of unbounded integers
in constant time applies to the latter result, but hashing should provide
a satisfactory implementation in practice.

\begin{figure}[H]
  \centering
  \begin{subfigure}{0.4\textwidth}
    \centering
  \begin{tikzpicture}
    \startdiagram{1}
    \diagslice{0}{0}{2}
    \diagslice{0}{1}{1}
    \diagslice{0}{2}{0}
    \finishdiagram
    \node at (1.5,-1.5) {$\neq$};
    \begin{scope}[xshift=3cm]
    \startdiagram{1}
    \diagslice{0}{0}{2}
    \diagslice{1}{1}{1}
    \diagslice{0}{2}{0}
    \finishdiagram
    \end{scope}
  \end{tikzpicture}
  \end{subfigure}
  \caption{Two non-equivalent string diagrams which are isotopic as maps}
  \label{fig:example-nonequivalent-sd}
\end{figure}
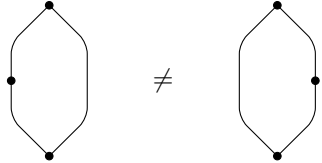

Our goal is to reuse this last result to solve the word problem for
connected string diagrams.  However, the word problems for string
diagrams and for planar maps do not match:
Figure~\ref{fig:example-nonequivalent-sd} shows two string diagrams
which are isotopic as planar maps but not equivalent as string
diagrams.

\subsection{Directed planar maps}

Maps are embeddings of undirected multigraphs. In this section, we
introduce an analoguous notion for directed multigraphs. A directed
multigraph is a set of vertices $V$ and a set of edges $E$, each edge
being associated to a pair of vertices $(s,t)$ (its source and
target).
A directed multigraph is connected if it is connected as an undirected multigraph.

\begin{defi}
A \textit{directed map} is a map $(\Omega,x,y)$ together with a choice of distinguished
darts $D \subseteq \Omega$ such that exactly one dart in each cycle of $x$ belongs
to $D$.
\end{defi}

\noindent Two directed maps are isomorphic when they are isomorphic as maps and
furthermore the bijection respects the distinguished darts. Similarly
to Figure~\ref{fig:example-nonequivalent-sd}, there are directed maps which
are isomorphic as undirected maps but not as directed maps.

Given a directed planar map $M$, we can define a planar map $\iota(M)$ by replacing each directed
edge by an undirected graph which encodes the direction of the original edge:
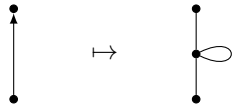
\begin{figure}[H]
  \centering
  \begin{tikzpicture}[scale=0.6]
    \def\drawswaps{1}
    \def\drawdartnumbers{0}
    \node[bnode] at (0,-1) (a) {};
    \node[bnode] at (0,1) (b) {};
    \draw[-latex] (a) -- (b);
    \node at (2,0) {$\mapsto$};

    \begin{scope}[xshift=4cm,scale=0.5]
      \node[bnode] at (0,-2) (a) {};
      \node[bnode] at (0,0) (b) {};
      \node[bnode] at (0,2) (c) {};

      \draw (a) -- (b) -- (c);
      \draw (b) .. controls (2,1) and (2,-1) .. (b);
    \end{scope}
  \end{tikzpicture}
\end{figure}

\begin{prop} \label{prop:map-equivalent-dmap-equivalent}
  Two directed planar maps $M$,$M'$ are isomorphic if and only
  if the undirected planar maps $\iota(M)$ and $\iota(M')$ are
  isomorphic.
\end{prop}

\begin{proof}
  If $M$ and $M'$ as isomorphic then clearly so are $\iota(M)$ and $\iota(M')$. Conversely, assume that $\iota(M)$ and $\iota(M')$ are isomorphic maps via an isomorphism $\phi$.
  Say that a vertex $v \in \iota(M)$ is a \emph{loop root} if a loop is rooted on $v$.
  Given the definition of $\iota$, the image $\iota(u)$ of a vertex $u \in M$ cannot be a loop root, as any loop on $u$ in $M$ is translated to non-loop edges in $\iota(M)$. Therefore, there is a bijection between the loop roots of $\iota(M)$ and the edges of $M$. As loop roots are
preserved by graph isomorphism, $\phi$ induces a bijection between the loop roots of $\iota(M)$ and $\iota(M')$, so we have a bijection $\psi$ between the edges of $M$ and $M'$. This bijection in turn determines a directed graph isomorphism between $M$ and $M'$.
For instance the source vertex of an edge can be recovered from its loop root $u$: follow the edge which comes after the loop, when browsing incident edges of $u$ in clockwise order. Similarily the target vertex can be recovered. Finally, as $\phi$ is a map isomorphism, the cyclic order of edges around vertices is preserved, so $\psi$ is a directed map isomorphism between $M$ and $M'$.
\end{proof}

\begin{cor} \label{coro:linear-time-dmap-iso}
  Testing whether two acyclic directed planar maps are isomorphic can be done in linear time.
\end{cor}

\begin{proof}
  The translation via $\iota$ can be computed in linear time so the problem reduces to
  deciding undirected planar map isomorphism, which is linear by Theorem~\ref{thm:linear-time-map-iso}.
\end{proof}

\begin{prop}
Two embeddings of connected directed multigraphs in the plane are
isotopic if and only if the corresponding directed maps are isomorphic. \qed
\end{prop}

\subsection{From string diagrams to maps} \label{sec:sd-to-dmap}

We translate any string diagram $D$ to a directed planar map $\gamma(D)$ by
replacing each vertex by the gadget below.  The original edges coming
from $D$ inherit their orientation from the string diagram (top to
bottom), and we add two dangling edges for each vertex. These
additional dangling edges are useful for vertices with only inputs or
only outputs by blocking any cyclic permutation of these edges around
the vertex.\footnote{These dangling edges are only useful for vertices with only inputs or only outputs but we choose to add them to all vertices for the sake of uniformity.}

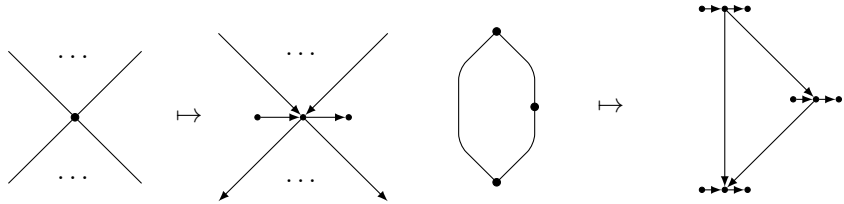
\begin{figure}[H]
  \centering
  \begin{subfigure}{0.45\textwidth}
    \centering
    \begin{tikzpicture}
      \node[bnode] at (0,0) (a) {};
      \node at (1,1) (b) {};
      \node at (1,-1) (c) {};
      \node at (-1,-1) (d) {};
      \node at (-1,1) (e) {};
      \node at (0,.8) {$\dots$};
      \node at (0,-.8) {$\dots$};
      \draw (b) -- (a) -- (d);
      \draw (e) -- (a) -- (c);

    \node at (1.5,0) {$\mapsto$};

    \begin{scope}[xshift=3cm,every node/.style={scale=0.7},scale=1.2]
      \def\drawswaps{0}
      \def\drawdartnumbers{0}
      
      \node[bnode] at (0,0) (a) {};
      \node[bnode] at (-.5,0) (h1) {};
      \node[bnode] at (.5,0) (h2) {};
      \node at (1,1) (b) {};
      \node at (1,-1) (c) {};
      \node at (-1,-1) (d) {};
      \node at (-1,1) (e) {};
      \node[scale=1.4] at (0,.7) {$\dots$};
      \node[scale=1.4] at (0,-.7) {$\dots$};
      \draw[-latex] (b) -- (a);
      \draw[-latex] (e) -- (a);
      \draw[-latex] (a) -- (d);
      \draw[-latex] (a) -- (c);
      \draw[-latex] (h1) -- (a);
      \draw[-latex] (a) -- (h2);
      \end{scope}      
    \end{tikzpicture}
  \end{subfigure}
  \begin{subfigure}{0.45\textwidth}
    \centering
    \begin{tikzpicture}
          \startdiagram{1}
    \diagslice{0}{0}{2}
    \diagslice{1}{1}{1}
    \diagslice{0}{2}{0}
    \finishdiagram
    \node at (1.5,-1.5) {$\mapsto$};

    \begin{scope}[xshift=3cm,yshift=-.2cm,every node/.style={scale=0.7},scale=1.2]

      \node[bnode] at (0,0) (a) {};
      \node[bnode] at (-.25,0) (a1) {};
      \node[bnode] at (.25,0) (a2) {};
      
      \node[bnode] at (1,-1) (b) {};
      \node[bnode] at (.75,-1) (b1) {};
      \node[bnode] at (1.25,-1) (b2) {};
      
      \node[bnode] at (0,-2) (c) {};
      \node[bnode] at (-.25,-2) (c1) {};
      \node[bnode] at (.25,-2) (c2) {};
      
      \draw[-latex] (a) -- (c);
      \draw[-latex] (b) -- (c);
      \draw[-latex] (a) -- (b);

      \draw[-latex] (a1) -- (a);
      \draw[-latex] (a) -- (a2);
      \draw[-latex] (b1) -- (b);
      \draw[-latex] (b) -- (b2);
      \draw[-latex] (c1) -- (c);
      \draw[-latex] (c) -- (c2);
      \end{scope}
    \end{tikzpicture}
  \end{subfigure}
  \label{fig:translation-sd-dmap}
  \caption{Translation of a string diagram to a directed map.}
\end{figure}
\begin{thm} \label{thm:equivalent-dmap-equivalent-sd}
  Any two connected diagrams are equivalent if and only if the induced directed maps are isomorphic.
\end{thm}

\begin{proof}
  Exchanges on connected diagrams preserve the translation to directed
  maps so any two equivalent connected diagrams are mapped to isomorphic
  directed maps.  For the converse direction, we can therefore assume
  that the two diagrams $D, D'$ are in right normal form. Let
  $\phi$ be the map isomorphism between the corresponding directed
  maps $\gamma(D), \gamma(D')$.  First, $\gamma(D)$ and $\gamma(D')$
  have the same number of vertices and so do $D$ and $D'$. Call $n$ the number of vertices of $D$.

  We prove by induction on $n$ that $D = D'$. We reuse the induction technique
  introduced in Section~\ref{sec:algorithm}: diagram $D$ contains a leaf or a face.

  If $D$ contains a leaf $l$, this leaf is mapped to a vertex
  $\gamma(l)$ connected to three edges. As $\phi(\gamma(l))$ is
  also connected to three edges, there is a leaf $l' \in D'$ such that
  $\gamma(l') = \phi(\gamma(l))$. Because $\phi$ is an isomorphism of
  directed maps, the orientations of $l$ and $l'$ are the
  same: they are both single-input or both single-output vertices.
  Moreover, they are connected to their parent vertices at the same position
  in their list of inputs or outputs, thanks to the auxilliary edges added
  in the translation.
  Consider the diagrams $E$ and $E'$ obtained from $D$ and $D'$
  by removing $l$ and $l'$ respectively. These diagrams are in right
  normal form. The isomorphism $\phi$ induces a map isomorphism
  between $\gamma(E)$ and $\gamma(E')$ so by induction $E = E'$.
  By Lemma~\ref{lemma:growing-leaf}, $D = D'$.

  If $D$ contains a face $f$, this face is mapped to a face
  $\gamma(f)$ in $\gamma(D)$. The face $\phi(\gamma(f))$ is itself the
  image of a face $f' \in D'$. Because $\phi$ preserves edge
  orientations, the mountain ranges of $f$ and $f'$ are equal. Let $e$
  be an eliminable edge in $f$ and let $e'$ be the preimage of $\phi(\gamma(e))$ in $f'$.
  By equality of the mountain ranges, $e'$ is also eliminable in $f'$.
  By Lemma~\ref{lemma:eliminable-normalized}, removing $e$ from $D$ 
  and $e'$ from $D'$ gives diagrams $F$ and $F'$ both in right normal form.
  Again we can apply the induction hypothesis to $F$ and $F'$, so $F = F'$,
  and therefore $D = D'$.
\end{proof}

\begin{cor} \label{coro:linear-time-sd-word-problem}
  The word problem for connected string diagrams can be solved in linear time.
\end{cor}

\begin{proof}
  The translation $\gamma$ from string diagrams to directed planar
  maps can be computed in linear time. The decision problem therefore
  reduces to the word problem for acyclic directed planar maps, which
  is solvable in linear time by
  Corollary~\ref{coro:linear-time-dmap-iso}.
\end{proof}

\section{Recumbent isotopy} \label{sec:recumbent-isotopy}

Joyal and Street's theorem relating diagram deformations to the axioms
of monoidal categories (Theorem~\ref{thm:joyal-street}) requires the
deformations to be recumbent.  This means that at each stage of the
deformation, the diagram's edges must remain upright, as shown in
Figure~\ref{fig:example-topological-diagrams}.  It was conjectured by
Selinger~\cite{selinger2011survey} that the recumbency condition can
be weakened. For this weakening, the requirement that all wires must
flow vertically can be dropped, but we must keep the requirement that
wires stay connected to their endpoints from the same side.
Figure~\ref{fig:rotate-leaf} shows a counter-example for the
conjecture without this last condition.
\begin{figure}
  \centering
  \begin{subfigure}{0.4\textwidth}
    \centering
    \begin{tikzpicture}
      \node[bnode] at (0,0) (a) {};
      \node[bnode] at (.5,1) (b) {};
      \node[bnode] at (.8,.5) (c) {};
      \node[bnode] at (1,.8) (d) {};

      \draw (a) .. controls (0,.8) and (.1,.8) .. (b) edge[bend right] (c);
            \draw (c) edge[bend right] (d);
    \end{tikzpicture}
    \caption{A recumbent plane diagram}
  \end{subfigure}
  \begin{subfigure}{0.4\textwidth}
    \centering
    \begin{tikzpicture}
      \node[bnode] at (0,0) (a) {};
      \node[bnode] at (.5,1) (b) {};
      \node[bnode] at (.8,.5) (c) {};
      \node[bnode] at (1,.8) (d) {};

      \draw (a) .. controls (-.2,.8) and (0,.8) .. (.1,.5) .. controls (.2,.3) and (.4,.5) .. (b) edge[bend right] (c);
      \draw (c) edge[bend right] (d);
    \end{tikzpicture}
    \caption{A locally recumbent plane diagram}
  \end{subfigure}
  \caption{Examples of topological diagrams.}
    \label{fig:example-topological-diagrams}
\end{figure}
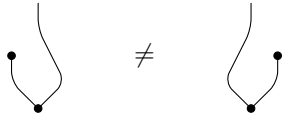
\begin{figure}[H]
  \centering
  \begin{tikzpicture}[scale=0.7]
    \startdiagram{2}
    \drawinitialstrands{2}
    \diagslice{0}{0}{1}
    \diagslice{0}{2}{0}
    \finishdiagram

    \node at (2,-.5) {$\neq$};

    \begin{scope}[xshift=4cm]
      \startdiagram{2}
      \drawinitialstrands{2}
      \diagslice{1}{0}{1}
      \diagslice{0}{2}{0}
      \finishdiagram
    \end{scope}
    
  \end{tikzpicture}
  \caption{Arbitrary planar isomorphism does not respect morphism equality.}
    \label{fig:rotate-leaf}
\end{figure}

We now show how our reduction from string diagrams to planar maps can be used to prove Selinger's conjecture,
generalizing Joyal and Street's Theorem~\ref{thm:joyal-street}.
To extend this result to disconnected diagrams, we only need to extend
the notion of directed map to disconnected cases.

\begin{defi}
  A \emph{disconnected planar map} is defined recursively as a tree, as follows.
  \begin{itemize}
  \item A \emph{face node} is a set of component nodes (possibly empty).
  \item A \emph{component node} is a planar map $m$, an outer face $f_0 \in m$ and face nodes for each face $f \neq f_0$ of $m$.
  \end{itemize}
  A disconnected planar map is given by its root face node, which has finite depth.
\end{defi}

As this definition mirrors that of the structural tree of a diagram (Definition~\ref{def:structural-tree}), it is straightforward to extend the translation of Section~\ref{sec:sd-to-dmap} to translate any diagram $D$ to a disconnected planar map.

Equivalence of disconnected planar maps is defined by pointwise
equivalence of the planar maps involved. By completeness of the
structural tree for string diagrams
(Theorem~\ref{thm:tree-completeness}), two string diagrams
are equivalent if and only if the corresponding disconnected planar
maps are equivalent. For this reason, Theorem~\ref{thm:jones} can be extended to the
disconnected case: two disconnected planar maps are equivalent if and
only if their embeddings in the plane are isotopic. We therefore
obtain the following theorem:

\begin{thm}
\label{thm:selinger}
Two string diagrams are equivalent if and only if their translations as disconnected planar maps from Section~\ref{sec:linear-time} are isotopic. \qed
\end{thm}

\noindent
This generalizes Joyal and Street's result in the way hinted by Selinger's conjecture: the isotopy is unconstrained, although some gadgets have been added to enforce the preservation of the order of inputs and outputs around vertices.

This generalization has a clean statement in terms of planar pivotal monoidal categories~\cite[Section~4.2]{selinger2011survey}. The coherence theorem for their graphical calculus is stated as follows~\cite[Theorem~4.14]{selinger2011survey}.

\begin{thmC}[Coherence for pivotal categories]
\label{thm:pivotalcoherence}
A well-formed equation between morphisms in the language of pivotal categories follows from the axioms of pivotal categories if and only if it holds in the graphical language up to planar isotopy.
\end{thmC}

\noindent
We write \textit{monoidal signature} for the generating data for a monoidal category which is free on objects and morphisms, and given a monoidal signature $\Sigma$, we write $\mathrm M(\Sigma)$ for the free monoidal category on $\Sigma$, and $\mathrm P(\Sigma)$ for the free pivotal category on $\Sigma$. Combining Theorems~\ref{thm:selinger} and~\ref{thm:pivotalcoherence} then yields the following.
\begin{cor}
\label{cor:pivotal}
Given a monoidal signature $\Sigma$, the obvious embedding functor $F: \mathrm M(\Sigma) \to \mathrm P(\Sigma)$ is faithful.
\end{cor}
\begin{proof}
For $A, B \in \Ob(\mathrm M(\Sigma))$ and morphisms \mbox{$f,g:A \to B$}, Theorem~\ref{thm:pivotalcoherence} says that $F(f) = F(g)$ just when the string diagrams for $f$ and $g$ are isotopic. But then by Theorem~\ref{thm:selinger}, we also have $f = g$, and hence faithfulness.
\end{proof}

\subsection{Acknowledgements}

We thank the members of the  \textit{S\'eminaire de cat\'egories sup\'erieures, polygraphes et homotopie} at IRIF, Paris for their feedback on these results, in particular Simon Forest and Samuel Mimram, and we are also grateful to Vincent Vidal, Éric Colin de Verdière and Arnaud de Mesmay for their feedback. Jamie Vicary acknowledges funding from the Royal Society and Antonin Delpeuch is supported by an EPSRC scholarship.

\bibliographystyle{alphaurl}
\bibliography{zotero}

\appendix

\section{Extension to disconnected diagrams} \label{app:disconnected-extension}

As the transformation described in Figure~\ref{fig:connected-via-boundary} is still applicable to disconnected diagrams, we consider a diagram $D$ without input or output edges ($D.S = 0$ and $D.W(D.N) = 0$).

\begin{defi} \label{defi:spot} For each level $h \in [D.N + 1]$, we
  define symbols $(s_{h,k})_{0 \leq k \leq D.N(h)}$. The symbol
  $s_{h,k}$ is the \emph{spot} at height $h$ and position $k$, which
  represents the empty space between the $k$th and $k+1$th edge
  at level $h$ (including diagram boundaries for the extrema).
  Similarly, we define symbols $(p_{h,k})_{1 \leq k \leq D.N(h)}$. The symbol
  $p_{h,k}$ represents the intersection of the $k$th edge that crosses
  level $h$ and level $h$ itself: we call this a \emph{place}.
  Vertices of the diagram are places too, and we represent the vertex
  between heights $h$ and $h+1$ as $v_h$.
\end{defi}

\begin{defi} \label{defi:adjacency-spots}
  Two spots $s_{h,k}$ and $s_{h+1,k'}$ are \emph{adjacent} when either $k = k'$ and $D.H(h) \geq k$
  or $k + D.\Delta(h) = k'$ and $D.H(h) + D.I(h) \leq k$.
\end{defi}

Graphically, two spots are adjacent if they lie in neighbouring slices and they are in the same region of the diagram seen as a planar graph.
We will formally define this notion of region by taking the connected closure of this adjacency relation.

\begin{defi}
  A \emph{face} is a connected component of spots for the adjacency relation defined above.
\end{defi}

\begin{defi} \label{defi:adjacency-places} Two places $p_{h,k}$ and
  $p_{h+1,k'}$ at consecutive levels are adjacent if they are on the same edge.
  Formally, this happens when
  either $k = k'$ and $D.H(h) > k$, or $k + D.\Delta(h) = k'$ and
  $D.H(h)+D.I(h) \leq k$.
  $D.H(h) \leq k < D.H(h) + D.I(h)$ and
  $D.H(h) \leq k' < D.H(h) + D.O(h)$.
  A place $p_{h,k}$ and a vertex $v_h'$ are adjacent when 
  Two places $p_{h,k}, p_{h,k'}$ at the same level are adjacent if
  $D.H(h) \leq k, k' < D.H(h) + D.I(h)$ or $D.H(h-1) \leq k, k' < D.H(h-1) + D.O(h-1)$.
\end{defi}

\begin{defi}
  A \emph{component} is a connected component of places for the adjacency relation defined above.
\end{defi}

\begin{lem}
  Any exchange $D \rightarrow D'$ induces bijections $\phi_F$ and $\phi_C$ between the faces
  and components of $D$ and $D'$.
\end{lem}

\begin{proof}
  Let $n$ and $n+1$ be the levels exchanged. By symmetry we can assume it is a right exchange.
  Let us define $\phi_C$ by mapping each spot in $D$ to a spot in $D'$, such that the
  adjacency relation is respected. Let $s_{h,k}$ be a spot.
  If $h \leq n$ or $h > n + 1$ (the spot lies in a slice that is untouched by the exchange) then $\phi_C(s_{h,k}) = s_{h,k}$.
  Otherwise, $h = n + 1$.
  If $k \leq D.H(n+1)$  (the spot lies to the left of both nodes exchanged) then $\phi_C(s_{n+1,k}) = s_{h,k}$ again.
  If $k > D.H(n)+D.I(n)$ (the spot lies to the right of both nodes exchanged) then $\phi_C(s_{n+1,k}) = s_{h,k-D.\Delta(n)+D.\Delta(n+1)}$.
  If $k > D.H(n+1)$ and $k < D.H(n+1) + D.I(n+1)$ (the spot lies in one of the input branches of the node at height $n+1$) then $\phi_C(s_{n+1,k}) = s_{n,k}$ (the spot just above).
  Similarly if $k > D.H(n)$ and $k < D.H(n) + D.O(n)$ then $\phi_C(s_{n+1,k}) = s_{n+2,k + D.\Delta(n+1)}$ (the spot just below).
  Finally, if $k \geq D.H(n+1) + D.I(n+1)$ and $k \leq D.H(n)$ then $\phi_C(s_{n+1,k}) = s_{n,k-D.\Delta(n)+D.\Delta(n+1)})$.
  In each of these cases one can check that $\phi_C$ preserves the adjacency relationship for spots.
  The mapping for places $\phi_F$ can be defined similarly.
\end{proof}

\begin{defi}
  Given a level $h$, a spot $s_{h,k}$ and a place $p_{h,k'}$ are \emph{neighbours} if $k = k'$ or $k + 1 = k'$.
  Furthermore, spots $s_{h,0}$ and $s_{h,D.W(h)}$ are \emph{neighbours of the boundary}.
\end{defi}

\begin{lem}
  There is a unique face containing spots which are neighbours of the boundary. We denote it by $f_0$.
\end{lem}

\begin{proof}
  Any neighbour of the boundary is adjacent to the neighbours of the boundary above and below it.
  By assumption, there is only one spot at the source and target levels. Therefore, all neighbours 
  of the boundary are connected together.
\end{proof}

\begin{defi}
  A component $c$ neighbours a face $f$ when there is a place
  $p \in c$ neighbouring a spot $s \in f$. We denote it by 
  $c \diamondsuit f$.
\end{defi}

Neighbourhood is preserved by exchanges, in the following sense:

\begin{lem} \label{lemma:neighbourhood-preservation}
  Let $d, d'$ be diagrams where $d'$ is obtained from $d$ by exchanges.
  The bijections between faces and components of $d$ and $d'$ induced
  by the exchanges respect the neighbourhood relation.
\end{lem}

\begin{proof}
  It suffices to show that any neighbourhood relation that holds between
  slices affected by an exchange also holds in the exchanged diagram. This
  can be achieved by simple inspection of the definition of neighbourhood
  and adjacency on spots and places.
\end{proof}

\begin{defi}
  Given a level $h$, two spots $s_{h,k}$ and $s_{h,k'}$ are \emph{over-connected} if they are connected
  by a path of spots which never go below level $h$. Similarly, over-connectivity is defined for places too.
\end{defi}

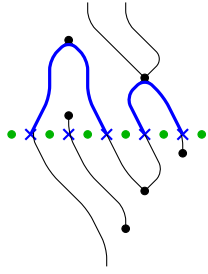
\begin{figure}
  \centering
  \begin{tikzpicture}[scale=0.5]
          \definecolor{dgreen}{RGB}{0,180,0}
    \def\defaultfacecolor{dgreen}
    
    \startdiagram{3}
    \drawinitialstrands{3}
    \diagslice{0}{0}{2}
    \diagslice{2}{2}{2}
    \diagslice{1}{0}{1}
    \enableplaces
    \enablefaces
    \disablefaces
    \diagslice{4}{1}{0}
    \diagslice{2}{2}{0}
    \diagslice{1}{1}{0}
    \finishdiagram

    \draw[blue,very thick,rounded corners] (place_3_1.center) -- (place_2_1.center) -- (place_1_1.center) -- (v0.center) -- (place_1_2.center) -- (place_2_2.center) -- (place_3_3.center);
    \draw[blue,very thick,rounded corners] (place_3_4.center) -- (place_2_3.center) -- (v1.center) -- (place_2_4.center) -- (place_3_5.center);
  \end{tikzpicture}
  \caption{Over-connectivity on an example}
\end{figure}

Note that over-connectivity is an equivalence relation on spots at the same level,
and similarly for places.

\begin{lem} \label{lemma:connected-above}
  Given a level $h$, assume that two distinct spots $s_{h,k}$ and $s_{h,k'}$
  are over-connected. Furthermore, assume that
  $\forall k < i < k'$, $s_{h,i}$ is over-connected to neither
  $s_{h,k}$ nor $s_{h,k'}$. Then the places $p_{h,k}$ and $p_{h,k'-1}$
  are over-connected.
\end{lem}

\begin{proof}
  By induction on the distance $h$ from the top of the diagram.  The
  property holds trivially for the topmost slice as no two spots are
  over-connected at this level. 

  Assuming it holds at level $h$ and consider spots $s_{h+1,k}$ and
  $s_{h+1,k'}$ over-connected. If they are both connected to the
  same spot $s_{h,k}$ then all places between them are over-connected, so the result holds.  Otherwise, they are connected to
  different spots $s_{h,k_0}$ and $s_{h,k_0'}$ respectively which
  are over-connected. Let $s_{h,i_1}, \dots, s_{h,i_q}$ be the spots
  between $s_{h,k_0}$ and $s_{h,k_0'}$ which are over-connected to
  $s_{h,k_0}$ and $s_{h,k_0'}$.
  Each of these spots are not connected to any spot at level $h+1$,
  so the places neighbouring them are all connected via the same morphism
  between level $h$ and $h+1$. In particular, $p_{h,i_1-1}$ and $p_{h,i_q}$ are
  neighbours.

  Apply the induction hypothesis to the pairs $(s_{h,k_0},s_{h,i_1})$ and
  $(s_{h,i_q}, s_{h,k_0'})$: $p_{h,k_0}$ and $p_{h,i_1-1}$ are
  over-connected and so are $p_{h,i_q}$ and $p_{h,k_0'-1}$. Therefore
  $p_{h,k_0}$ and $p_{h,k_0'-1}$ are over-connected.
\end{proof}

\begin{lem} \label{lemma:def-enclosing} Let $f$ be a face,
  $f \neq f_0$. There is a component $c$ such that for each level $h$
  containing a spot in $f$, the place to the left of the first spot in
  $f$ at level $h$ and the place to the right of the last spot in $f$
  at level $h$ are both in $c$. Such a component $c$ is therefore
  unique. We say that $c$ \emph{encloses} $f$, denoted by $f \prec c$.
\end{lem}

\begin{proof}
  First, consider the highest level $h_0$ where $f$ occurs.
  All spots in $f$ at $h_0$ are not connected to any spot at the higher
  level, so their neighbouring places are all connected to the morphism
  above. Let $c$ be their common component.
  For any further level $h$ we prove the result by induction.
  Consider the first and last spots $s_{h,k}$, $s_{h,k'}$ which belong to $f$
  at level $h$.
  We show that $p_{h,k-1}$ belongs to $c$. Showing that so does $p_{h,k'}$ is
  similar.
  Let $s_{h-1,k_0}$ be the leftmost spot neighboured by $s_{h,k}$ at level $h-1$.
  If $s_{h-1}$ is the first spot $s_{h,k_0}$ in $f$ at $h-1$, then by induction
  $p_{h-1,k_0-1}$ belongs to $c$ and is connected to $p_{h,k-1}$, so $p_{h,k-1}$
  belongs to $c$.
  Otherwise, let $s_{h-1,i_1}, \dots, s_{h-1,i_p}$ be the spots in $f$ to the left
  of $s_{h-1,k_0}$.
  We can apply Lemma~\ref{lemma:connected-above} to $s_{h-1,i_p}$ and obtain
  that $p_{h-1,i_p}$ and $p_{h-1,k_0-1}$ are connected. Furthermore, the $s_{h-1,i_1}, \dots
  s_{h-1,i_p}$ are not neighbours of any spots at level $h$, so the edges separating them
  are all connected to the same vertex between $h$ and $h-1$. So $p_{h-1,i_p}$
  and $p_{h-1,i_1-1}$ are neighbours, and finally $p_{h-1,k_0 -1}$ and $p_{h-1,i_1-1}$ are connected.
  By induction, $p_{h-1,i_1-1}$ belongs to $c$, so so does $p_{h-1,k_0-1}$.
\end{proof}

\begin{defi}
  Given a face $f$, a spot $s \in f$ is maximal if it is at the highest level
  where spots of $f$ occur. Similarly, it is minimal if it is at the lowest level
  where spots of $f$ occur.
\end{defi}

Note that a face can have multiple maximal or minimal spots.

\begin{lem}
  Maximal and minimal spots in a face only neighbour the enclosing component, or the boundary in the
  case of the root face.
\end{lem}

\begin{proof}
  This is a direct consequence of the proof of Lemma~\ref{lemma:def-enclosing}.
\end{proof}

\begin{defi}
  Let $f$ be a face. For each component $c$ neighbour of $f$ such that $c$ does not enclose $f$, we say that $f$ encloses $c$, denoted by $c \prec f$.
\end{defi}

\noindent The enclosure relation is preserved by exchanges as follows:

\begin{lem} \label{lemma:exchange-bij-respect-enclosed}
  $\phi_F$ and $\phi_C$ respect $\prec$, i.e.
  $f \prec c \Leftrightarrow \phi_F(f) \prec \phi_C(c)$ and
  $c \prec f \Leftrightarrow \phi_C(c) \prec \phi_F(f)$.
\end{lem}

\begin{proof}
  By Lemma~\ref{lemma:neighbourhood-preservation} and because
  $c \prec f \Leftrightarrow c \diamondsuit f \land \neg (f \prec c)$,
  it is enough to show preservation of $f \prec c$.
  
  Let $d \rightarrow d'$ be a right exchange, $f$ be a face in $d$
  enclosed by $c$.  Let $s_h \in f$ be a maximal spot in $f$, and
  $s_l \in f$ be a minimal spot in $f$.  If $s_h$ is untouched by the
  exchange, then it is still maximal in $\phi_F(f)$, and $\phi_C(c)$
  is still the only component neighboured by $\phi_F(f)$ at $s_h$'s
  slice, so we have $\phi_F(f) \prec \phi_C(c)$.  Similarly, if $s_l$
  is untouched by the exchange, $\phi_F(f) \prec \phi_C(c)$.  If both
  $s_l$ and $s_h$ are touched by the exchange, then they are equal in
  $d$ and $f$ neighbours only $c$ in $d$. By
  Lemma~\ref{lemma:neighbourhood-preservation}, $\phi_F(f)$ neighbours
  only $\phi_C(c)$. As $\phi_F(f)$ is not the root face in $d'$,
  $\phi_F(f) \prec \phi_C(c)$.
\end{proof}

We next introduce an order on the faces enclosed by a component $c$.
Let $N(c)$ be the right normal form of $c$, seen as a standalone
diagram.  The right reduction from $c$ to $N(c)$ induces a bijection
between the faces of $c$ and those of $N(c)$.

Given two faces $f,f'$ in $N(c)$, consider the leftmost maximal spots
$s_{h,k}, s_{h',k'}$ of $f$ and $f'$. We order $f$ and $f'$ by
lexicographic order on the pairs $(h,k), (h',k')$. This defines an
order $<$ on faces of $N(c)$ and therefore on faces of $c$.

\begin{defi} \label{def:structural-tree}
  We inductively define the \emph{structural tree} of faces and
  components.  Given a face $f$,
  $T(f) = \{ T(c) | c \text{ component enclosed by } f \}$.  Given a
  component $c$, let $f_1, \dots, f_n$ be the set of faces enclosed by
  $c$, ordered with the order defined above. We set
  $T(c) = (N(c), T(f_1), \dots, T(f_n))$. Finally, the structural tree $T(D)$ of the entire diagram is
  $T(f_0)$.
\end{defi}

\noindent To make sure that this tree is finite, we must make sure that
none of its nodes is a child of itself.

\begin{lem} \label{lemma:prec-well-founded}
  $\prec$ is well-founded.
\end{lem}

\begin{proof}
  A diagram contains a finite number of components and faces.
  It is therefore enough to show that given a component $c$, it is
  impossible that $c \prec \dots \prec c$.
  We first show that if $c \prec f \prec c'$, then at each level $h$
  where $f$ appears, then for any place $p_{h,k} \in c$ there are
  places $p_{h,a}, p_{h,b} \in c'$ with $a < k < b$. This is a simple
  consequence of Lemma~\ref{lemma:def-enclosing}.
  Then, by induction, we extend this to the transitive closure of $\prec$,
  which shows the result.
\end{proof}

\begin{lem}
  $T(D)$ is invariant under exchanges.
\end{lem}

\begin{proof}
  By Lemma~\ref{lemma:exchange-bij-respect-enclosed}, $\prec$ is invariant
  under exchanges. The order on the faces enclosed by a given component
  is also invariant as it is defined on the right normal form of the component.
\end{proof}

\subsection{Completeness of the structural tree} \label{app:completeness-structural-tree}

We show that the structural tree $T(D)$ of a diagram is a complete
invariant for exchanges:

\begin{thm} \label{thm:tree-completeness}
  Two diagrams $D$, $D'$ with no source and target edges are equivalent if and only if $T(D) = T(D')$.
\end{thm}

\noindent To prove this theorem we introduce a few notions of diagram surgery, to manipulate
components and faces.

\begin{defi}
  Given a diagram $D$ and a spot $s \in D$, the \emph{injection} of a closed diagram $D'$
  at $s$, denoted by $I^D_s(D')$, is obtained by inserting $D'$ in place of $s$ in $D$.
  Concretely, this means that the vertices of $D'$ are inserted at the slice of $s$,
  shifted to the right by the number of edges to the left of $s$. Formally, $I^D_s(D')$ is defined
  as follows:
  \begin{align*}
    & I^D_s(D').S = D.S \\
    & I^D_s(D').N = D.N + D'.N \\
    & I^D_s(D').H(h) = 
  \begin{cases}
    D.H(h)       & \quad \text{if } h < a \\
    D'.H(h - a)+k  & \quad \text{if } a \leq h < b \\
    D.H(h - D'.N)   & \quad \text{if } b \leq h
  \end{cases} \\
  & I^D_s(D').I(h) =
    \begin{cases}
    D.I(h)       & \quad \text{if } h < a \\
    D'.I(h - a)  & \quad \text{if } a \leq h < b \\
    D.I(h - D'.N)   & \quad \text{if } b \leq h
    \end{cases} \\
   & I^D_s(D').O(h) =
    \begin{cases}
    D.O(h)       & \quad \text{if } h < a \\
    D'.O(h - a)  & \quad \text{if } a \leq h < b \\
    D.O(h - D'.N)   & \quad \text{if } b \leq h
  \end{cases}
  \end{align*}
  where $s = s_{a,k}$ and $b = a + D'.N$.
\end{defi}

By abuse of language, if $x \in D'$ is a place, spot, face or component, then we
denote again by $x$ the corresponding place, spot, face or component
in $I^D_s(D')$, as this is unambiguously defined.

\begin{figure}[b]
  \centering
  \begin{subfigure}{0.3\textwidth}
    \centering
    \begin{tikzpicture}[scale=0.5]
      \startdiagram{1}
      \diagslice{0}{0}{1}
      \diagslice{0}{1}{2}
      \diagslice{1}{1}{2}
      \diagslice{0}{2}{0}
      \diagslice{0}{1}{0}
      \finishdiagram
      \node[yspot] at (spot_2_1) {};
      \node[below of=spot_2_1,node distance=.3cm] {$s$};
    \end{tikzpicture}
    \caption{The diagram $D$ with a spot $s \in D$}
  \end{subfigure}
  \hspace{5cm}
  
  \begin{subfigure}{0.3\textwidth}
    \centering
    \begin{tikzpicture}[scale=0.5]
        \startdiagram{1}
        
  \diagslice{0}{0}{1}
  \diagslice{1}{0}{1}
  \diagslice{0}{2}{1}
  \diagslice{0}{1}{0}
  \finishdiagram
    \end{tikzpicture}
    \caption{The diagram $D'$}
    \end{subfigure}
  \begin{subfigure}{0.45\textwidth}
    \vspace{-2cm}
    \centering
    \begin{tikzpicture}[scale=0.5]
      \startdiagram{1}
      \diagslice{0}{0}{1}
      \diagslice{0}{1}{2}

  \diagslice{1}{0}{1}
  \diagslice{2}{0}{1}
  \diagslice{1}{2}{1}
  \diagslice{1}{1}{0}
      
      \diagslice{1}{1}{2}
      \diagslice{0}{2}{0}
      \diagslice{0}{1}{0}
      \finishdiagram
    \end{tikzpicture}
    \caption{The diagram $I^D_s(D')$}
  \end{subfigure}
  \caption{Injection of a diagram in a face}
\end{figure}

\begin{lem} \label{lemma:injection-spot-in-face}
  Let $D$ be a diagram and $s, s' \in D$ be spots in the same face $f \in D$. For all closed diagram $D'$,
  $I^D_s(D') \simeq I^D_{s'}(D')$.
\end{lem}

\begin{proof}
  Let us assume $s$ and $s'$ are adjacent. Let $v$ be the vertex between
  the two slices containing $s$ and $s'$. The vertices of $D'$ in $I_s(D')$ can be successively
  exchanged with $v$, leading to $I^D_{s'}(D')$, which shows that $I^D_s(D') \simeq I^D_{s'}(D')$.
  By induction, this can be repeated for any adjacency path between two spots in the same face.
\end{proof}

\noindent Bearing in mind that this is only defined up to exchange, we can therefore write $I^D_f(D')$
to inject $D'$ anywhere in the face $f$.

\begin{lem} \label{lemma:injection-exchange}
  Injection respects exchanges on the outer and inner diagrams:
  If $D \simeq D'$ and $C \simeq C'$, then for any face $f \in D$
  and its corresponding face $f' \in D'$, $I^D_f(C) \simeq I^{D'}_{f'}(C')$.
\end{lem}

\begin{proof}
  Any exchange $C \rightarrow_R C'$ translates into a single exchange
  $I^D_f(C) \rightarrow_R I^D_f(C')$.  So, by induction, if $C \simeq
  C'$, then $I^D_f(C) \simeq I^D_f(C')$.
  To show that injection respects equivalence on the outer diagram, let $s \in f$ and consider a single rewriting step $D \rightarrow_R D'$.  If $s$ is not in the slice
  between the two vertices $u$ and $v$ being exchanged in $D$, then
  it corresponds to a spot $s' \in D'$. We have $s' \in f'$ and
  $I^D_s(C) \rightarrow_R I^{D'}_{s'}(C)$ in one step again. Otherwise,
  $D'.N$ exchanges are required to move $u$ past $D'$, one to exchange
  $u$ and $v$, and $D'.N$ again to move $v$ past $D'$. So $I^D_s(C)
  \simeq I^{D'}_s(C)$. So injections are compatible with exchanges both on the inner and outer diagram.
\end{proof}

\begin{defi}
  Let $D$ be a diagram and $c \in D$ be a component.
  The \emph{erasure} of $c$ in $D$, denoted by $D - c$, is the diagram obtained by removing from $D$
  any vertex from $c$ or its sub-components.
\end{defi}

\begin{lem} \label{lemma:acyclic-gathering}
  Let $D$ be a diagram and $c \in D$ be an acyclic component.
  Then there is a face $f \in D - c$ such that $D \simeq I^{D-c}_f(c)$.
\end{lem}

\begin{proof}
  Pick a vertex $r \in c$: we will consider $c$ as a tree rooted in $r$. By induction on this tree, we are going to gather all vertices around $r$, meaning that the heights of these vertices in the diagram form an interval.
  
  Say that a vertex $v \in c$ is collapsed if the set of diagram heights of
  the vertices in its subtree form an interval. For any $v \in c$, we
  show that $D$ is equivalent to a diagram $D'$ where $v$ is collapsed
  and such that the vertical order of vertices which are
  outside this subtree is preserved in $D'$.

  If $v$ is a leaf, it is always collapsed.  Consider the case where
  $v$ has children $u_1, \dots, u_n$. Because $c$ is acyclic, it is
  possible to exchange each child $u_i$ with any vertex on a slice between $u_i$
  and $v$, so we can assume that the vertical positions of $v$ and
  $u_1, \dots, u_n$ form an interval. By induction, each $u_i$ can be
  successively collapsed without changing the vertical order of
  vertices which are not in the subtree of $u_i$. Once this is done,
  the vertical position of vertices in the subtrees of the $u_i$ and $v$
  form an interval, so $v$ is collapsed. By doing so we have preserved the vertical ordering of vertices outside the subtree of $v$.

  Therefore, there is a $D'$ where $r$ is collapsed. Let $c'$ be the component corresponding to $c$ in $D'$. Let $f' \in D'$ be the face enclosing $c'$ in $D'$. We have $I^{D'-c'}_{f'}(c') = D'$. By invariance of injection up to exchanges (Lemma~\ref{lemma:injection-exchange}), $I^{D'-c'}_{f'}(c') \simeq I^{D-c}_f(c)$ for the corresponding face $f \in D-c$.
\end{proof}

\begin{defi}
  Let $D$ be a diagram and $f \in D$ be a face that does not neighbour the boundary.  The \emph{erasure} of
  $f$ in $D$, denoted by $D - f$, is the diagram obtained by removing
  all spots in $f$ and descendant faces.
  Formally, let $P(h,i,j) = |\{ s_{h,k} | i \leq k < j, s_{h,k} \in f' \prec^* f \}|$.
  Then $D - f$ is defined as follows:
\medskip

  \begin{align*}
    & (D - f).S = D.S \\
    & (D - f).N = D.N \\
    & (D - f).H(h) = D.H(h) - P(h,0,D.H(h)+1) \\
  \end{align*}
  \begin{align*}
    & (D - f).I(h) = \begin{cases}
      1 \text{ if $D.I(h) = 0$ and $P(h,D.H(h),D.H(h)) = 1$} \\
      D.I(h) - P(h,D.H(h)+1,D.H(h)+D.I(h)+1)
      \text{~otherwise}
    \end{cases} \\
    & (D - f).O(h) = \begin{cases}
      1 \text{ if $D.O(h) = 0$ and $P(h+1,D.H(h),D.H(h)+1) = 1$} \\
      D.O(h) - P(h+1,D.H(h)+1,D.H(h)+D.O(h)+1)
      \text{~otherwise}
      \end{cases}
  \end{align*}
\end{defi}

\bigskip

\noindent One can check that this defines a valid diagam.

\begin{figure}[H]
  \centering
  \begin{subfigure}{0.45\textwidth}
    \centering
    \begin{tikzpicture}[scale=0.5]
      \startdiagram{1}
      \diagslice{0}{0}{3}
      \diagslice{0}{0}{1}
      \diagslice{2}{0}{1}
      \diagslice{0}{2}{1}
      \diagslice{1}{1}{0}
      \diagslice{1}{2}{1}
      \diagslice{0}{2}{0}
      \node at (spot_6_1) {\small $f$};
      \finishdiagram
    \end{tikzpicture}
    \caption{A face $f$ in diagram $D$}
  \end{subfigure}
  \begin{subfigure}{0.45\textwidth}
    \centering
    \begin{tikzpicture}[scale=0.5]
      \startdiagram{1}
      \diagslice{0}{0}{2}
      \diagslice{0}{0}{1}
      \diagslice{1}{1}{1} 
      \diagslice{0}{2}{1}
      \diagslice{0}{1}{1}
      \diagslice{0}{2}{1}
      \diagslice{0}{1}{0}
      \finishdiagram
    \end{tikzpicture}
    \caption{The diagram $D - f$}
  \end{subfigure}  
\end{figure}

\begin{lem} \label{lemma:injection-functoriality}
  Let $s \in D$ be a spot. Then for any closed diagram $c$ and face $f
  \in c$, $I^D_s(c - f) = I^D_s(c) - f$.
\end{lem}

\begin{proof}
  This can be checked directly from the definitions of injections and erasures.
\end{proof}

\begin{lem} \label{lemma:exchange-erasure-lifting}
  Let $D$ be a diagram and $f \in D$ be a face.  For any diagram $D'
  \simeq D - f$, there is a diagram $D_0 \simeq D$ such that $D' = D_0 - f$.
\end{lem}

\begin{proof}
  By inspection of the definition of $D - f$ one can see that any
  exchange on $D - f$ can be lifted back to an exchange at the same
  height on $D$ (the converse is false). By induction, this defines $D_0$.
\end{proof}

\begin{lem} \label{lemma:collapsing}
  Let $D$ be a diagram and $c \in D$ be an arbitrary component (not necessarily acyclic this time).
  Then there is a spot $s \in D - c$ such that $D \simeq I^{D - c}_s(c)$.
\end{lem}

\begin{proof}
  Let $f_1, \dots, f_k$ be all the faces enclosed by $c$.  Consider
  $D' = D - f_1 \dots - f_k$. Let $c'$ be the component corresponding
  to $c$ in $D'$: as a diagram, $c' = c - f_1 \dots - f_k$. As $c'$
  does not enclose any face, $c'$ is acyclic. By
  Lemma~\ref{lemma:acyclic-gathering}, we can gather $c'$ in one spot:
  there is $s \in D' - c'$ such that $D' \simeq I^{D' - c'}_s(c')$. (In fact,
  because all the faces removed from $D$ to obtain $D'$ are enclosed
  by $c$, $D' - c' = D - c$.)  Then, by
  Lemma~\ref{lemma:exchange-erasure-lifting}, there is a $D_0 \simeq
  D$ such that $D_0 - f_1 \dots - f_k = I^{D' - c'}_s(c')$.  By Lemma~\ref{lemma:injection-functoriality}, $I^{D' - c'}_s(c')
  = I^{D' - c'}_s(c - f_1 \dots - f_k) = I^{D' - c'}_s(c) - f_1 \dots - f_k$. Therefore, we obtain
  $D_0 - f_1 \dots - f_k = I_s(c) - f_1 \dots - f_k$ and finally $D_0 = I^{D'-c'}_s(c) = I^{D-c}_s(c)$.
\end{proof}

We can now prove Theorem~\ref{thm:tree-completeness}, showing the completeness of the structural tree for exchanges.

\begin{proof}
  Let $C,D$ be diagrams such that $T(C) = T(D)$.

  To each node $n$ of $T(C)$ we can associate diagrams $C_n$ (respectively $D_n$)
  obtained by erasing vertices not contained in the subtree below $n$ in $C$ (respectively $D$).
  We show by induction on $n$ that $C_n \simeq D_n$.

  If $n$ is a leaf face node, then both $C_n$ and $D_n$ are empty diagrams and are therefore equivalent.
  If $n$ is a leaf component node, then both $C_n$ and $D_n$ are acyclic connected diagrams with identical right normal forms, so they are equivalent.

  If $n$ is an internal face node, let $\{ c_1, \dots, c_m \}$ be its child components. By induction their corresponding diagrams in $C$ and $D$ are pairwise equivalent. We can apply Lemma~\ref{lemma:collapsing} for each of them and express both $C$ and $D$ as iterated injections of the $c_i$ in the empty diagram: therefore $C_n \simeq D_n$.

  If $n$ is an internal component node, let $(f_1, \dots, f_m)$ be its child faces. Again, by induction their corresponding diagrams in $C$ and $D$ are pairwise equivalent. Moreover, the components corresponding to $n$ in $C$ and $D$ have the same right normal form $F$. We can therefore obtain both $C$ and $D$ as iterated injections of the $f_i$ in the faces of $F$, in the designated order. Therefore $C_n \simeq D_n$, which completes the proof.
\end{proof}

\end{document}